\documentclass[a4paper,12pt,english]{article}
\usepackage[T1]{fontenc}
\usepackage{geometry}
\usepackage{amssymb,amsmath}
\usepackage{hyperref}
\usepackage{graphicx}
\usepackage{cite,mcite}

\usepackage[usenames,dvipsnames]{color}

\newcommand{\executeiffilenewer}[3]{%
\ifnum\pdfstrcmp{\pdffilemoddate{#1}}%
{\pdffilemoddate{#2}}>0%
{\immediate\write18{#3}}\fi%
}

\global\parskip 6pt

\begin{document}
\newgeometry{left=1.9cm,right=1.9cm,bottom=3cm}
\pagestyle{empty}
\begin{flushright}    
  {MPP-2015-178}
\end{flushright}

\begin{center}
{\Large{\bfseries Quartic AdS Interactions in Higher-Spin Gravity\\ from Conformal Field Theory
}}
\vskip 0.05\textheight

{\bf Xavier Bekaert\footnote{xavier.bekaert@lmpt.univ-tours.fr}, Johanna Erdmenger\footnote{jke@mpp.mpg.de}, Dmitry Ponomarev\footnote{dmitry.ponomarev@physik.uni-muenchen.de} and Charlotte Sleight\footnote{csleight@mpp.mpg.de}}\\

\vskip 0.5cm
\emph{${}^{1} \:$Laboratoire de Math\'{e}matiques et Physique Th\'{e}orique}\\ 
\emph{Unit\'{e} Mixte de Recherche 7350 du CNRS}\\  
\emph{F\'{e}d\'{e}ration de Recherche 2964 Denis Poisson}\\ 
\emph{Universit\'{e} Fran\c{c}ois Rabelais, Parc de Grandmont, 37200 Tours, France}
\vskip 0.5cm
\emph{${}^{2,\;4}\:$Max-Planck-Institut f\"{u}r Physik (Werner-Heisenberg-Institut)}\\
\emph{F\"{o}hringer Ring 6, D-80805 Munich, Germany}
\vskip 0.5cm
\emph{${}^{3}\:$Arnold Sommerfeld Center for Theoretical Physics}\\ 
\emph{Ludwig-Maximilians University Munich}\\
\emph{Theresienstr. 37, D-80333 Munich, Germany}
\vskip 0.5cm

\end{center}


\begin{center} 
	{\bf Abstract }  
\end{center} 
\begin{quotation}
Clarifying the locality properties of higher-spin gravity is a pressing task, but notoriously difficult due to the absence of a 
weakly-coupled flat regime. The simplest non-trivial case where this question can be addressed is the quartic self-interaction of the AdS scalar field present in the higher-spin multiplet. We investigate this issue in the context of the holographic duality between the minimal bosonic higher-spin theory on AdS$_4$ and the free $O\left(N\right)$ vector model in three dimensions. In particular, we determine the exact explicit form of the derivative expansion of the bulk scalar quartic vertex. The quartic vertex is obtained from the field theory four-point function of the operator dual to the bulk scalar, by making use of our previous results for the Witten diagrams of higher-spin exchanges. This is facilitated by establishing the conformal block expansions of both the boundary four-point function and the dual bulk Witten diagram amplitudes. We show that the vertex we find satisfies a generalised notion of locality. 
\end{quotation}

\newpage
\restoregeometry

\pagestyle{plain}

\tableofcontents

\section{Introduction}
\numberwithin{equation}{section}
\setcounter{footnote}{0}
Despite significant progress in understanding the structure of higher-spin 
gravity theories during the last decades, 
interactions of higher-spin gauge fields are not fully understood. The standard approach to address higher-spin interactions is to use the Noether procedure, which amounts to a perturbative reconstruction
of interactions in a way such that they are compatible with higher-spin gauge transformations, the latter of which may also be 
deformed \cite{Berends:1984rq}. There is a vast literature on cubic 
interactions, see e.g. \cite{Bengtsson:1983pd,Berends:1984wp,Fradkin:1986qy,Fradkin:1987ks,Metsaev:1993ap,Metsaev:1993mj,Bekaert:2005jf,Boulanger:2005br,Metsaev:2005ar,Boulanger:2006gr,Metsaev:2006ui,Fotopoulos:2008ka,Zinoviev:2008ck,Boulanger:2008tg,Manvelyan:2009vy,Bekaert:2010hp,Manvelyan:2010wp,Manvelyan:2010jr,Sagnotti:2010at,Manvelyan:2010je,Fotopoulos:2010ay,Vasilev:2011xf,Joung:2011ww,Joung:2012rv,Boulanger:2012dx}, which in particular gives a complete classification
of local cubic interactions in any dimensions \cite{Metsaev:2005ar}. However, this procedure meets essential technical difficulties
at the quadratic order in a coupling constant, the order at which one extracts quartic vertices.
 The first source of complication is that the number of possible 
candidates for a vertex grows considerably with the order of a vertex, even if the spin of the external fields is fixed. For example, there is only one 
cubic vertex for scalar fields that is non-trivial on the free mass-shell, while the number of independent
non-trivial on-free-shell quartic
vertices is infinite. As we shall demonstrate, these can be parametrised
by two integer parameters taking an infinite set of values. Another issue that complicates the Noether procedure
for quartic vertices, is that at this order the consistency condition becomes non-linear in deformations. In particular, this implies that the general solution for a quartic vertex cannot be constructed as a linear combination of quartic vertices involving fields with fixed spins, as is the case for cubic interactions.
 Instead, such vertices should be considered all together. One of 
the consequences of this analysis is that higher-spin interactions cannot be consistently deformed to
a quartic level unless one includes an infinite tower of fields with unbounded spin. 
This spectacular property was expected since the formulation of the higher-spin interaction problem
\cite{Fronsdal:1978rb}
and comes from the structure of the higher-spin algebra, which is
to large extent unique \cite{Fradkin:1986ka,Boulanger:2013zza}. This complicates the problem
even further. Quartic vertices in higher-spin theory have been studied in \cite{Vasiliev:1989yr,Metsaev:1991mt,Fotopoulos:2010ay,Polyakov:2010sk,Taronna:2011kt,Dempster:2012vw,Buchbinder:2015apa}, and some of these results indicate that local solutions of the consistency conditions in flat space at this order are problematic.

Let us emphasise that locality of interactions is in essence the main constraint of the Noether procedure. In particular,
it has been shown that in relaxing the condition of locality any cubic vertex, consistent till cubic order, can always be completed consistently to all orders \cite{Barnich:1993vg}.
On the other hand, as the analysis of cubic vertices indicates \cite{Metsaev:2005ar}, the minimal number of derivatives in 
a consistent vertex grows with the spin of fields forming the vertex. Together with the fact that higher-spin interactions require fields of arbitrarily high spin, this implies that higher-spin interactions are already non-local at the cubic level. 
In turn, these cubic vertices contribute to the consistency condition for quartic ones. The standard expectation is that non-localities should be present even in quartic interactions of lower-spin fields within higher-spin theories.
To summarise, the Noether procedure without locality admits infinitely many solutions, while
enforcing locality does not seem to admit
any solutions at all.

In the discussion so far, locality was understood in a strong sense: That is,  requiring a finite number of derivatives in the action. A natural way to reconcile locality and the Noether procedure for higher-spin interactions
 could be to extend the definition of locality in such a way that it allows terms with
infinitely many derivatives in the action. At the same time, this weaker definition of locality should rule out
 vertices that give rise to clearly non-local behaviour in the theory. 
It is reasonable to expect that if the coefficients in the derivative expansions of vertices decrease fast enough with the number of derivatives, then such a theory is physically indistinguishable from a local one even if the number of derivatives in the action is
infinite.
More precisely, one can call a vertex local if its scattering amplitude is analytic. In this paper we give some supporting arguments for the extension of this definition to anti-de Sitter (AdS) spacetime.
 For a recent discussion on the locality of field redefinitions in higher-spin theory see \cite{Vasiliev:2015wma}.

Instead of solving the Noether procedure, an alternative approach to the higher-spin interaction problem is to use AdS/CFT to extract bulk interactions from
the conjectured dual boundary conformal field theory (CFT). A decade ago, it was conjectured \cite{Sezgin:2002rt,Klebanov:2002ja}
 that unbroken Vasiliev's higher-spin theory \cite{Vasiliev:1990en,Vasiliev:1992av,Vasiliev:2003ev}
is dual to the free $O(N)$ vector model. This duality has been verified at tree level for 
3-point Witten diagrams in AdS$_4$ \cite{Giombi:2009wh,Giombi:2010vg,Giombi:2012ms} (see also \cite{Petkou:2003zz,Sezgin:2003pt} for an early test),
and at the level of one-loop vacuum energy \cite{Giombi:2013fka,Giombi:2014iua,Giombi:2014yra,Beccaria:2014xda}.
Moreover, all $n$-point functions of the free CFT${}_3$ have been identified
with suitable invariants in Vasiliev's theory \cite{Colombo:2012jx,Didenko:2012tv,Gelfond:2013xt,Didenko:2013bj}. 
Beyond these checks, it remains an open question whether or not it is Vasiliev's theory in particular that is the bulk higher-spin dual of the $O\left(N\right)$ vector model. In any case, one can use holography as a constructive approach to define an interacting
higher-spin gravity theory in AdS.
 Since the dual conformal field theory is free, all boundary correlators are calculable. By equating these to 
 the associated Witten diagrams in the bulk, in principle one can extract interactions of the
 bulk theory order by order.\footnote{Another approach taken in the literature is to use the holographic RG flow \cite{Douglas:2010rc,*Zayas:2013qda,*Sachs:2013pca,*Leigh:2014tza,*Leigh:2014qca,*Mintun:2014gua} (see also \cite{Koch:2014aqa}). While the holographic duality is manifest by construction in these works, the structure of the bulk higher-spin interactions and their locality properties remain to be clarified. For another approach to studying the latter issue, see e.g. \cite{Sarkar:2014dma}.} It is the goal of this paper to uncover the quartic vertex
 involving scalar fields of the bulk higher-spin theory in this manner, and to study its locality properties.
 
Although extracting just the quartic self-interaction of the scalar in higher-spin theory does not fully address the question of quartic higher-spin interactions, the resulting vertex can already be used to probe such issues as locality in 
higher-spin theories. It can also be employed to test the higher-spin AdS/CFT
conjecture at the more non-trivial level of four-point functions, by checking whether or not the vertex we reconstruct holographically matches the one obtained from Vasiliev's equations. To do so, what remains
is to extract the quartic vertex of scalar fields in Vasiliev's theory. So far only the cubic interactions have been extracted in explicit form from Vasiliev's equations \cite{Sezgin:2003pt,Kristiansson:2003xx,Giombi:2009wh,Giombi:2010vg,Chang:2011mz,Kessel:2015kna}.

Another interesting issue that we would like to address is bulk locality in AdS/CFT in
general. While bulk locality above the AdS length scale
is to be expected \cite{Heemskerk:2009pn,Fitzpatrick:2012yx,Fitzpatrick:2014vua}, locality at distances much smaller than this scale is more miraculous.
Criteria for a CFT to have a local bulk dual at the these scales have been proposed in \cite{Heemskerk:2009pn,Fitzpatrick:2010zm,ElShowk:2011ag,Fitzpatrick:2012cg}. The higher-spin holographic duality provides a convenient framework in which they can be investigated, for in this case both the bulk and boundary 
theories are simple enough to make the necessary objects calculable. 

Let us now explain in more detail the approach taken in this paper. We work in the context of the type A minimal bosonic higher-spin theory on AdS$_4$, whose spectrum consists of a parity even scalar and an infinite tower of even spin gauge fields. With the appropriate boundary conditions, this is conjectured to be dual, in the large $N$ limit, to the singlet sector of the \emph{free} massless scalar $O\left(N\right)$ vector model in three dimensions. The spectrum of single-trace operators in the boundary theory consists of the scalar singlet
\begin{equation}
{\cal O} = \phi^a \phi^a,
\end{equation}
of dimension $\Delta = d-2 = 1$, and an infinite set of even spin singlet conserved currents
\begin{equation}
{\cal J}_{i_1 ... i_s} = \phi^a \partial_{\left(i_1\right.} ...\; \partial_{\left.i_s\right)}  \phi^a + ...\,, \qquad s = 2, 4, 6, ... \label{hsc}
\end{equation}
of dimension $\Delta_s = d+s-2 = s+1$. The fields $\phi^a$, where $a= 1, 2, ..., N$, are the free real scalar fields in the fundamental representation of $O(N)$. Under the duality, each spin-$s$ conserved current \eqref{hsc} is dual to a spin-$s$ gauge field $\varphi_{\mu_1 ...\mu_s}$ in AdS$_4$. The scalar singlet ${\cal O} $ is dual to the bulk parity-even scalar, which we denote by $\varphi_0$. The singlet sector also contains multi-trace operators, which are dual to multi-particle states in the bulk. For example, double-trace operators of the schematic form
\begin{align}
\mathcal{O}^{(2)}_{n,i_1 ... i_s}\left(x\right) = \Box^n\big({\cal O}\left(x\right)\big) \partial_{\left(i_1\right.} ...  \partial_{\left.i_s\right)}{\cal O}\left(x\right) + ... 
\end{align}
are bi-linear in the single-trace scalar 
operator ${\cal O}$, with dimension $\Delta_{n,s} = 2\Delta + 2n+s=2+2n+s$. These are dual to two-particle states of the scalar in AdS, and will be of relevance in the present work.

\begin{figure}[h]
 \centering
\includegraphics[width=0.9\linewidth]{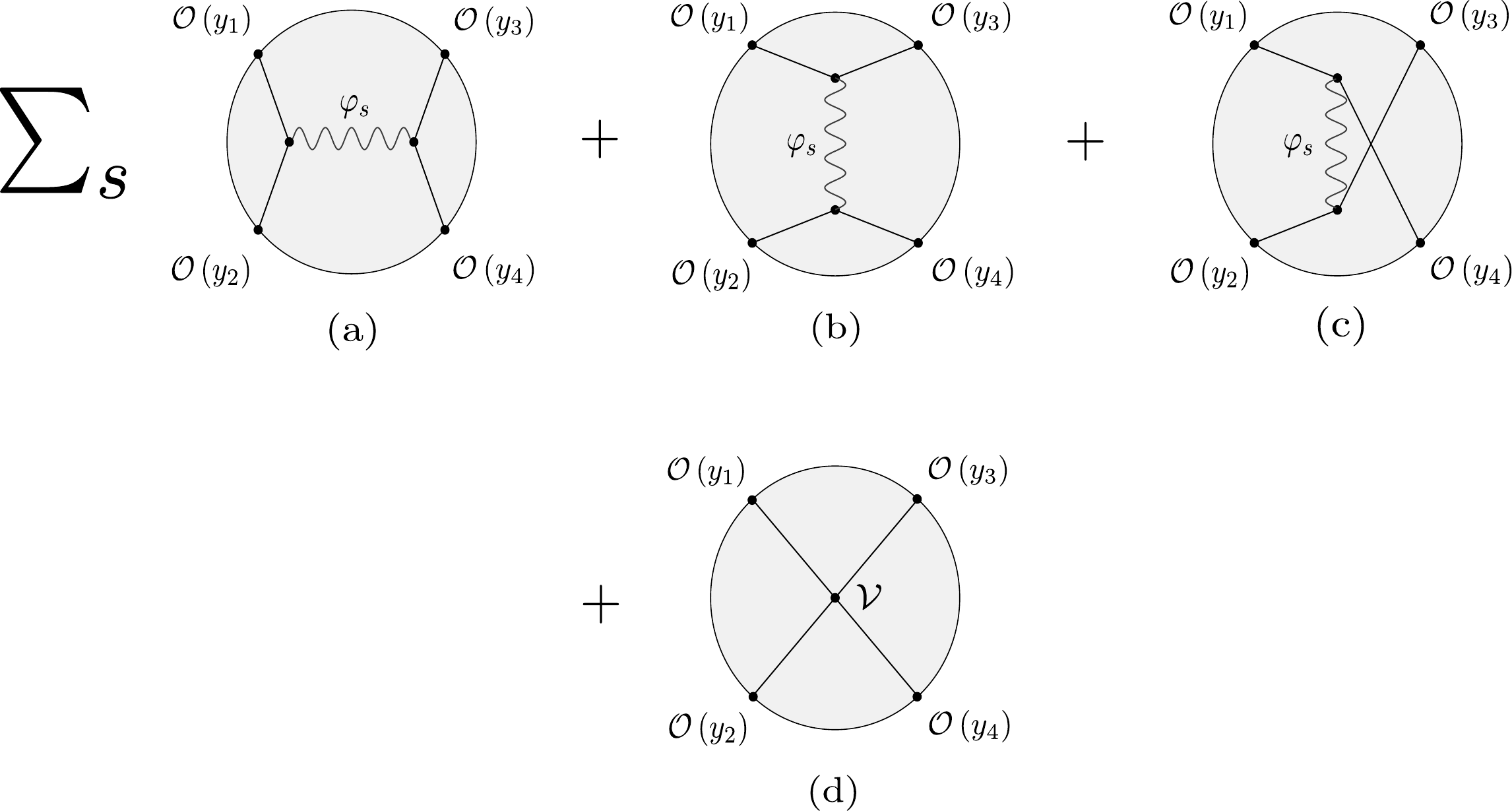}
 \caption{Four-point Witten diagrams contributing to the holographic computation of the connected scalar singlet ${\cal O}$ four-point function. Diagrams (a), (b) and (c) are exchanges of massless spin-$s$ fields between two pairs of the real scalar $\varphi_0$, in the s-, t- and u-channel respectively. The contact diagram (d) is the amplitude associated to the quartic vertex ${\cal V}$ of $\varphi_0$, which we seek to establish by matching with the dual CFT result. }\label{fig::adscft}
 \end{figure}

Correlation functions of the singlet operators may be computed holographically via Witten diagrams involving their dual fields in AdS$_4$. It is this identification of boundary correlation functions with bulk Witten diagrams that we use in this paper to probe the nature of quartic interactions in higher-spin theory on AdS$_4$: In general, the bulk tree-level computation of four-point functions receives contributions from exchange Witten diagrams mediated by cubic interactions, and contact Witten diagrams generated by quartic interactions. While the quartic interactions in higher-spin theory on AdS space are thus far unknown in the metric-like formulation, with the knowledge of the bulk cubic action the exchange diagrams can be computed. Further, when the boundary CFT is free, correlation functions are straightforward to compute via Wick contractions without the need to resort to holographic methods.  Therefore, together with the results for the four-point exchange Witten diagrams, by employing the AdS/CFT correspondence one should be able to infer the bulk quartic interactions that are responsible for the remaining contact amplitudes. In this work we focus on the simplest case of establishing the quartic vertex of the parity even scalar $\varphi_0$ in this manner, and the relevant bulk Witten diagrams are displayed in figure \ref{fig::adscft}. With the appropriate boundary conditions, the latter diagrams comprise the holographic computation of the connected part of the scalar singlet operator ${\cal O}$ in the free scalar $O\left(N\right)$ vector model.

In identifying four-point correlation functions with Witten diagrams to extract bulk quartic vertices, both quantities should be put on an equal footing to facilitate their comparison. An effective framework is to organise both bulk and boundary quantities according to the irreducible $SO\left(d+1,1\right)$  representations of the intermediate states. In the bulk, this is the isometry group of Euclidean AdS$_{d+1}$, while on the boundary this is the $d$-dimensional Euclidean conformal group. In conformal field theory, this is achieved using the operator product expansion (OPE), which allows the decomposition of a given four-point function into conformal blocks. In a given channel, each conformal block $G_{\Delta_s,s}$ represents the contribution to the four-point function of a given primary field (+ all of its descendants) present in the OPE, transforming in the spin-$s$ and dimension $\Delta_s$ representation of the conformal group. For a four-point function of identical scalars ${\cal O}$ of dimension $\Delta$, its conformal block expansion can be expressed in the following contour-integral form \cite{Dobrev:1975ru} in the direct, i.e. (12)(34), channel
\begin{equation}
\langle {\cal O}\left(y_1\right){\cal O}\left(y_2\right){\cal O}\left(y_3\right){\cal O}\left(y_4\right) \rangle = \frac{1}{\left(y^{2}_{12}y^2_{34}\right)^{\Delta}}\left\{1+ \sum\nolimits_{s} \int^{\infty}_{-\infty} d\nu\; f_{s}\left(\nu\right) G_{\tfrac{d}{2}+i\nu,s}\left(u,v\right)\right\},\label{introcb}
\end{equation}
where $u$ and $v$ are cross-ratios: $u = \frac{y^2_{12}y^2_{34}}{y^2_{13}y^2_{24}}$, $v = \frac{y^2_{14}y^2_{23}}{y^2_{13}y^2_{24}}$, and $y_{ij} = y_i - y_j$. The function $f_{s}\left(\nu\right)$ encodes the contributions of the spin-$s$ operators in the $ {\cal O} {\cal O}$ OPE: for each spin-$s$ primary operator present, $f_{s}\left(\nu\right)$ contains a pole located at the value of $\nu$ where the dimension $\tfrac{d}{2}+i\nu$ of the conformal block $G_{d/2+i\nu,s}\left(u,v\right)$ coincides with that of the operator. The residue of $f_{s}\left(\nu\right)$ at this pole gives the square of the operator's OPE coefficient, as in the conventional representation of the conformal block expansion.
\begin{figure}[h]
 \centering
\includegraphics[width=0.9\linewidth]{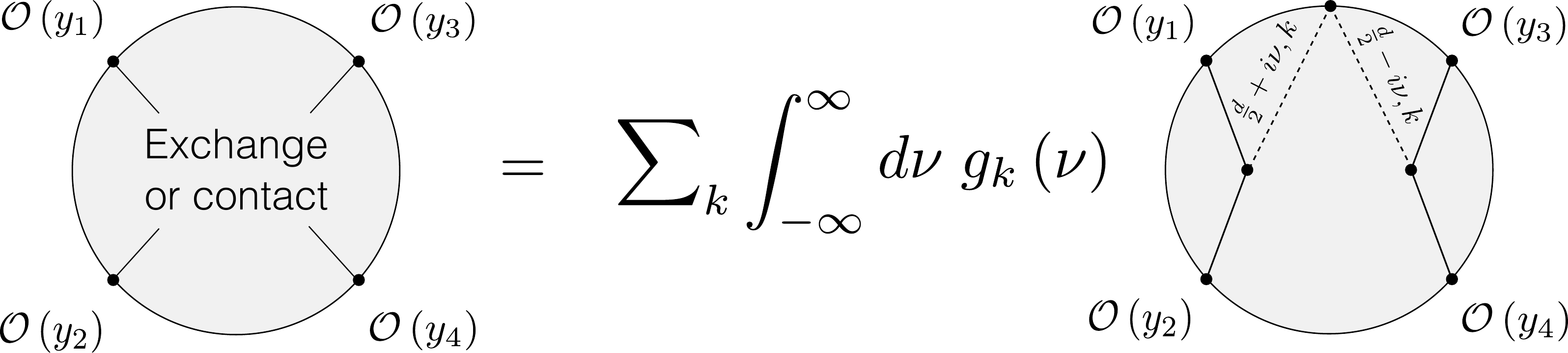}
 \caption{The split representation of four-point bulk amplitudes: A given four-point exchange or contact Witten diagram can be decomposed into products of two three-point diagrams involving two of the original external fields and a spin-$k$ field of dual operator dimension $\tfrac{d}{2} \pm i\nu$, integrating over the common boundary point. Analogous to the conformal block expansion of a CFT correlator \eqref{introcb}, the function $g_{k}\left(\nu\right)$ encodes the intermediate states.}\label{fig::splitint}
 \end{figure}
 
We refer to the analogous form for the bulk four-point Witten diagrams as their \emph{split representation}. This is a decomposition of the four-point bulk amplitudes into products of three-point Witten diagrams, and is illustrated in figure \ref{fig::splitint}.\footnote{While this approach provides the bulk analogue of the conformal block expansion, it does not identify explicitly a bulk quantity that directly corresponds to a boundary conformal block. Such an object was recently defined in \cite{Hijano:2015zsa}.} This approach was employed in \cite{Penedones:2010ue, Paulos:2011ie, Costa:2014kfa} to evaluate scalar and spin-1 exchanges, and the graviton exchange in AdS between two pairs of scalars.\footnote{Historically, the latter exchanges were first computed in \cite{Freedman:1998bj,*Chalmers:1998wu,*Liu:1998ty,*DHoker:1998gd,*DHoker:1998mz,*Chalmers:1999gc,*DHoker:1999pj}. In the context of higher-spins, see \cite{Francia:2007qt,*Francia:2008hd,*Sagnotti:2010jt}.} In the present context of theories describing massless higher-spin fields on AdS, the split representation for the four-point exchange amplitude of a massless spin-$s$ field between two pairs of real scalars on AdS$_{d+1}$ was established in \cite{Bekaert:2014cea}. The computation of exchange diagrams in the latter publication was intended as the first step towards determining the quartic vertex of the scalar in higher-spin theory holographically, and we complete this task in the present paper: By evaluating the three-point amplitudes in the split representations of the Witten diagrams in figure \ref{fig::adscft}, the latter can be converted into boundary conformal block expansions in the contour integral form \eqref{introcb}, allowing for direct comparison with the CFT result. 

As explained previously, it is well known that interactions in higher-spin theory are in general unbounded in their number of derivatives. The quartic vertex of the scalar in higher-spin theory on AdS therefore takes the general form 
\begin{equation}
{\cal V}\left(x\right) = \sum_{m,s} a_{m,s} \left(\varphi_0\left(x\right) \nabla_{\mu_1} ... \nabla_{\mu_s} \varphi_0\left(x\right) + ...\; \right) \Box^m\left(\varphi_0\left(x\right) \nabla^{\mu_1} ... \nabla^{\mu_s} \varphi_0\left(x\right) + ...\; \right), \label{intquart}
\end{equation}
but its properties and behaviour of the coefficients $a_{m,s}$ have thus far remained elusive. Using \eqref{intquart} as an ansatz to complete the total bulk amplitude in figure \ref{fig::adscft} for the dual CFT correlator, we determine the coefficients $a_{m,s}$ required to be consistent with the holographic duality. This requires two further steps that build upon the previous work \cite{Bekaert:2014cea} on the four-point exchange amplitudes: the split representation (and subsequent conformal block expansion) of the most general four-point contact amplitude associated to a quartic vertex of the form \eqref{intquart} must be determined, as well as the conformal block expansion of the scalar single-trace operator four-point function. The latter is only known in $d=4$ CFTs \cite{Dolan:2000ut} in the literature, whereas for the former only partial results are known in certain limits \cite{Heemskerk:2009pn}. While our choice of using the contour integral representation \eqref{introcb} for conformal block expansions is rather unconventional, it is extremely useful for extracting the coefficients in the quartic ansatz \eqref{intquart}. This is because it reduces the comparison between the bulk and boundary conformal block expansions to a matching of the functions $f_{s}\left(\nu\right)$ for each spin $s$.
 
Before the quartic vertex can be determined holographically in this manner, the explicit couplings associated to the cubic vertices appearing in the exchange diagrams need to be uncovered. For the computation of the exchange amplitudes in \cite{Bekaert:2014cea}, these couplings were left arbitrary. We therefore begin in section \ref{sec::cubic} by reviewing the free quadratic Fronsdal action in AdS$_{d+1}$, and its cubic extension to the $0$-$0$-$s$ interactions in the type A minimal bosonic theory. Beyond the level of three-point diagrams, these cubic interactions generate exchanges of the massless spin-$s$ fields. In section \ref{subsec::cubiccoupoing} we fix the couplings of the latter interactions holographically in general dimensions, by matching with corresponding three-point functions in the $d$-dimensional free $O\left(N\right)$ vector model. In section \ref{sec::4ptwitten} we move onto the computation of the tree-level four-point Witten diagrams with four external scalar fields, establishing their split representations and subsequent conformal block expansions. We begin in section \ref{sec::4ptwittenexch} by reviewing our previous work \cite{Bekaert:2014cea} on four-point exchanges of a single massless spin-$s$ field in AdS$_{d+1}$. We then focus on the exchanges in AdS$_4$, and establish their conformal block expansions in the contour integral form \eqref{introcb}. In section \ref{subsec::contactamp} we derive the analogous expansion for a four-point contact amplitude associated to a general quartic vertex \eqref{intquart} of the scalar. To do so we introduce a convenient basis of local quartic vertices, in which any quartic contact interaction of the scalar can be expanded. By computing the conformal block expansion of the amplitudes associated to the basis vertices, the result for a general contact diagram follows. 

In section \ref{sec::CFT} we turn to the CFT side of the story, where in general dimensions we determine the contour integral form \eqref{introcb} of the conformal block expansion of the scalar single-trace operator four-point function. This requires the knowledge of the OPE coefficients of the operators in the scalar singlet OPE, which for the double-trace operators were so far absent in the literature -- with the exception of the $d=4$ case derived in \cite{Dolan:2000ut}. To extend this result to general dimensions, we take the direct approach of extracting the OPE coefficients from the computation of the two-point functions of the double-trace operators, and their three-point functions with the scalar single-trace operator ${\cal O}$. We determine the latter by deriving the explicit form of double-trace operators $\mathcal{O}^{(2)}_{n,s}$ built from two scalar single-trace operators in appendix \ref{appendix::doubletrace}. This result for the double-trace operators completes those already available in the literature \cite{Anselmi:1999bb,Mikhailov:2002bp,Penedones:2010ue,Fitzpatrick:2011dm} for their explicit form.

In section \ref{sec::quartic} we combine the results for the conformal block expansions of the bulk Witten diagrams and the dual CFT four-point function derived in the preceding section, to extract the quartic vertex of the scalar. In particular, we determine a generating function for the coefficients in its derivative expansion \eqref{intquart}. In section \ref{sec::locality} we then probe the nature of the vertex, studying the amplitude of its four-point Witten diagram in order to quantify its locality properties. To do this, we draw on the similarity of Mellin amplitudes to flat space amplitudes. We also comment on the role of holography in the locality of interactions in higher-spin theory duals to CFTs.

\section{Fixing the cubic action}
\label{sec::cubic}
We begin by refining the cubic part of the higher-spin action required to study the quartic vertex of the bulk scalar field holographically. In particular, we fix the couplings of the cubic vertices that mediate the spin-$s$ exchanges in figure \ref{fig::adscft}, by comparing with the corresponding three-point functions in the free conformal scalar $O\left(N\right)$ vector model in $d$-dimensions. This also serves as a simple demonstration of the logic we apply later at the quartic order: We determine interactions in higher-spin theory assuming the validity of the holographic duality, by matching with relevant dual CFT correlation functions. We also take the opportunity to introduce notation.

\subsection{Fronsdal action and cubic vertices of higher-spin gauge fields on AdS}
In this paper we work in Euclidean anti-de Sitter space, which we refer to in the sequel as AdS. We label points in the bulk of AdS by $x^{\mu}$ with $\mu = 0, 1, ..., d$, while points on the conformal boundary $\partial$AdS will be denoted by $y^i$ with $i = 1, ..., d$. 

The theory we are concerned with is the interacting minimal bosonic higher-spin theory on AdS$_4$, whose spectrum consists of a parity even scalar and an infinite tower of gauge fields of even spins $s = 2, 4, 6, ... $.

At the free level, minimal bosonic higher-spin theory is governed by the Frondsal action \cite{Fronsdal:1978rb, Fronsdal:1978vb}\footnote{For the rest of the paper, when expressing tensor contractions through generating functions, setting the auxiliary vector to zero is left implicit.} 
\begin{align}
S_{2} = \frac{s!}{2} \sum\limits^{\infty}_{s=0} \int \sqrt{\left|g\right|} \, d^{d+1}x\; \varphi_{s}\left(x,\partial_u\right)\left(1-\frac{1}{4} \,u^2\, \partial_u \cdot \partial_u \right) \mathcal{F}_{s}\left(x, u, \nabla, \partial_u \right) \varphi_s\left(x, u\right) \Big|_{u=0},\label{FronsdalAdS}
\end{align}
where ${\cal F}_{s}(x,u,\nabla,\partial_u)$ is the Fronsdal operator \cite{Metsaev:1999ui, Mikhailov:2002bp}
\begin{align} \label{Fronsdaltensor}
{\cal F}_{s}(x,u,\nabla,\partial_u)
=
\Box- m^2_s-u^2(\partial_u\cdot \partial_u)
-\;&(u\cdot \nabla)\left((\nabla\cdot\partial_u)-\frac{1}{2}(u\cdot \nabla)
(\partial_u\cdot \partial_u)
\right),
\end{align}
\begin{equation}
m_s^2\equiv
s^2+s(d-5)-2(d-2)
, \notag
\end{equation}
and $\varphi_{s}(x,u)$ is a generating function for the off-shell spin-$s$ Fronsdal field, which is a rank-$s$
symmetric double-traceless tensor 
\begin{equation} \label{gfield}
\varphi_{s}(x,u)\equiv\frac{1}{s!}\, \varphi_{\mu_1\mu_2\dots \mu_s}(x)\,u^{\mu_1}u^{\mu_2}\dots u^{\mu_s},
\qquad
(\partial_u\cdot \partial_u)^2\varphi_{s}(x,u)
= 0.
\end{equation}
In the above, and throughout this paper, we have adopted the use of constant auxiliary vectors to effectively manage symmetric tensor fields. We use $u^{\mu}$ for symmetric tensor fields in the bulk of in AdS, and $z^i$ with $z^2=0$ for symmetric and traceless boundary fields. Our use of this formalism is further explained in appendix \ref{appendix::notation}. 

The quadratic action \eqref{FronsdalAdS} is invariant with respect to the linearised gauge transformations
\begin{equation}
\label{fronsdalgaugetr}
\delta_{0}\varphi_{s}(x,u)=(u\cdot\nabla)\varepsilon_{s-1}(x,u),
\end{equation}
where $\varepsilon_{s-1}(x,u)$ is a generating function for a rank-$\left(s-1\right)$ symmetric and traceless
gauge parameter 
\begin{equation} \label{gaugepara}
\varepsilon_{s-1}(x,u)\equiv\frac{1}{(s-1)!}\, \varepsilon_{\mu_1\mu_2\dots \mu_{s-1}}(x)\,u^{\mu_1}u^{\mu_2}\dots u^{\mu_{s-1}},
\qquad
(\partial_u\cdot \partial_u)\varepsilon_{s-1}(x,u)
= 0.
\end{equation}
To move towards a Lagrangian description of the interacting theory, generally the Noether procedure is applied to determine the possible interactions that are consistent with the gauge symmetries of the theory. For the collection of bulk Witten diagrams we are concerned with in figure \ref{fig::adscft}, the relevant part of the cubic action takes the following form\footnote{I.e. it contains the $0$-$0$-$s$ interactions that mediate the four-point exchange diagrams.}
\begin{align}
S_3 = S_2 -  \sum\limits^{\infty}_{s=0} s!\; g_s \int \sqrt{\left|g\right|}\, d^{d+1}x \; \varphi_s\left(x,\partial_u\right) J_{s}\left(x,u\right), \label{scalaraction}
\end{align}
modulo vertices that vanish on the free mass shell. The $g_s$ are the coupling constants of the $0$-$0$-$s$ cubic interaction, and $J_{s}$ is a bulk spin-$s$ conserved current bi-linear in $\varphi_0$. It has the form \cite{Bekaert:2010hk}
\begin{align} \label{current}
J_{s}\left(x,u\right) = \sum^{s}_{k=0} \frac{\left(-1\right)^k}{k!\left(s-k\right)!} \left(u \cdot \nabla \right)^{s-k} \varphi_0\left(x\right)\left(u \cdot \nabla \right)^{k} \varphi_0\left(x\right) + \Lambda\; u^2 \left( ... \right),
\end{align}
where the second term proportional to the cosmological constant $\Lambda$ is pure trace and vanishes in the flat-space limit, $\Lambda \rightarrow 0$.\footnote{Everywhere else in the paper, we have set $\Lambda = 1$.} 

The Noether approach of determining higher-spin interactions has been successful up to cubic order in the fields, as it has led to the establishment of all consistent cubic vertices. However, beyond this order much less is known in the metric-like formulation. Further, solving the Noether procedure at the cubic level is not sufficient to fix the couplings $g_s$ of the action \eqref{scalaraction}, and thus far their explicit form has not yet been determined.\footnote{See however \cite{Kessel:2015kna} in AdS$_3$ for the $\lambda = \tfrac{1}{2}$ theory, with the general case in $d$-dimensions to appear in \cite{Kessel}. Note that the $\lambda =1$ theory in AdS$_3$ corresponds to the present case of a CFT dual with a free boson.}

The cubic vertex in \eqref{scalaraction} mediates the exchange of higher-spin gauge fields between two pairs of the scalar $\varphi_0$, and is thus required in the computation of the four-point exchange diagrams (a), (b) and (c) in figure \ref{fig::adscft}. Before the quartic interaction of the scalar can be studied holographically through the contact diagrams (d) in figure \ref{fig::adscft}, it is therefore crucial that the couplings $g_s$ are fixed. In the following section we determine the explicit form of these cubic couplings as dictated by the holographic duality, through matching the associated three-point Witten diagrams to the corresponding three-point functions in the free scalar $O\left(N\right)$ vector model. Moreover, the couplings are established for general field theory dimension $d \ge 3$.

\subsection{Fixing the cubic couplings in AdS using holography}
\label{subsec::cubiccoupoing}
The goal of this section is to fix the cubic coupling $g_s$ associated to the cubic interactions
\begin{equation} \label{cubic}
\mathcal{V}^{(3)}_s = s!\; g_{s} \int_{\text{AdS}} \sqrt{\left|g\right|}\; d^{d+1}x\; \varphi_{s}\left(x,\partial_u\right) J_{s}\left(x,u\right),
\end{equation}
according to the duality between the type A minimal bosonic higher-spin theory on AdS$_{d+1}$ and the free scalar $O\left(N\right)$ vector model in $d$ dimensions. This thus fully determines the relevant part of the cubic action \eqref{scalaraction}, which we require to complete the computation of the exchange diagrams in section \ref{sec::4ptwittenexch}.

For the duality with the \emph{free} scalar vector model, we impose on the parity even scalar $\varphi_0$ the higher-spin symmetry preserving boundary condition such that it is dual to the scalar single-trace operator ${\cal O}$ in the free theory, of dimension $\Delta_{{\cal O}} = d-2$. Then, since the vertex \eqref{cubic} is unique up to terms that vanish on the free mass shell,\footnote{For $n$-point Witten diagrams involving only contact interactions, all fields involved are on shell and thus such trivial vertices do not contribute to the amplitude.} the associated three-point Witten diagram gives the holographic computation of the following three-point function in the free scalar vector model\footnote{Throughout this paper, we signify equalities that employ the holographic duality (i.e. those which identify a purely CFT quantity with its bulk counterpart) by $\overset{\text{AdS / CFT}}=$. In other words, such an equality stands for the Gubser-Klebanov-Polyakov-Witten prescription.} 
\begin{align} \label{3ptcoupl}
\langle \mathcal{O}\left(y_1\right) \mathcal{O}\left(y_2\right)\mathcal{J}_s\left(y_3, z\right) \rangle  \overset{\text{AdS / CFT}}= s!\; g_{s} \int_{\text{AdS}} \sqrt{\left|g\right|}\; d^{d+1}x \;\Pi_{d+s-2, s}\left(x,\partial_{u}; y_3, z \right) J_{s}\left(x, u; y_1, y_2\right),
\end{align}
where ${\cal J}_s$ is a boundary spin-$s$ conserved current dual to the bulk spin-$s$ gauge field $\varphi_s$. The right-hand-side (RHS) of \eqref{3ptcoupl} gives the bulk computation of the CFT three-point function on the left-hand-side (LHS), where $\Pi_{d+s-2, s}$ is the bulk-to-boundary propagator of a spin-$s$ gauge field \cite{Mikhailov:2002bp} and $J_{s}\left(x, u; y_1, y_2\right)$ is the bulk current \eqref{current}, with bulk-to-boundary propagators $\Pi_{d-2,0}\left(x,y_1\right)$ and $\Pi_{d-2,0}\left(x,y_2\right)$ for the scalar $\varphi_0$ inserted.\footnote{We use the notation $\Pi_{\Delta,s}$ for the bulk-to-boundary propagators, corresponding to spin-$s$ bulk fields dual to CFT operators of spin-$s$ and dimension $\Delta$.}  Recall that $z$ is an auxiliary vector that encodes traceless and symmetric boundary fields (see appendix \ref{appendix::notation} for details.)

The above three-point Witten diagram was computed in the metric-like formulation in \cite{Bekaert:2014cea} (with auxiliary results in \cite{Costa:2014kfa}), and for convenience in later sections we give the result here for a spin-$s$ bulk-to-boundary propagator $\Pi_{\Delta_s, s}$ with generic dimension $\Delta_s$ (i.e. it is not necessarily the case that $\Delta_s = s+d-2$ )
\begin{align} \label{3pt}
&g_{s} \int_{\text{AdS}} \sqrt{\left|g\right|}\; d^{d+1}x\; s!\; \Pi_{\Delta_s, s}\left(x,\partial_{u}; y_3, z \right) J_{s}\left(x, u; y_1, y_2\right) \\ \nonumber
&\hspace*{4.5cm}= g_{s}\;\frac{ b\left(\Delta_s,s\right)}
{\left(y^2_{12}\right)^{d-2+\frac{s-\Delta_s}{2}}\left(y^2_{13}\right)^{\frac{\Delta_s-s}{2}}\left(y^2_{23}\right)^{\frac{\Delta_s-s}{2}}} \left( \frac{y_{13} \cdot z}{y^2_{13}} -  \frac{y_{23} \cdot z}{y^2_{23}} \right)^s,
\end{align}
where
\begin{align}\label{b}
b\left(\Delta_s,s\right) = \frac{2^{2 s-\frac{5}{2}}\Gamma \left(d-2+\frac{s-\Delta_s}{2} \right) \Gamma \left(\frac{d}{2}-1+\frac{\Delta_s+s-2}{2}\right)\Gamma \left(\frac{s+\Delta_s }{2}\right)^2}{\pi ^{\frac{d}{4}} \Gamma \left(\frac{d}{2}-1\right) \Gamma (d-2) \Gamma (s+\Delta_s )} \sqrt{\frac{\Gamma (\Delta_s -1) (\Delta_s +s-1)}{\Gamma \left(\Delta_s +1-\frac{d}{2}\right)}}.
\end{align}
In the above, we fixed the normalisation such that the bulk-to-boundary propagators give two-point functions of unit norm: 
\begin{align} \label{2ptnorm}
&\langle \mathcal{J}_s\left(y_1, z_1\right)\mathcal{J}_s\left(y_2, z_2\right) \rangle = \frac{1}{\left(y^2_{12}\right)^{d+s-2}} \left(z_1 \cdot z_2 - 2\,\frac{\left(z_1\cdot y_{12}\right) \left(z_2 \cdot y_{12}\right)}{y^2_{12}}\right)^s,
\end{align}
where $\mathcal{J}_0\left(y\right) = \mathcal{O}\left(y\right)$ and $y_{ij} = y_i - y_j$.

The free theory three-point correlation function on the LHS of \eqref{3ptcoupl} is straightforward to determine via Wick contractions. With the normalisation \eqref{2ptnorm}, it is \cite{Diaz:2006nm}
\begin{align} \label{diazdorn}
&\langle \mathcal{O}\left(y_1\right) \mathcal{O}\left(y_2\right) \mathcal{J}_{s}\left(y_3, z\right) \rangle = \frac{2^{\tfrac{s+3}{2}}}{\sqrt{s!} \sqrt{N}} \frac{\left(\frac{d}{2}-1\right)_s}{\sqrt{\left(d+s-3\right)_s}}\left( \frac{y_{13} \cdot z}{y^2_{13}} -  \frac{y_{23} \cdot z}{y^2_{23}} \right)^s \frac{1}
{\left(y^2_{12}\right)^{\tfrac{d}{2}-1}\left(y^2_{13}\right)^{\tfrac{d}{2}-1}\left(y^2_{23}\right)^{\tfrac{d}{2}-1}},
\end{align}
where $\left(a\right)_{r} = \Gamma\left(a+r\right)/\Gamma\left(a\right)$ is the rising Pochhammer symbol.

The holographic duality implies that equation \eqref{3ptcoupl} identifying the bulk three-point Witten diagram \eqref{3pt} with $\Delta_s = s+d-2$, and the CFT correlator \eqref{diazdorn}, holds. This dictates that the cubic couplings for the interaction \eqref{cubic} have the following explicit form:  
\begin{align} \label{couplgend}
\boxed{g_{s} =  \frac{2^{\tfrac{3d-s-1}{2}}\pi^{\tfrac{d-3}{4}}\Gamma\left(\frac{d-1}{2}\right)\sqrt{\Gamma\left(s+\tfrac{d}{2}-\tfrac{1}{2}\right)}}{\sqrt{N}\sqrt{s!}\,\Gamma\left(d+s-3\right)}. }
\end{align}
In particular, for $d=3$ we have 
\begin{align} \label{coupl3d}
g_{s} =  \frac{2^{4-\frac{s}{2}}}{\sqrt{N}\,\Gamma\left(s\right)}.
\end{align}
This is consistent with the known vanishing of the cubic scalar self coupling in AdS$_4$ \cite{Sezgin:2003pt}, which itself provided a non-trivial check of the duality by comparing with earlier CFT results in \cite{Petkou:1994ad}.

These results for the cubic couplings can now be used to complete the computation \cite{Bekaert:2014cea} of the corresponding exchange Witten diagrams in type A minimal bosonic higher spin theory on AdS$_{d+1}$, which we do in the following section. In the latter publication, the $0$-$0$-$s$ cubic couplings were left arbitrary.

\section{Four-point Witten diagrams}
\label{sec::4ptwitten}
\subsection{Exchange diagrams}
\label{sec::4ptwittenexch}
The tree-level amplitude for the exchange of a single massless higher-spin field between two real scalars on an AdS background was originally computed for arbitrary cubic coupling in \cite{Bekaert:2014cea}, for general space-time dimension. In particular, the exchanges were decomposed into products of three-point amplitudes, which we refer to as their \emph{split representation}. This is briefly reviewed for a single spin-$s$ exchange for even $s$\footnote{For odd spins the exchange is zero, since there are no non-trivial conserved currents of odd spin that are bi-linear in \emph{real} scalar fields. Indeed, in the present context of the \emph{minimal} bosonic higher-spin theory, there are only gauge fields of even spin in the spectrum.} before focusing on those in AdS$_4$, where certain useful features emerge which are not present in the general dimensional case. We further show how the exchange amplitudes can then be expressed as a conformal block expansion \eqref{introcb} on the boundary of AdS. This way of representing the four-point bulk amplitudes will prove significant in determining the bulk quartic vertex of the scalar in higher-spin theory holographically in section \ref{sec::quartic}, as it allows us to effectively compare with the dual CFT.

The exchange of a spin-$s$ gauge field takes place in the s-, t- and u-channels. It is mediated by the cubic interaction \eqref{cubic}, whose couplings $g_s$ we fixed holographically in section \ref{subsec::cubiccoupoing}. The exchange diagrams are depicted in figures \ref{fig::adscft}. (a), (b) and (c), and in the following we focus on the computation of the s-channel exchange amplitude associated to figure \ref{fig::adscft} (a). Applying the usual recipe for tree-level bulk Witten diagrams, the amplitude takes the form
\begin{align} \label{sexch}
&\mathcal{A}^{\text{exch.}}_{s}\left(y_1,y_2; y_3,y_4\right)\\ \nonumber 
&\hspace*{0.5cm}= g^{2}_{s} \int_{\text{AdS}} \sqrt{\left|g\right|}\; d^{d+1}x_1 \int_{\text{AdS}} \sqrt{\left|g\right|}\; d^{d+1}x_2\; \Pi_{s}\left(x_1,\partial_{u_1};x_2,\partial_{u_2}\right) J_{s}\left(x_1, u_1; y_1, y_2\right) J_{s}\left(x_1, u_2; y_3, y_4\right)
\end{align}
where $\Pi_{s}$ is the bulk-to-bulk propagator for a massless spin-$s$ field. The t- and u-channel exchanges follow analogously from the s-channel case, and we comment on them briefly towards the end of this subsection.

\begin{figure}[h]
 \centering
\includegraphics[width=0.85\linewidth]{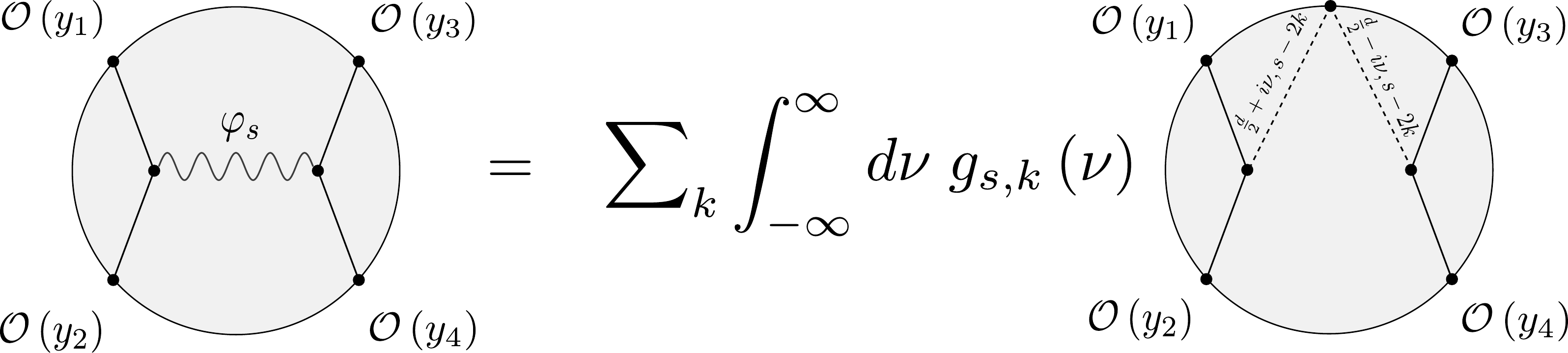}
 \caption{The split representation of exchange Witten diagrams: In expressing the bulk-to-bulk propagators in a basis of harmonic functions \eqref{split}, the exchange amplitude decomposes into products of two three-point Witten diagrams. These involve two of the original external fields and a field of dual operator dimension $\tfrac{d}{2} \pm i\nu$, integrated over their common boundary point.}\label{fig::splitexchsch}
 \end{figure}

To derive the split representation of the exchange amplitude \eqref{sexch}, it is conducive to express the massless spin-$s$ bulk-to-bulk propagator in a basis of bi-tensorial eigenfunctions $\Omega_{\nu,s}$ of the Laplace operator, where $\nu \in \mathbb{R}$ and spin-$s$ label the representation of $SO\left(d,2\right)$. The function $\Omega_{\nu,s}$ can be written as a product of two spin-$s$ bulk-to-boundary propagators of dimensions $\tfrac{d}{2}\pm i\nu$, integrated over the common boundary point
 (see \cite{Penedones:2007ns,Costa:2014kfa} and references therein)\footnote{In accordance with their effect on the form of the exchange amplitude, this way of representing bulk-to-bulk propagators in terms of products of bulk-to-boundary propagators is also referred to as their split representation. Early literature on this form for bulk-to-bulk propagators includes \cite{Fronsdal:1974ew, Dobrev:1998md,Leonhardt:2003qu,Leonhardt:2003sn}, where it is motivated group theoretically. In \cite{Costa:2014kfa} the split representations of the traceless part of massive spin-$s$ bulk-to-bulk propagators and the graviton propagator were established. The split form of the scalar and spin-1 propagators was derived in \cite{Penedones:2010ue,Paulos:2011ie}.}
 \begin{equation} \label{split}
\Omega_{\nu,s}\left(x_1,u_1;x_2,u_2\right) = \frac{\nu^2}{\pi\, s!\,\left(\tfrac{d}{2}-1\right)_{s}}\, \int_{\partial \text{AdS}} \; d^{d}y\;\Pi_{\tfrac{d}{2} + i\nu, s}(x_1,u_1; y, \hat{\partial}_z)\,\Pi_{\tfrac{d}{2} - i\nu, s}\left(x_2,u_1; y, z \right)\,.
\end{equation}
 Owing to this property, the exchange amplitude then decomposes into products of three-point Witten diagrams, which are straightforward to evaluate. This approach is illustrated schematically in figure \ref{fig::splitexchsch}.\footnote{In this sense, expressing four-point bulk amplitudes in a basis of eigenfunctions of the Laplace operator is the analogue in the bulk of decomposing CFT four-point functions in terms of eigenfunctions of the quadratic conformal Casmir, viz. the conformal block expansion.}
 
This method of computing higher-spin four-point exchange amplitudes was taken in \cite{Bekaert:2014cea}, for three different gauges of the spin-$s$ field. For reasons that will become clear shortly, in AdS$_4$ a convenient gauge to use for the spin-$s$ bulk-to-bulk propagator is the one that makes the trace structure manifest (see sections 3.5 and 4.3 of \cite{Bekaert:2014cea}). In this gauge, the propagator reads
\begin{equation} \label{mtg}
\Pi_{s}\left(x_1,u_1;x_2,u_2\right)  = \sum\limits^{\left[\tfrac{s}{2}\right]}_{k=0} \int^{\infty}_{-\infty} d\nu \;g_{s,k}\left(\nu\right) \left(u^2_1\right)^k  \left(u^2_2\right)^k \Omega_{\nu, s-2k}\left(x_1,u_1;x_2,u_2\right) ,
\end{equation}
where the $\left(u^2\right)^k$ represent $k$ symmetric products of the background metric. The coefficients $g_{s,k}$ are given by
\begin{align}
\notag
g_{s,0}(\nu)&=\frac{1}{(\tfrac{d}{2}+s-2)^2+\nu^2},\\ 
\label{answertrgauge}
g_{s,k}(\nu)&=-\,\frac{\left(\frac{1}{2}\right)_{k-1}}{2^{2k+3}\; k!}\frac{(s-2k+1)_{2k}}{(\tfrac{d}{2}+s-2k)_k (\tfrac{d}{2}+s-k-\tfrac{3}{2})_k}
\frac{\left(\frac{\frac{d}{2}+s-2k+i\nu}{2}\right)_{k-1}\left(\frac{\frac{d}{2}+s-2k-i\nu}{2}\right)_{k-1}}{\left(\frac{\frac{d}{2}+s-2k+1+i\nu}{2}\right)_{k}\left(\frac{\frac{d}{2}+s-2k+1-i\nu}{2}\right)_{k}}, \quad k\ne 0.
\end{align}
As opposed to bulk-to-boundary propagators, bulk-to-bulk propagators are not on-shell (i.e. are not traceless and transverse). In the split representation, this fact manifests itself in contributions from harmonic functions of lower spin. For the massless spin-$s$ propagator in the manifest trace gauge, these are given by the $k>0$ terms in \eqref{mtg}.

In the above gauge, the s-channel exchange amplitude \eqref{sexch} decomposes as
\begin{align} \label{exch}
&\mathcal{A}^{\text{exch.}}_{s}\left(y_1,y_2; y_3,y_4\right)\\ \nonumber 
&\hspace*{2cm}= g^{2}_{s} \sum\limits^{\left[\tfrac{s}{2}\right]}_{k=0} \int^{\infty}_{-\infty} d\nu \;g_{s,k}\left(\nu\right)  \int_{\text{AdS}} \sqrt{\left|g\right|}\; d^{d+1}x_1\; \int_{\text{AdS}} \sqrt{\left|g\right|}\; d^{d+1}x_2\; s!^2\; \Omega_{\nu, s-2k}\left(x_1,\partial_{u_1}; x_2,\partial_{u_2}\right) \\ \nonumber
& \hspace*{4cm} \times \left(\partial_{u_1} \cdot \partial_{u_1} \right)^k J_{s}\left(x_1, u_1; y_1, y_2\right) \left(\partial_{u_2} \cdot \partial_{u_2} \right)^k J_{s}\left(x_2, u_2; y_3, y_4\right),
\end{align}
where $ \left(\partial_{u} \cdot \partial_{u} \right)^k J_{s}$ denotes the $k$-th trace of $J_{s}$. The complete evaluation of \eqref{exch} in general dimensions is given in \cite{Bekaert:2014cea}, but at this point we depart from the general dimensional analysis to focus on the particular case of AdS$_4$.

The suitability of the gauge choice \eqref{mtg} to AdS$_4$ becomes apparent when we recall that the bulk scalar $\varphi_0$ is conformal in this particular dimension. As a consequence, there is some freedom which allows us to take the currents $J_{s}$ to be traceless on-shell, i.e. $ \left(\partial_{u} \cdot \partial_{u} \right)^k J_{s} = 0$ for $k > 0$ when the scalar fields are on their mass shell. This provides a significant simplification, because when the currents are traceless just the $k=0$ term in \eqref{exch} contributes. With this choice the exchange amplitude in AdS$_4$ takes the form
\begin{align} \label{exchd4}
\mathcal{A}^{\text{exch.}}_{s}\left(y_1,y_2; y_3,y_4\right) &= \int^{\infty}_{-\infty} d\nu \; \frac{1}{\nu^2+\left(s-\tfrac{1}{2}\right)^2} \frac{\nu^2}{\pi} \int_{\partial \text{AdS}} d^{3}y \; \frac{1}{s!\left(\tfrac{1}{2}\right)_{s}} \\ \nonumber 
& \hspace*{2cm} \times g_{s} \int_{\text{AdS}}\sqrt{\left|g\right|}\; d^{4}x_1\;s!\; \Pi_{\tfrac{3}{2} + i\nu, s}(x_1,\partial_{u_1}; y, \hat{\partial}_z) J_{s}\left(x_1, u_1; y_1, y_2\right) \\ \nonumber 
& \hspace*{2cm} \times g_{s} \int_{\text{AdS}} \sqrt{\left|g\right|}\; d^{4}x_2\;s!\; \Pi_{\tfrac{3}{2} - i\nu, s}\left(x_2,\partial_{u_2}; y, z \right) J_{s}\left(x_2, u_2; y_3, y_4\right),
\end{align}
where we have inserted the representation \eqref{split} for $\Omega_{\nu, s}$. The amplitude can then be expressed in terms of boundary quantities by using the result \eqref{3pt} for the three-point Witten diagrams, with $\Delta_s = \tfrac{3}{2} \pm i\nu$:\footnote{Note that here the normalisation of the spin-$s$ bulk-to-boundary propagator is different to that used in \eqref{3pt}. It is given by equation (3.53) in \cite{Bekaert:2014cea}, and is required to be compatible with the spin-$s$ bulk-to-bulk propagator \eqref{mtg}. All normalisations used here are consistent with the canonical normalisation of the kinetic term in the cubic action \eqref{scalaraction}.}
\begin{align} \nonumber
&\mathcal{A}^{\text{exch.}}_{s}\left(y_1,y_2; y_3,y_4\right) = g^2_s \;\int^{\infty}_{-\infty} d\nu \; \frac{\nu^2+\left(s+\frac{1}{2}\right)^2}{\nu^2+\left(s-\tfrac{1}{2}\right)^2}\; \frac{b\left(\tfrac{3}{2}+i\nu,s\right)  b\left(\tfrac{3}{2}-i\nu,s\right)\Gamma \left(\frac{1}{2}-i \nu \right) \Gamma \left(i \nu +\frac{1}{2}\right) }{4 \pi ^4  \Gamma (-i \nu ) \Gamma (i \nu)} \\ \label{3ptprod}
& \times \frac{1}{s!\left(\tfrac{1}{2}\right)_{s}} \int_{\partial \text{AdS}} d^{3}y \; \frac{(y^2_{20}\;(y_{10} \cdot \hat{\partial}_z) -  y^2_{10}\;(y_{20} \cdot \hat{\partial}_z) )^s}
{\left(y^2_{12}\right)^{\frac{2s-2 i \nu+1}{4}}\left(y^2_{10}\right)^{\frac{2s-2 i \nu+3}{4}}\left(y^2_{20}\right)^{\frac{2s-2 i \nu+3}{4}}} 
 \frac{\left(y^2_{40}\;\left(y_{30} \cdot z\right) -  y^2_{30}\;\left(y_{40} \cdot z\right) \right)^s}
{\left(y^2_{34}\right)^{\frac{2s+2 i \nu+1}{4}}\left(y^2_{30}\right)^{\frac{2s+2 i \nu+3}{4}}\left(y^2_{40}\right)^{\frac{2s+2 i \nu+3}{4}}},
\end{align}
where here $y_{i0} = y_i - y$, and the bulk three-point amplitude normalisations $b\left(\frac{3}{2} \pm i\nu,s\right)$ were defined by equation \eqref{b} in section \eqref{subsec::cubiccoupoing}.
 Note that the second line is a product of three-point functions of unit norm, integrated over their common boundary point.

To make contact with the dual CFT four-point function, it will be useful to express the amplitude \eqref{3ptprod} in terms of conformal blocks. It is well known that a boundary integral of the form shown in second line of \eqref{3ptprod} can be expressed in terms of conformal blocks corresponding to a given representation of the conformal group and its shadow \cite{Dolan:2011dv,Costa:2014kfa}. In our case, the precise relation is
\begin{align}\label{integcb}
&\frac{1}{s!\left(\tfrac{1}{2}\right)_{s}} \int_{\partial \text{AdS}} d^{3}y \; \frac{(y^2_{20}\;(y_{10} \cdot \hat{\partial}_z) -  y^2_{10}\;(y_{20} \cdot \hat{\partial}_z) )^s}
{\left(y^2_{12}\right)^{\frac{2s-2 i \nu+1}{4}}\left(y^2_{10}\right)^{\frac{2s-2 i \nu+3}{4}}\left(y^2_{20}\right)^{\frac{2s-2 i \nu+3}{4}}} 
 \frac{\left(y^2_{40}\;\left(y_{30} \cdot z\right) -  y^2_{30}\;\left(y_{40} \cdot z\right) \right)^s}
{\left(y^2_{34}\right)^{\frac{2s+2 i \nu+1}{4}}\left(y^2_{30}\right)^{\frac{2s+2 i \nu+3}{4}}\left(y^2_{40}\right)^{\frac{2s+2 i \nu+3}{4}}} \\ \nonumber
& = \frac{1}{y^2_{12}y^2_{34}}\left[\: K_{i \nu, s}\;G_{\frac{3}{2}+i\nu}\left(u,v\right) \quad + \quad (\nu \leftrightarrow -\nu) \: \right],
\end{align}
where 
\begin{align}
K_{i \nu, s} =  \frac{ \pi ^{5/2} 2^{-2 (i \nu +s-1)}\Gamma (-i \nu ) \Gamma \left(s-i \nu +\frac{1}{2}\right) \Gamma \left(s+i \nu +\frac{1}{2}\right)}{\Gamma \left(\frac{1}{2}-i \nu \right) (2 i \nu +2 s+1) \Gamma \left(\frac{2 s-2 i \nu +3}{4}\right)^2 \Gamma \left(\frac{2 s+2 i \nu +1}{4} \right)^2},
\end{align}
and $G_{3/2+i\nu,s}\left(u,v\right)$ is a direct (i.e. (12)(34)) channel conformal block, of dimension $3/2+i\nu$ and spin-$s$. The second term with $\nu \leftrightarrow -\nu$ gives the contribution from the shadow representation, whose contribution to the bulk amplitude is identical owing to the integration over $\nu$. Using this relationship between the boundary integral in \eqref{3ptprod} and conformal blocks, the s-channel exchange amplitude in AdS$_4$ can be expressed as the following conformal block expansion in the direct-channel 
\begin{align} \label{cbexch}
&\mathcal{A}^{\text{exch.}}_{s}\left(y_1,y_2; y_3,y_4\right) = \frac{1}{y^{2}_{12}y^{2}_{34}} \frac{2^{8-s}}{N\Gamma\left(s\right)^2}\int^{\infty}_{-\infty} d\nu \; \frac{1}{ \nu ^2+(s-\frac{1}{2})^2} \,\kappa_s(\nu)\, G_{\tfrac{3}{2}+i \nu,s}\left(u,v\right),
\end{align}
where we have inserted the explicit expression \eqref{coupl3d} for the cubic coupling $g_s$, and the factor
\begin{equation}
\label{kappa}
\kappa_s(\nu)=
\frac{2^{-2 i \nu +2s-3}\, \Gamma \left(i \nu +\frac{1}{2}\right) \Gamma \left(\frac{2 s-2 i \nu +1}{4} \right)^2\Gamma \left(\frac{2 s+2 i \nu +3}{4} \right)^2}{\pi ^{5/2} \Gamma (i \nu ) (2 i \nu +2 s+1)}
\end{equation}
in the integrand will keep appearing when representing various relevant amplitudes as contour integrals.

This completes the computation of the amplitude for the exchange of a massless spin-$s$ field on AdS$_4$ between two pairs of the real scalar $\varphi_0$ in the s-channel, with the result expressed as a conformal block expansion \eqref{cbexch} in the direct-channel. For the t- and u-channel exchanges (figures \ref{fig::adscft}. (b) and \ref{fig::adscft}. (c)), the process follows much in the same way. In fact, since the external scalars are not distinct, explicit computation of their amplitudes can be avoided: the t- and u-channel amplitudes can be obtained by permuting the external legs of the s-channel result \eqref{cbexch}. Namely, to acquire the t-channel amplitude one should exchange $y_2 \leftrightarrow y_3$ in \eqref{cbexch}, and the result is thus given by 
\begin{align} \label{cbexcht}
&\mathcal{A}^{\text{exch.}}_{s}\left(y_1,y_3; y_2,y_4\right) = \frac{1}{y^{2}_{13}y^{2}_{24}} \frac{2^{8-s}}{N\Gamma\left(s\right)^2} \int^{\infty}_{-\infty} d\nu \; \frac{1}{ \nu ^2+(s-\frac{1}{2})^2}
\,\kappa_s(\nu)\,G_{\tfrac{3}{2}+i \nu,s}\left(\tfrac{1}{u},\tfrac{v}{u}\right),
\end{align}
which is a (13)(24) crossed-channel expansion. Likewise, for the u-channel we exchange: $y_2 \leftrightarrow y_4$, giving the amplitude $\mathcal{A}^{\text{exch.}}_{s}\left(y_1,y_4; y_2,y_3\right)$. The latter amplitude is thus a (14)(23) crossed-channel expansion conformal blocks $G_{3/2+i \nu,s}\left(v,u\right)$.

\subsection{Contact diagrams}
\label{subsec::contactamp}
In this section we show how four-point Witten diagrams associated to quartic contact interactions of the scalar $\varphi_0$ can be evaluated. In particular, as with the exchange diagrams in the previous section, we demonstrate how to derive the split representation of the amplitudes, and subsequently their conformal block expansions in a single channel. The results derived below for a contact diagram associated to a general quartic vertex will be instrumental in section \ref{sec::quartic}, where by matching the four-point Witten diagrams with external $\varphi_0$ to the corresponding CFT four-point function, we determine the quartic vertex of $\varphi_0$ in AdS$_4$.
\begin{figure}[h]
 \centering
\includegraphics[width=0.6\linewidth]{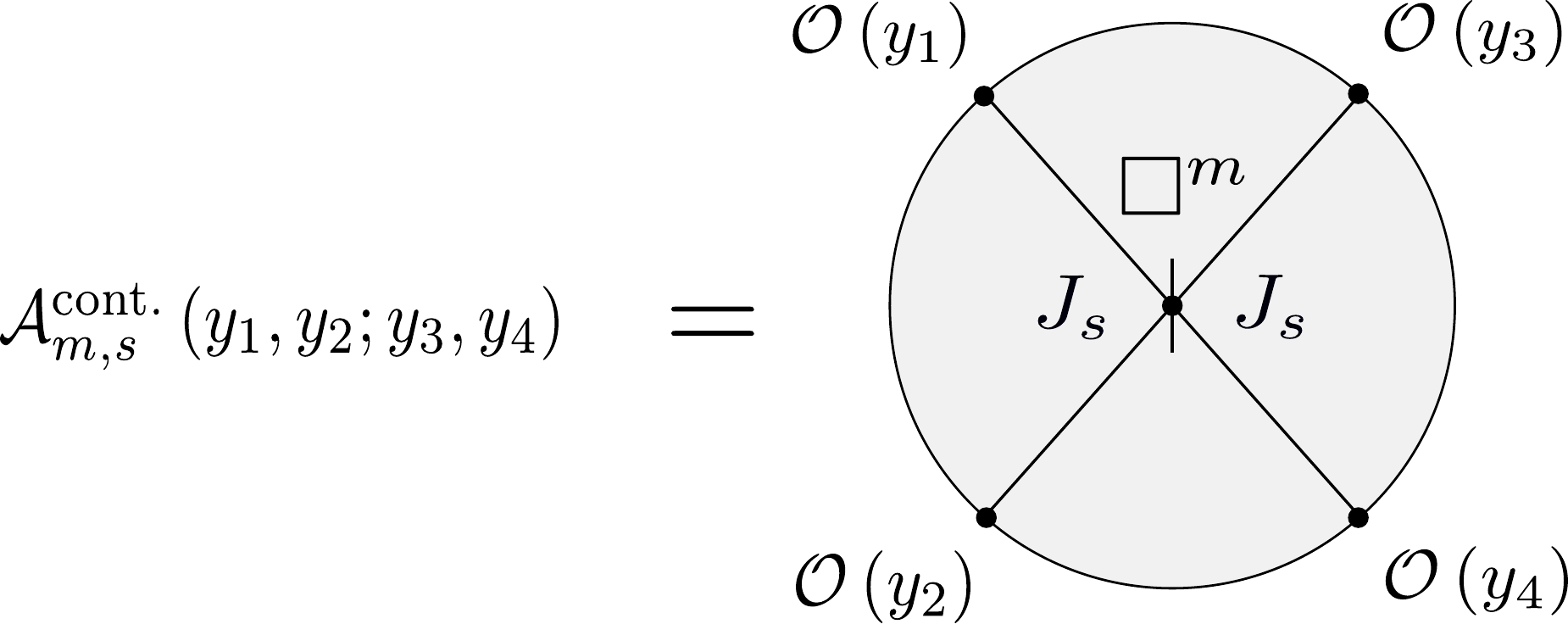}
 \caption{Four-point contact diagram generated by a quartic vertex ${\cal V}_{m,s}$ of the basis \eqref{quartbasis}. The vertical line through the interaction point serves to illustrate that, for the given labeling of external legs, 
the quartic interaction corresponds to gluing a bulk current $J_s$, associated the two external legs $y_{1,2}$, to
another bulk current $J_s$, associated with the two remaining external legs $y_{3,4}$, with a power of the Laplacian $\Box^m$
in between them, c.f. \eqref{::contactsch}. In accordance with the Feynman rules, the remaining diagrams contributing to the total contact amplitude for this vertex can be obtained from this amplitude by permuting the external legs.}\label{fig::contactsch}
 \end{figure} 
 
In general, the computation of a four-point contact Witten diagram associated to an arbitrary quartic vertex is very involved, owing to the manipulations required of non-commuting covariant derivatives in AdS acting on the fields in the vertex. However, a given quartic contact interaction of $\varphi_0$ can be expressed in terms of a basis of particular local quartic vertices, whose corresponding amplitudes in the split representations can easily be evaluated. The basis is given by the set
\begin{align}
\mathcal{V}_{m,s} = J_{s}\left(x, \partial_u\right) \Box^m \big(J_{s}\left(x, u\right)\big), \qquad  s = 2k,\quad k \ge m \ge 0,\quad k, m \in \mathbb{N}, \label{quartbasis}
\end{align}
where $J_{s}$ is the traceless spin-$s$ conserved current bi-linear in the scalar $\varphi_0$, introduced in the previous section. In appendix \ref{appendix::quarticbasis}, we show that in AdS$_4$ they account for all independent vertices quartic in the scalar. A general quartic vertex of $\varphi_0$  can therefore be expressed in the form
\begin{equation}
{\cal V}\left(x\right) = \sum\nolimits a_{m,s}\; {\cal V}_{m,s}\left(x\right), \qquad  a_{m,s} \in \mathbb{R} \label{quartansatz}
\end{equation}
for some coefficients $a_{m,s}$. The efficacy of working in this basis to compute four-point contact Witten diagrams is explained in the following.

Consider the Witten diagram associated to the vertex $\mathcal{V}_{m,s}$, for the following permutation of the external $y_i$ (as shown in figure \ref{fig::contactsch})
\begin{align}\label{::contactsch}
\mathcal{A}^{\text{cont.}}_{m,s}\left(y_1,y_2; y_3,y_4\right) = \int_{\text{AdS}} \sqrt{\left|g\right|}\; d^{4}x\; J_{s}\left(x, \partial_u; y_1, y_2\right) \Box^m \big(J_{s}\left(x, u; y_3, y_4\right)\big).
\end{align}
The first step towards establishing the split representation (i.e. factorisation into products of three-point Witten diagrams) of the amplitude, is to introduce a second bulk integral in a way that pair-wise separates the external points $y_i$ between the two integrals. This can be achieved by inserting a Dirac delta function. How it is inserted, dictates the way in which the external points $y_i$ are distributed between the two bulk integrals. For reasons that will become clear shortly, for the given permutation of the external points it is most strategic to place the delta function such that the pairs $\left(y_1, y_2\right)$ and $\left(y_3, y_4\right)$ are separated. To wit,
\begin{align}
&\mathcal{A}^{\text{cont.}}_{m,s}\left(y_1,y_2; y_3,y_4\right)\\ \nonumber
& = \int_{\text{AdS}} \sqrt{\left|g\right|}\; d^{4}x_1\; J_{s}\left(x_1, \partial_{u_1}; y_1, y_2\right)  \int_{\text{AdS}} \sqrt{\left|g\right|}\; d^{4}x_2\;  \Box^m \big(J_{s}\left(x_2, \partial_{u_2}; y_3, y_4\right)\big) \left(u_1 \cdot u_2\right)^s \delta^{4}\left(x_1-x_2\right).
\end{align}

The suitability of the basis $\mathcal{V}_{m,s}$ emerges when one inserts the completeness relation \eqref{complete1} for the harmonic functions $\Omega_{\nu, \ell}$. Since the currents $J_{s}$ are conserved \emph{and} traceless, only the harmonic function with $\ell=s$ contributes and one obtains
\begin{align} 
&\mathcal{A}^{\text{cont.}}_{m,s}\left(y_1,y_2; y_3,y_4\right)\\ \nonumber
& = \int^{\infty}_{-\infty} d\nu\int_{\text{AdS}} \sqrt{\left|g\right|}\; d^{4}x_1\; J_{s}\left(x_1, \partial_{u_1}; y_1, y_2\right)  \int_{\text{AdS}} \sqrt{\left|g\right|}\; d^{4}x_2\;  \Box^m \big(J_{s}\left(x_2, \partial_{u_2}; y_3, y_4\right)\big) \Omega_{\nu, s}\left(x_1, u_1; x_2, u_2\right) \\ \nonumber
& = \int^{\infty}_{-\infty} d\nu \left(-1\right)^m \left(\nu^2 + s + \tfrac{9}{4}\right)^m \frac{\nu^2}{\pi} \int_{\partial \text{AdS}} d^{3}y \; \frac{1}{s!\left(\tfrac{1}{2}\right)_{s}} \int_{\text{AdS}} \sqrt{\left|g\right|}\; d^{4}x_1\; J_{s}\left(x_1, \partial_{u_1}; y_1, y_2\right)  \Pi_{\tfrac{3}{2} + i\nu, s}(x_1,u_1; y, \hat{\partial}_z) \\ \nonumber
& \hspace*{7cm} \times  \int_{\text{AdS}} \sqrt{\left|g\right|}\; d^{4}x_2\;  J_{s}\left(x_2, \partial_{u_2}; y_3, y_4\right)  \Pi_{\tfrac{3}{2} - i\nu, s}\left(x_2,u_2; y, z \right),
\end{align}
where in the second equality we used the equation of motion \eqref{harmeom} for $\Omega_{\nu, s}$, and inserted its representation \eqref{split} in terms of spin-$s$ bulk-to-boundary propagators. 

The diagram can now easily be evaluated using the known results \eqref{3pt} for the three-point Witten diagrams, just as in the previous section. The resulting conformal block expansion is the direct channel expansion
\begin{align} \label{contactpwe}
\mathcal{A}^{\text{cont.}}_{m,s}\left(y_1,y_2; y_3,y_4\right)= \frac{1}{y^{2}_{12}y^{2}_{34}} \int^{\infty}_{-\infty} d\nu \left(-1\right)^m\left(\nu^2 + s + \tfrac{9}{4}\right)^m \kappa_s(\nu)\,
G_{\tfrac{3}{2}+i \nu,s}\left(u,v\right),
\end{align}
which can be obtained using \eqref{integcb}, as with the exchange amplitudes in section \eqref{sec::4ptwittenexch}.

To complete the full four-point amplitude associated to the vertex $\mathcal{V}_{m,s}$, in accordance with the Feynman rules the contact diagrams corresponding to the remaining permutations of the external $y_i$ have to be taken into account. As for the exchange diagrams in section \ref{sec::4ptwittenexch}, since the external scalars are not distinct their amplitudes can be obtained from the result \eqref{contactpwe} with $y_2 \leftrightarrow y_3$ and $y_2 \leftrightarrow y_4$, and are given by $\mathcal{A}^{\text{cont}}_{m,s}\left(y_1,y_3; y_2,y_4\right)$ and $\mathcal{A}^{\text{cont}}_{m,s}\left(y_1,y_4; y_2,y_3\right)$ respectively. In the same way as for the exchanges, these amplitudes are also crossed-channel (13)(24) and (14)(23) expansions in terms of conformal blocks: $G_{3/2+i \nu,s}\left(\tfrac{1}{u},\tfrac{v}{u}\right)$ and $G_{3/2+i \nu,s}\left(v,u\right)$, respectively. The total contact amplitude for the vertex ${\cal V}_{m,s}$ is thus
\begin{equation}
{\cal A}^{\text{cont.}}_{m,s}(y_1,y_2;y_3,y_4)+{\cal A}^{\text{cont.}}_{m,s}(y_1,y_3;y_2,y_4)+{\cal A}^{\text{cont.}}_{m,s}(y_1,y_4;y_3,y_2).
\end{equation}
\section{Scalar singlet four-point function}
\label{sec::CFT}
\begin{figure}[h]
 \centering
\includegraphics[width=0.8\linewidth]{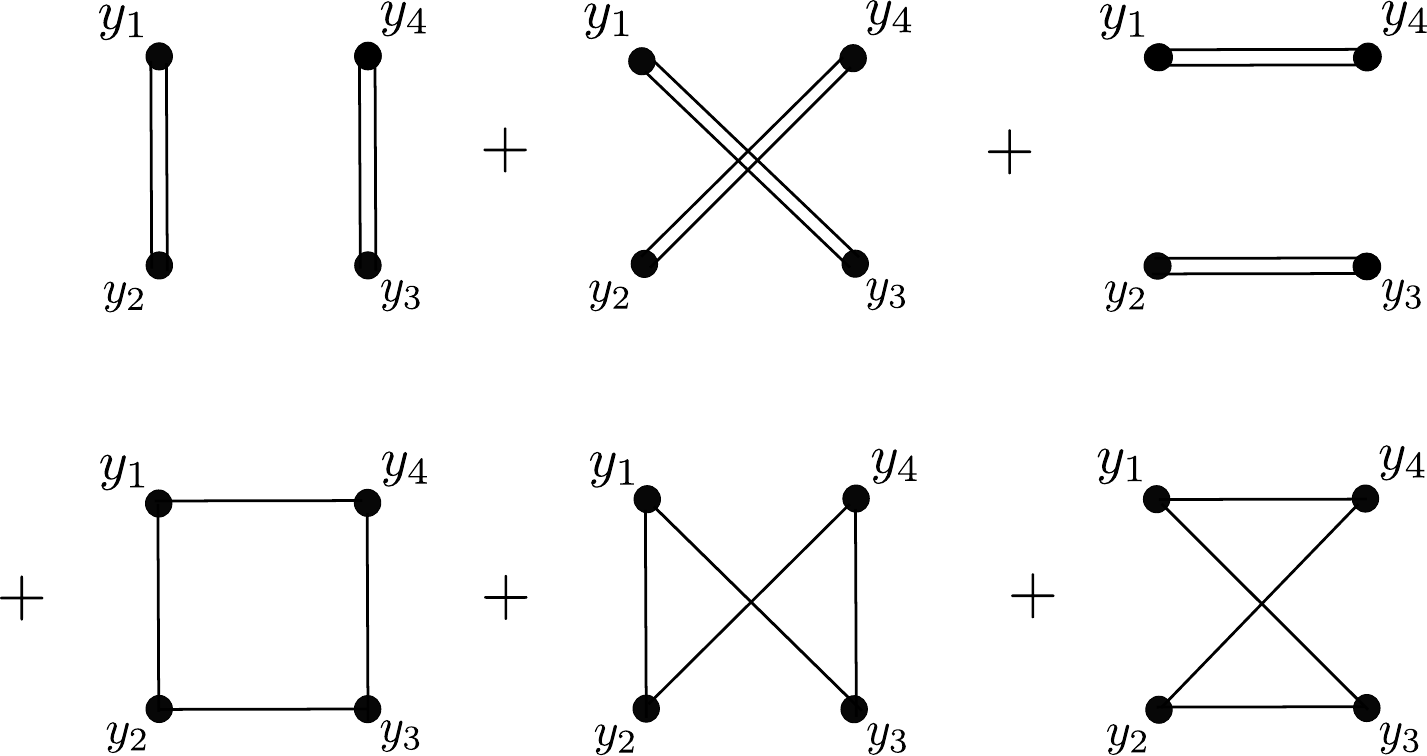}
 \caption{Contributions to the full scalar single-trace operator four-point function \eqref{full} using Wick contractions. The first line constitutes the disconnected $O\left(N^0\right)$ terms, while the second line comprises the connected part of the correlator, which is $O\left(1/N\right)$.}\label{fig::wick}
 \end{figure}

The four-point Witten diagrams displayed in figure \ref{fig::adscft} constitute the holographic computation of the connected part of the scalar-singlet four-point function in the free scalar $O\left(N\right)$ vector model. With normalisation ${\cal O} = \tfrac{1}{\sqrt{2N}}\phi^a\phi^a$, this is the $\mathcal{O}\left(1/N\right)$ part of the full scalar single-trace operator four-point function \cite{Dolan:2000ut} 
\begin{align}\label{full}
&\langle \mathcal{O}\left(y_1\right) \mathcal{O}\left(y_2\right) \mathcal{O}\left(y_3\right) \mathcal{O}\left(y_4\right) \rangle \\ \nonumber
& \hspace*{2.5cm}=   \frac{1}{\left(y^2_{12} y^{2}_{34}\right)^{d-2}} \left\{1 +  u^{d-2} + \left(\frac{u}{v}\right)^{d-2} + \frac{4}{N}\left(u^{\frac{d}{2}-1}+\left(\frac{u}{v}\right)^{\frac{d}{2}-1}+u^{\frac{d}{2}-1}\left(\frac{u}{v}\right)^{\frac{d}{2}-1}\right)\right\}, \nonumber
\end{align}
which is straightforward to obtain in the free field theory via Wick contractions. We express the full four-point function \eqref{full} diagrammatically in figure \ref{fig::wick}.

Since ultimately we will be comparing the connected part of \eqref{full} with the bulk Witten diagram computations of section \ref{sec::4ptwitten}, both the bulk and boundary results for the scalar single-trace operator four-point function should be put on an equal footing in order to ease the identification. To avoid having to re-sum infinite series of conformal blocks from the bulk amplitudes to reach a finite Laurent polynomial in $\sqrt{u}$ and $\sqrt{v}$ as in \eqref{full}, we instead derive the conformal block decomposition of the field theory result. As with the Witten diagrams, we employ the contour integral representation \eqref{introcb} of the conformal block expansion. In the direct channel, this reads
\begin{align}\label{cpwecft}
&\langle \mathcal{O}\left(y_1\right) \mathcal{O}\left(y_2\right) \mathcal{O}\left(y_3\right) \mathcal{O}\left(y_4\right) \rangle =   \frac{1}{\left(y^2_{12} y^{2}_{34}\right)^{d-2}} \left\{1+ \int^{\infty}_{-\infty}d\nu\; \sum\nolimits_s f_s\left(\nu\right) G_{\tfrac{d}{2}+i\nu,s}\left(u,v\right) \right\},
\end{align}
where from now all sums over $s$ will be implicitly over the even spins $s = 0, 2, 4, \dots$ The first term in the braces corresponds to the contribution of the identity operator.  The function $f_s\left(\nu\right)$ contains poles that ensure only the spin-$s$ operators present in the scalar singlet OPE contribute, and encodes their OPE coefficients. The goal of this section is to establish the function $f_s\left(\nu\right)$. As a first step, we determine those OPE coefficients in the scalar single-trace operator product expansion (OPE) which are not yet known explicitly:

The OPE of the scalar singlet $\mathcal{O}$ takes the schematic form \cite{Ferrara:1971zy,Ferrara:1971vh,Ferrara:1972cq}
\begin{align} \label{ope}
\mathcal{O}\mathcal{O} \; \sim \; \mathbb{I} &+ \sum\nolimits_s c_s\; \mathcal{J}_s + \sum\nolimits_{n, s} c_{n,s}\; \mathcal{O}^{(2)}_{n, s}\:+\:\text{descendants},
\end{align}
where $\mathbb{I}$ is the identity operator, $c_s$ is the OPE coefficient of the spin-$s$ single-trace conserved current  $\mathcal{J}_s$ and $c_{n,s}$ is the OPE coefficient of the double-trace operator $\mathcal{O}^{(2)}_{n, s}$. The conserved currents $\mathcal{J}_s$ take the schematic form 
\begin{equation}
{\cal J}_{i_1 ... i_s} = \phi^a \partial_{\left(i_1\right.} ...\; \partial_{\left.i_s\right)}  \phi^a + ..., \quad \partial^{i_1}{\cal J}_{i_1 ... i_s} = 0,
\end{equation}
while the double-trace operators are bi-linear in the single-trace scalar 
operator ${\cal O}$:
\begin{align}
\mathcal{O}^{(2)}_{n,i_1 ... i_s}\left(x\right) = \Box^n\big({\cal O}\left(x\right)\big) \partial_{\left(i_1\right.} ...  \partial_{\left.i_s\right)}{\cal O}\left(x\right) + ... .
\end{align}
The explicit form for a general double-trace operator built from two scalar single-trace operators in conformal field theory is derived in appendix \ref{appendix::doubletrace}. 

In the direct channel, the explicit conformal block decomposition of \eqref{full} is thus 
\begin{align} \label{cbfull}
&\langle \mathcal{O}\left(y_1\right) \mathcal{O}\left(y_2\right) \mathcal{O}\left(y_3\right) \mathcal{O}\left(y_4\right) \rangle \\ \nonumber
&\hspace*{4cm}=  \frac{1}{\left(y^{2}_{12}y^{2}_{34}\right)^{d-2}}\left\{ 1 + \sum\nolimits_{s} c^{2}_{s} \;G_{s+d-2, s}\left(u,v\right) + \sum\nolimits_{n, s} c^{2}_{n, s}\;G_{\Delta_{n,s}, s}\left(u,v\right) \right\},
\end{align}
where $\Delta_{n,s} = 2(d-2)+2n+s$ is the dimension of the spin-$s$ double-trace operator $\mathcal{O}^{(2)}_{n, s}$, and the spin-$s$ conserved current $\mathcal{J}_s$ has dimension $\Delta_s = s+d-2$.  The function $f_s\left(\nu\right)$ of the corresponding contour integral representation \eqref{cpwecft} can be determined with the knowledge of the explicit form of $c_s$ and $c_{n,s}$.

The OPE coefficients $c_s$ of the spin-$s$ conserved currents $\mathcal{J}_s$ are known \cite{Dolan:2000ut, Diaz:2006nm}, and are given by 
\begin{equation}
c_s = \frac{2^{\tfrac{s+3}{2}}}{\sqrt{s!} \sqrt{N}} \frac{\left(\frac{d}{2}-1\right)_s}{\sqrt{\left(d+s-3\right)_s}}. \label{cbcc}
\end{equation}

For the double-trace operators $\mathcal{O}^{(2)}_{n, s}$, so far the OPE coefficients have been determined only for $d=4$ in \cite{Dolan:2000ut}. However, the OPE coefficient of an operator $\mathcal{O}_k$ in the OPE of operators $\mathcal{O}_i$ and $\mathcal{O}_j$ can be determined with the knowledge of the coefficient of the three-point function $\langle \mathcal{O}_i\mathcal{O}_j\mathcal{O}_k \rangle$ and the two-point function $\langle \mathcal{O}_k\mathcal{O}_k \rangle$. In the following section we take this approach to compute the OPE coefficients $c_{n,s}$ of the double-trace operators in general field theory dimension $d$, thus completing the conformal block decomposition \eqref{cbfull} of the scalar single-trace operator four-point function \eqref{full} in general dimensions.

\subsection{OPE coefficients of double-trace operators}
The aim of this section is to compute the OPE coefficients $c_{n,s}$ of the double-trace operators $\mathcal{O}^{(2)}_{n, s}$ in the OPE of the scalar singlet $\mathcal{O}$. In a free theory, it is feasible to compute the coefficients using the relation
\begin{equation}
c_{n,s} = \frac{C_{\mathcal{O}\mathcal{O}\mathcal{O}^{(2)}_{n, s}}}{\sqrt{C_{\mathcal{O}^{(2)}_{n, s}}}},
\end{equation}
where $C_{\mathcal{O}\mathcal{O}\mathcal{O}^{(2)}_{n, s}}$ is the coefficient of the three-point function $\langle \mathcal{O}\mathcal{O}\mathcal{O}^{(2)}_{n, s} \rangle$ and $C_{\mathcal{O}^{(2)}_{n, s}}$ is the coefficient of the two-point function $\langle\mathcal{O}^{(2)}_{n, s}\mathcal{O}^{(2)}_{n, s} \rangle$ (see below for their explicit forms). The reason for this is that in a free theory the computation of correlation functions is relatively straightforward using Wick contractions, with the knowledge of the explicit form of the operators concerned in terms of the fundamental fields.

Using conformal symmetry, it is possible to show that the three- and two-point functions take the form
\begin{equation}
\langle \mathcal{O}\left(y_1\right)\mathcal{O}\left(y_2\right)\mathcal{O}^{(2)}_{n, s}\left(y_3, z\right) \rangle = C_{\mathcal{O}\mathcal{O}\mathcal{O}^{(2)}_{n, s}} \frac{\left(y^2_{12}\right)^{n}}
{\left(y^2_{13}\right)^{d-2+n}\left(y^2_{23}\right)^{d-2+n}} \left( \frac{y_{13} \cdot z}{y^2_{13}} -  \frac{y_{23} \cdot z}{y^2_{23}} \right)^s,
\end{equation}
and 
\begin{equation}
\langle \mathcal{O}^{(2)}_{n, s}\left(y_1, z_1\right)\mathcal{O}^{(2)}_{n, s}\left(y_2, z_2\right) \rangle = C_{\mathcal{O}^{(2)}_{n, s}} \frac{1}{\left(y^2_{12}\right)^{\Delta_{n,s}}} \left(z_1 \cdot z_2 - 2\frac{\left(z_1\cdot y_{12}\right) \left(z_2 \cdot y_{12}\right)}{y^2_{12}}\right)^s.
\end{equation}
The coefficients $C_{\mathcal{O}\mathcal{O}\mathcal{O}^{(2)}_{n, s}}$ and $C_{\mathcal{O}^{(2)}_{n, s}}$ can then be determined by comparing with the result obtained using Wick contractions. Since the explicit form of the double trace operators (appendix \ref{appendix::doubletrace}) is very involved, we were not able to calculate the coefficients explicitly for general $n$ and $s$. We conjecture their form for any $n$ and $s$ based on the explicit results we could obtain for any $n$ when $s=0$, and for any $s$ when $n=0,1$ (see appendix \ref{appendix::cn0}). This leads to the following expression for the double-trace OPE coefficients
\begin{align} \label{dtope}
&c^2_{n, s} = \frac{\left[\left(-1\right)^s+1\right] 2^{s}\left(\tfrac{d}{2}-1\right)^{2}_n \left(d-2\right)^2_{s+n}}{s! n! \left(s+\tfrac{d}{2}\right)_n \left(d-3 + n \right)_n \left(2d + 2n +s-5\right)_{s} \left(\tfrac{3d}{2}-4 + n +s\right)_n} \\ \nonumber 
&\hspace*{5cm} \times \left(1+\left(-1\right)^n\frac{4}{N} \frac{\Gamma\left(s\right)}{2^{s}\Gamma\left(\frac{s}{2}\right)} \frac{\left(\frac{d}{2}-1\right)_{n+\tfrac{s}{2}}}{\left(\frac{d-1}{2}\right)_{\tfrac{s}{2}}\left(d-2\right)_{n+\tfrac{s}{2}}}\right).
\end{align}
Although we were unable to derive the OPE coefficients in full generality, our conjecture above for any $n$ and $s$ reproduces the results available in $d=4$ \cite{Dolan:2000ut}, and agrees with the coefficients at $N = \infty$ in general $d$ \cite{Fitzpatrick:2011dm}. These provide two supplementary important checks of its validity. 
This formula is instrumental in our holographic reconstruction of the quartic self-interaction of the scalar field in higher-spin gravity on AdS$_4$, dual to the free vector model CFT$_3$.

\subsection{Contour integral representation}
\label{subsec::cftcpwe}
In this subsection we complete the derivation of the conformal block expansion \eqref{cpwecft} of the scalar single-trace operator four-point function in the form of a contour integral. We use the knowledge of the explicit form \eqref{cbcc} of the OPE coefficients $c_s$ of the single-trace higher-spin conserved currents, and those $c_{n,s}$ of the double-trace operators obtained in the previous section. For concision, we focus on the specific case of interest: $d=3$.

Essentially, the task is to find a function $f_s\left(\nu\right)$ which gives precisely the conformal block expansion $\eqref{cbfull}$ when the contour integral over $\nu$ in \eqref{cpwecft} is evaluated. Therefore $f_s\left(\nu\right)$ must contain poles in $\nu$, within the prescribed contour, where the dimension $\tfrac{d}{2}+i\nu$ of the conformal blocks matches that of the spin-$s$ operators $\mathcal{J}_s$ and $\mathcal{O}^{(2)}_{n,s}$ in the scalar singlet OPE \eqref{ope}.

Since the conformal block $G_{\frac{d}{2}+i\nu,s}\left(u,v\right)$ decays exponentially as Im$\left(\nu\right) \rightarrow -\infty$, we close the contour in the lower-half plane. Accordingly, $f_s\left(\nu\right)$ must have the following pole structure for the correct operator spectrum:
\begin{enumerate}
\setlength\itemsep{1.1em} 
\item Spin-$s$ conserved current $\mathcal{J}_s$ of dimension $\Delta_s = s+d-2$\: \vspace*{0.3cm} \\ 
\hspace*{1cm}$\rightarrow$\: single pole at $\nu = - i\left(s+\tfrac{d}{2}-2\right)$.
\item Spin-$s$ double-trace operator $\mathcal{O}^{(2)}_{n,s}$ of dimension $\Delta_{n,s} = 2\left(d-2\right) +2n+s$, $n = 0, 1, 2, ... $\vspace*{0.3cm}  \\ 
\hspace*{1cm} $\rightarrow$ single poles at $\nu = - i\left(2\left(d-2\right) +2n+s - \tfrac{d}{2}\right)$.
\end{enumerate}

It will be useful to decompose $f_s\left(\nu\right)$ as
\begin{equation}
 f_s\left(\nu\right) = f_{\mathcal{J}_s}\left(\nu\right) +  f_{\mathcal{O}^{(2)}_s}\left(\nu\right),
 \end{equation}
where $f_{\mathcal{J}_s}\left(\nu\right)$ and $f_{\mathcal{O}^{(2)}_s}$ generate the contributions of $\mathcal{J}_s$ and the spin-$s$ double-trace operators $\mathcal{O}^{(2)}_{n,s}$, respectively.

The functions $f_{\mathcal{J}_s}\left(\nu\right)$ and $f_{\mathcal{O}^{(2)}_s}\left(\nu\right)$ are not unique, as they can be modified in ways that do not alter the pole structure within the contour, and the corresponding residues. This freedom can be used to ease comparison with the conformal block expansions of the four-point Witten diagrams derived in section \ref{sec::4ptwitten}, allowing us to cast the functions into the form
\begin{equation} \label{constr}
f\left(\nu\right) = p_s\left(\nu\right)\,\kappa_s(\nu)\,,
\end{equation}
where $p_s\left(\nu\right)$ is an even function of $\nu$, as is the case for the Witten diagrams (see for example the four-point amplitudes \eqref{cbexch} and \eqref{contactpwe}). We demonstrate how this can be achieved in the following, for the contribution of the double-trace operators.

\subsubsection*{Double-trace operator contribution}
To represent the contribution of the double-trace operators, the function $f_{\mathcal{O}^{(2)}_s}\left(\nu\right)$ must be chosen such that 
\begin{align} \label{dtf}
\int^{\infty}_{-\infty} d\nu \; f_{\mathcal{O}^{(2)}_s}\left(\nu\right) G_{\tfrac{d}{2}+i\nu,s}\left(u,v\right) =  \sum\limits^{\infty}_{n=0}c^{2}_{n,s}G_{\Delta_{n,s},s}\left(u,v\right),
\end{align}
where we close the contour in the lower-half plane. 

As explained above,  $f_{\mathcal{O}^{(2)}_s}\left(\nu\right)$ must have poles at $\nu = - i\left(2\left(d-2\right) +2n+s - \tfrac{d}{2}\right)$ for $n = 0, 1, 2, ...$ These poles can be neatly packaged in the gamma function $\Gamma\left(\frac{4\left(d-2\right)+2s-d-2i\nu}{4}\right)$, which has poles precisely at those values of $\nu$. The most direct way to find a function which satisfies \eqref{dtf}, is then to multiply the double-trace conformal block coefficient \eqref{dtope} by $\Gamma\left(-n\right)$ and a factor to neutralise the effect of its residue.\footnote{The residue of $\Gamma\left(z\right)$ at $z=-n$ is $\frac{\left(-1\right)^n}{n!}$, so in this case the neutralising factor is $\left(-1\right)^n n!$}. To obtain a function of $\nu$ one then simply replaces $n \rightarrow - \left(\frac{4\left(d-2\right)+2s-d-2i\nu}{4}\right)$. For $d=3$, this direct method gives the following function
\begin{align} \label{tildef}
\tilde{f}_{\mathcal{O}^{(2)}_s}\left(\nu\right) =  \tilde{p}_{\mathcal{O}^{(2)}_s}\left(\nu\right)\,\kappa_s(\nu)\,,
\end{align}
with $\kappa_s$ given by \eqref{kappa} and $\tilde{p}_{\mathcal{O}^{(2)}_s}$ by
\begin{align} \label{directp}
 &\tilde{p}_{\mathcal{O}^{(2)}_s}\left(\nu\right) = \left[1+\left(-1\right)^s\right]\frac{\pi ^{5/2} 2^{5-s} \Gamma \left(s+\frac{3}{2}\right)}{ \Gamma\left(s+1\right) \Gamma \left(\frac{2 s +1-2 i \nu}{4}\right) \Gamma \left(\frac{2 s +1+2 i \nu}{4}\right) \Gamma \left(\frac{2 s +3-2 i \nu}{4} \right) \Gamma \left(\frac{2 s +3+2 i \nu}{4} \right)}
 \\ \nonumber
 & +\frac{1}{N} \left[1+(-1)^s\right] \frac{ \pi ^3 2^{6-2s} \Gamma \left(s+\frac{3}{2}\right) \csc \left(\frac{\pi}{4} (1+2 i \nu )\right) \csc \left(\frac{\pi}{4} (2 i \nu -2 s+1)\right)}{\Gamma \left(\frac{s}{2}+1\right)^2\Gamma \left(\frac{3-2i \nu}{4}\right) \Gamma \left(\frac{3+2i \nu}{4}\right) \Gamma \left(\frac{2 s +1-2 i \nu}{4}\right) \Gamma \left(\frac{2 s+1+2 i \nu }{4} \right) \Gamma \left(\frac{2 s+3-2 i \nu }{4} \right) \Gamma \left(\frac{2 s +3+2 i \nu}{4} \right)},
\end{align}
where we used the identity $\Gamma\left(z\right)\Gamma\left(1-z\right) = \pi \,\text{cosec} \left(\pi z\right)$ to make the behaviour of $\tilde{p}_{\mathcal{O}^{(2)}_{s}}\left(\nu\right)$ under the transformation $\nu \leftrightarrow -\nu$ manifest. It is then clear that the only terms preventing the symmetry condition $p_{\mathcal{O}^{(2)}_s}(-\nu)=p_{\mathcal{O}^{(2)}_s}(\nu)$ of \eqref{constr} from being satisfied are the cosecant functions in the numerator of the $O\left(1/N\right)$ part. Since the only poles of \eqref{tildef} in the lower-half plane are single poles at $\nu = - i\left(\tfrac{1}{2}+2n+s\right)$, the discrepancy can be rectified by evaluating the cosecant functions at the location of the poles:
\begin{align}
\csc \left(\tfrac{\pi}{4} (1+2 i \nu )\right) \csc \left(\tfrac{\pi}{4} (2 i \nu -2 s+1)\right) \Big|_{\nu = - i(2\Delta +2n+s - \tfrac{d}{2})} = \frac{\left(-1\right)^{\tfrac{s}{2}}}{\sqrt{2}}.
\end{align}
The conformal block coefficient function $f_{\mathcal{O}^{(2)}_{s}}\left(\nu\right)=p_{\mathcal{O}^{(2)}_{s}}\left(\nu\right) \,\kappa_s(\nu)$ with the desired symmetry \eqref{constr} therefore has
\begin{align} \label{directp}
 &p_{\mathcal{O}^{(2)}_{s}}\left(\nu\right) = \left[1+\left(-1\right)^s\right]\frac{\pi ^{\frac{3}{2}}\; 2^{s+4} \Gamma \left(s+\frac{3}{2}\right)}{ \Gamma\left(s+1\right) \Gamma \left(s+\tfrac{1}{2}+ i \nu \right) \Gamma \left(s+\tfrac{1}{2}- i \nu \right)}
 \\ \nonumber
 & \hspace*{0.25cm}+\frac{1}{N} \left[1+(-1)^s\right] \frac{ \left(-1\right)^{\tfrac{s}{2}} \pi ^{\frac{3}{2}} 2^{s+4} \Gamma \left(s+\frac{3}{2}\right)\Gamma \left(\frac{s}{2}+\frac{1}{2}\right)}{\sqrt{2}\,\Gamma \left(\frac{s}{2}+1\right)\Gamma\left(s+1\right)\Gamma \left(\frac{3}{4}-\frac{i \nu }{2}\right) \Gamma \left(\frac{3}{4}+\frac{i \nu }{2}\right) \Gamma \left(s+\tfrac{1}{2}+ i \nu \right) \Gamma \left(s+\tfrac{1}{2}- i \nu \right)},
\end{align}
where simplifications were also made using the identity: $\Gamma\left(z\right)\Gamma\left(z+\frac{1}{2}\right) = \sqrt{\pi}\, 2^{1-2z}\Gamma\left(2z\right)$.
\subsubsection*{Higher-spin conserved currents}
Just as for the double-trace operators above, one can derive a function $f_{{\cal J}_s}\left(\nu\right)=p_{{\cal J}_s}\left(\nu\right)\,\kappa_s(\nu)$ that generates the contribution \eqref{cbcc} of a spin-$s$ conserved current to the conformal block expansion \eqref{cbfull} of the scalar single-trace operator four-point function. The corresponding $p_{{\cal J}_s}\left(\nu\right)$ that satisfies the symmetry property \eqref{constr} is
\begin{align} \label{pwhs}
&p_{{\cal J}_s}\left(\nu\right) = \frac{\pi\;  2^{8-s}}{N} \frac{1}{\nu ^2+(s-\tfrac{1}{2})^2}\frac{1}{\Gamma \left(\frac{2 s-2 i \nu +1}{4}\right)^2 \Gamma \left(\frac{2 s+2 i \nu +1}{4} \right)^2}. 
\end{align}
Notice that the factor $\Gamma \left(\frac{2 s-2 i \nu +1}{4}\right)^2$ in the denominator cancels that in the numerator of $\kappa_s\left(\nu\right)$. The function $f_{{\cal J}_s}\left(\nu\right)=p_{{\cal J}_s}\left(\nu\right)\kappa_s(\nu)$ therefore only has one pole in the lower-half plane at $\nu = -i\left(s-1/2\right)$, corresponding to a spin-$s$ conserved current.

\section{Uncovering the quartic vertex} 
\label{sec::quartic}
\subsection{Summary: CFT interpretation of Witten diagrams}
\label{sec::CFTWitten}
Before the scalar quartic vertex in higher-spin theory on AdS$_4$ can be investigated through the comparison of the CFT$_3$ results in section \ref{subsec::cftcpwe} with the four-point bulk amplitudes in section \ref{sec::4ptwitten}, it is instructive to study the latter as objects in conformal field theory. That is, the operator contributions in their conformal block expansions.

\subsubsection*{Exchange Witten diagrams}
Let us focus on the s-channel exchange of a massless spin-$s$ field (figure \ref{fig::adscft}. (a)), whose conformal block expansion \eqref{cbexch} in the direct channel was derived as a contour integral in section \ref{sec::4ptwittenexch}. The discussion for the t- and u-channel exchanges is analogous, and we comment on them briefly towards the end of this subsection. We take the same prescription for the contour as for the CFT conformal block expansions in section \ref{subsec::cftcpwe}, closing in the lower-half plane. The four-point exchange amplitude \eqref{cbexch}, interpreted as a CFT quantity, therefore receives contributions from the following operators:
\begin{enumerate}
\setlength\itemsep{1.25em} 
\item Single pole at: $\nu = -i \left(s-\frac{1}{2}\right)$ $\rightarrow$ spin-$s$ conserved current ${\cal J}_s$.
\item Double poles at: $\nu = - i\left(2\left(d-2\right) +2n+s - \tfrac{d}{2}\right)$, $n = 0, 1, 2, ... $\vspace*{0.2cm}  \\ \hspace*{3cm}$\rightarrow$ spin-$s$ double-trace operator $\mathcal{O}^{(2)}_{n,s}$ with anomalous dimensions. 
\end{enumerate}
The cubic coupling $g_s$ was fixed in section \ref{subsec::cubiccoupoing} according to the OPE coefficient of ${\cal J}_s$ in the operator product expansion \eqref{ope} (see also equation  \eqref{diazdorn}). Therefore, as can be checked explicitly from the split representation, the corresponding contribution of ${\cal J}_s$ to the s-channel exchange is identical to that in the direct channel conformal block decomposition of the dual CFT four-point function \eqref{cbfull}. For the double-trace operator contributions the story is clearly not the same, as indicated by the anomalous dimensions: In a free theory, such as the dual free scalar $O\left(N\right)$ vector model, these are not present. The explicit conformal block expansion of the s-channel exchange thus takes the form 
\begin{align} \label{schcb}
&\mathcal{A}^{\text{exch.}}_{s}\left(y_1,y_2; y_3,y_4\right) = \frac{1}{y^{2}_{12}y^{2}_{34}} \left\{ c^2_s\;G_{s+d-2\,,\, s}\left(u,v\right) + \sum\limits^{\infty}_{n=0} d^{\text{exch.}}_{n,s}\;G_{\Delta_{n,s} +\gamma_{n,s}\,,\, s}\left(u,v\right) \right\},
\end{align}
where the $\gamma_{n,s}$ represent the anomalous dimensions of the double-trace operators $\mathcal{O}^{(2)}_{n,s}$. This result is consistent with observations in the literature \cite{Liu:1998th,Heemskerk:2009pn} for lower-spin exchanges, that an exchange Witten diagram does not simply correspond in the CFT to the exchange of the operator dual to the exchanged bulk field: There are additional contributions from double-trace operators built from the single-trace operators that are dual to the external fields of the exchange Witten diagram. 

As explained at the end of section \ref{sec::4ptwittenexch}, the t- and u-channel exchanges have (13)(24) and (14)(23) crossed-channel expansions, which are obtained from the s-channel result by exchanging $y_2 \leftrightarrow y_3$ and $y_2 \leftrightarrow y_4$ respectively. Consequently, as with the conformal block expansion \eqref{schcb} of the s-channel exchange, the contributions from the operator ${\cal J}_s$ in the conformal block expansion of these amplitudes coincide with those in the corresponding (13)(24) and (14)(32) crossed channel expansions of the CFT four-point function \eqref{full}.

Before moving onto the analysis of the contact diagrams, let us make a final comment on the operator contributions to the spin-$s$ exchange. It is important to note that in general it is not the case that only spin-$s$ representations of the conformal group contribute to the four-point exchange of a bulk spin-$s$ field. In the present context of higher-spin theory on AdS$_4$, this is only possible since the conformal scalar present in this dimension allows the conserved currents in the cubic vertex \eqref{cubic} to be traceless. As a consequence, only spin-$s$ conformal blocks appear in the amplitude \eqref{schcb}. This can be seen from the split representation \eqref{exch} of the exchange in general dimensions, where just the leading $\left(k=0\right)$ term contributes if the current is traceless. In general dimensions the terms for $k > 0$ would be non-vanishing, because generally the scalar is not conformal. These terms would generate contributions to the amplitude from double-trace operators of spin lower than $s$.

\subsubsection*{Contact Witten diagrams}
We now perform the CFT analysis of the contact Witten diagrams considered in section \ref{subsec::contactamp}, which give all possible contact amplitudes for the scalar $\varphi_0$ in AdS$_4$, since they are generated by quartic vertices belonging to the basis \eqref{quartbasis}. A similar analysis can be found in \cite{Heemskerk:2009pn}, where a complete basis of local quartic interactions was presented and the conformal block expansions of corresponding four-point contact diagrams were determined in the Regge limit. 

The contour integral representation of the contact amplitude \eqref{contactpwe} produced by a generic quartic vertex ${\cal V}_{m,s}$ in the basis (\ref{quartbasis}), has double poles located at $\nu = - i\left(2\left(d-2\right) +2n+s - \tfrac{d}{2}\right)\text{, }$ $n = 0, 1, 2,\dots$ in the lower-half $\nu$-plane. Its conformal block expansion thus receives contributions only from spin-$s$ double-trace operators with anomalous dimensions, 
\begin{align} \label{contactcb}
&\mathcal{A}^{\text{cont.}}_{m,s}\left(y_1,y_2; y_3,y_4\right) = \frac{1}{y^{2}_{12}y^{2}_{34}} \sum\limits^{\infty}_{n=0} d^{\text{cont.}}_{n,s,m}\;G_{\Delta_{n,s} +\gamma_{n,s}\,,\, s}\left(u,v\right).
\end{align}
As with the exchange amplitudes, the conformal block expansions of the remaining contact amplitudes are crossed channel expansions obtained from \eqref{contactcb} by exchanging $y_2 \leftrightarrow y_3$ and $y_2 \leftrightarrow y_4$.  They therefore also receive contributions only from double-trace operators in their respective channels, with the same coefficients $d^{\text{cont.}}_{n,s,m}$ and anomalous dimensions $\gamma_{n,s}$.

\subsection{Combining bulk contributions from all channels}
\label{sec::combining}
\begin{figure}[h]
 \centering
\includegraphics[width=0.9\linewidth]{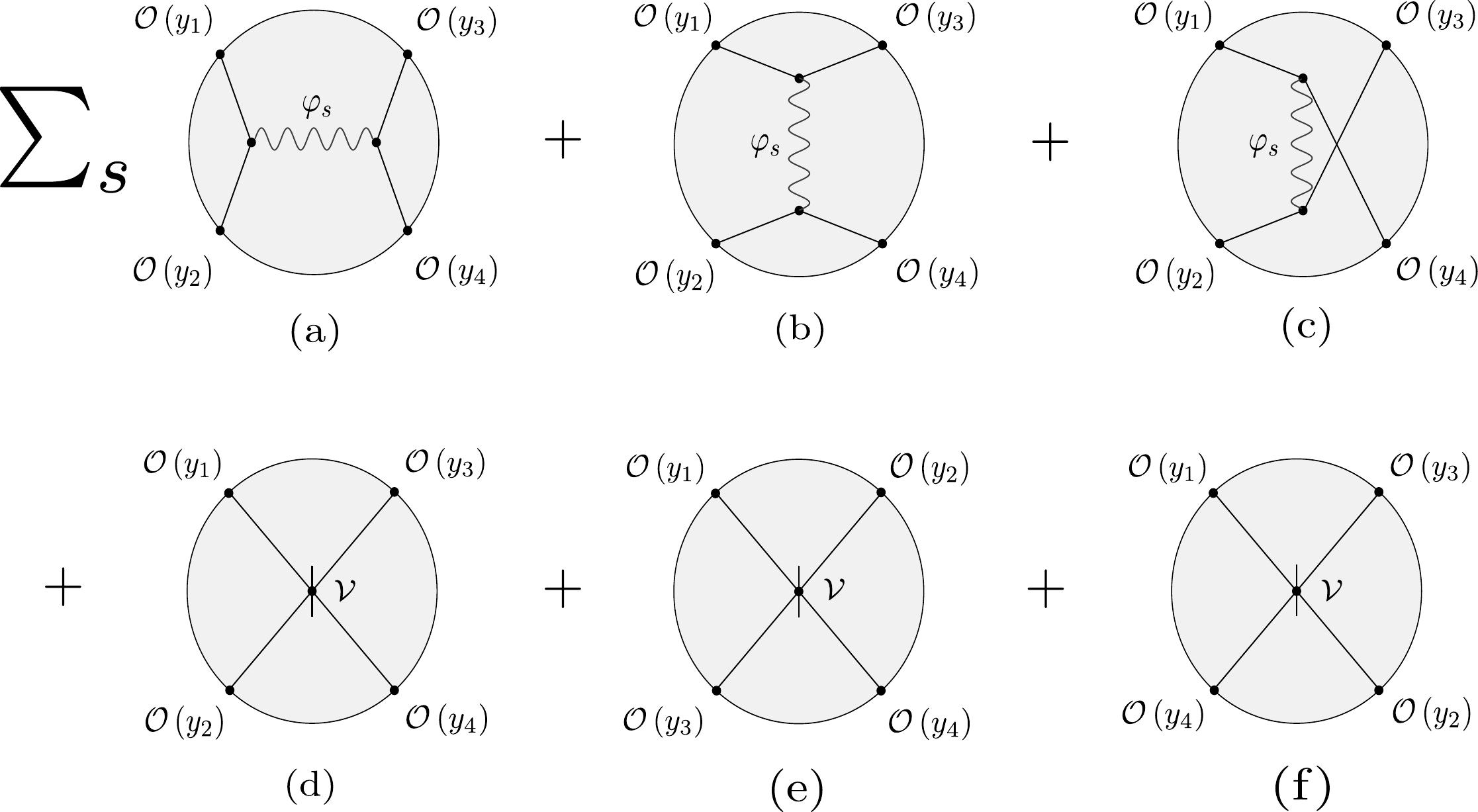}
 \caption{Four-point bulk Witten diagrams involving external scalars $\varphi_0$, which give the holographic computation of the connected part of the dual scalar single-trace operator four-point function. Contact diagrams (d), (e) and (f) are generated by the quartic vertex ${\cal V}$ that we seek to establish, and represent the distinct permutations of the external legs. In the preceding sections, the amplitudes of the diagrams in the first, second and third columns of this figure have been expressed as (12)(34) direct-, (13)(24) and (14)(23) crossed-channel expansions in conformal blocks, respectively. }\label{fig::adscft2}
 \end{figure}

At the level of four-point functions, holography requires
\begin{align} \label{qv1}
&\sum\nolimits_s {\cal A}^{\text{exch.}}_s(y_1,y_2;y_3,y_4)+{\cal A}^{\text{exch.}}_s(y_1,y_3;y_2,y_4)+{\cal A}^{\text{exch.}}_s(y_1,y_4;y_3,y_2)
\nonumber\\ &+{\cal A}^{\text{cont.}}(y_1,y_2;y_3,y_4)+{\cal A}^{\text{cont.}}(y_1,y_3;y_2,y_4)+{\cal A}^{\text{cont.}}(y_1,y_4;y_2,y_3)\\ \nonumber
&\hspace*{9cm}\overset{\text{AdS / CFT}}=\langle {\cal O}(y_1) {\cal O}(y_2) {\cal O}(y_3) {\cal O}(y_4) \rangle_{\text{conn.}},
\end{align}
where the $\mathcal{A}^{\text{cont.}}$ are the contributions from quartic contact interaction that we seek, and the RHS is the connected, i.e. ${\cal O}\left(1/N\right)$, part of the scalar single-trace operator four-point function \eqref{full}. The bulk Witten diagrams that comprise the LHS of equation \eqref{qv1} are given in figure \ref{fig::adscft2}.

To extract the quartic vertex ${\cal V}\left(x\right)$ from equation the contact diagrams in \eqref{qv1}, we make an ansatz of the form
\begin{equation}
{\cal V}
\left(x\right) = \sum\nolimits_{m,s} a_{m,s}\; {\cal V}_{m,s}\left(x\right), \qquad  a_{m,s} \in \mathbb{R} \label{quartansatz}
\end{equation}
built from vertices in the basis \eqref{quartbasis}, which accounts for all possible independent quartic interactions of $\varphi_0$ on AdS$_4$. The corresponding four-point contact diagrams (d), (e) and (f) in figure \ref{fig::adscft2} can be evaluated in terms of conformal blocks using the results for the contact amplitudes of the individual basis elements established in section \ref{subsec::contactamp}. For example, for the amplitude of diagram \ref{fig::adscft2}. (d)
\begin{equation}\label{ansatzamp}
{\cal A}^{\text{cont.}}(y_1,y_2;y_3,y_4) = \sum_{m,s}a_{m,s}\;{\cal A}^{\text{cont.}}_{m,s}(y_1,y_2;y_3,y_4),
\end{equation}
with ${\cal A}^{\text{cont.}}_{m,s}(y_1,y_2;y_3,y_4)$ given by \eqref{contactpwe}. Combined with the results for the conformal block expansions of the CFT four-point function \eqref{cpwecft} and the exchange diagrams in section \eqref{sec::4ptwittenexch}, one can then solve equation \eqref{qv1} for the coefficients $a_{m,s}$. However, there is one issue preventing us from extracting the vertex directly in this manner, which we discuss in the following.

From the derivation of the exchange amplitudes in section \ref{sec::4ptwittenexch}, the s-channel exchange is expressed in a direct channel decomposition of conformal blocks, and is given by equation \eqref{cbexch}. By simply exchanging the external points of \eqref{cbexch}, we could obtain results for the t- and u-channel exchanges. In this way, the t- and u-channel exchange amplitudes are not expressed as direct channel decompositions like for the s-channel, but rather (13)(24) and (14)(23) crossed channel expansions such as \eqref{cbexcht}. Thus, at the current stage we are able 
to evaluate all the exchange diagrams contributing to \eqref{qv1}, yet they are expressed
in terms of conformal blocks in different channels. This creates an essential difficulty in extracting the vertex, since the conformal block expansion of the CFT result on the RHS of \eqref{qv1} is, as derived in section \ref{subsec::cftcpwe}, an expansion only in a single channel. At present there is no straightforward means to relate arbitrary collections of conformal blocks in different channels. This is essentially the key technical problem of the conformal bootstrap, which so far is addressed mainly numerically or analytically in certain limits.

In \cite{Hoffmann:2000tr, ElShowk:2011ag} conformal block decomposition of the scalar
exchange in the s-channel has been obtained in terms of crossed channel conformal blocks. In principle one can 
try to generalise these results to higher spin exchanges, to express them as conformal block expansions in one and the same channel. However, the split representation of the bulk-to-bulk propagators is not conducive to deriving a crossed-channel decomposition of a s-channel exchange, since the construction naturally delivers s-channel amplitudes as direct channel expansions. Fortunately, as we shall explain in the next subsection, in our case dealing with the conformal block decomposition of the exchanges in different channels can be avoided.

Before explaining how this can be achieved, let us emphasise that the same difficulty also appears for the contact Witten diagrams: Consider one of the vertices in the basis (\ref{quartbasis})
\begin{align}
\mathcal{V}_{m,s} = J_{s}\left(\varphi_0,\varphi_0, \partial_u\right) \Box^m J_{s}\left(\varphi_0,\varphi_0, u\right),
 \label{contactchannels}
\end{align}
where the fact that $J_s$ is bilinear in scalar fields $\varphi_0$ is made explicit. According to the
Feynman rules, the bulk-to-boundary propagators sourced
at different boundary points should be attached to the vertex in a Bose-symmetric way. This implies that
the vertex (\ref{contactchannels}) results not only in the amplitude
\begin{align}
 \label{contactchannel1}
&{\cal A}^{\text{cont.}}_{m,s}(y_1,y_2;y_3,y_4) \\ \nonumber
&\hspace*{1.5cm}\equiv \int_{\text{AdS}} \sqrt{\left|g\right|}d^{d+1}x J_{s}\left(\Pi_{d-2,0}(y_1,x),\Pi_{d-2,0}(y_2,x), \partial_u\right) \Box^m J_{s}\left(\Pi_{d-2,0}(y_3,x),\Pi_{d-2,0}(y_4,x), u\right),
\end{align}
but also in the amplitudes ${\cal A}^{\text{cont.}}_{m,s}(y_1,y_3;y_2,y_4)$ and ${\cal A}^{\text{cont.}}_{m,s}(y_1,y_4;y_3,y_2)$. The complete result is therefore
\begin{equation}
\label{qv3}
{\cal A}^{\text{cont.}}_{m,s}(y_1,y_2;y_3,y_4)+{\cal A}^{\text{cont.}}_{m,s}(y_1,y_3;y_2,y_4)+{\cal A}^{\text{cont.}}_{m,s}(y_1,y_4;y_3,y_2).
\end{equation}
The contact amplitude (\ref{contactchannel1}) was evaluated explicitly in section \ref{subsec::contactamp}, where it was expressed in terms of direct-channel conformal blocks \eqref{contactpwe}. By exchanging $y_2  \leftrightarrow y_3$ and $y_2  \leftrightarrow y_4$ in the direct-channel result \eqref{contactpwe} for \eqref{contactchannel1}, we could obtain the remaining amplitudes ${\cal A}^{\text{cont.}}_{m,s}(y_1,y_3;y_2,y_4)$ and ${\cal A}^{\text{cont.}}_{m,s}(y_1,y_4;y_3,y_2)$ as expansions in terms of (13)(24) and (14)(23) cross-channel conformal blocks, respectively. In this way, the total contact amplitude \eqref{qv3} is therefore also an expansion in a mixture of the three channels. To try and express \eqref{qv3} as an expansion in just a single channel, in principle one could try and rearrange amplitudes ${\cal A}^{\text{cont.}}_{m,s}(y_1,y_3;y_2,y_4)$ and ${\cal A}^{\text{cont.}}_{m,s}(y_1,y_4;y_3,y_2)$ at the level of the bulk integrand: This would involve taking \eqref{contactchannel1} with $y_2  \leftrightarrow y_3$ and $y_2  \leftrightarrow y_4$, in attempt to express them as linear combinations of ${\cal A}^{\text{cont.}}_{m,s}(y_1,y_2;y_3,y_4)$ through integrating by parts and using the equations of motion. Their direct channel expansions could then be obtained using \eqref{contactpwe}. However, this would require the knowledge of the explicit complete form of traceless conserved currents $J_s$ in AdS, as well as cumbersome algebra involving covariant derivatives. As with the equivalent issue for exchange diagrams explained above, these complicated manipulations can be avoided.
 
\subsection{Solving for the vertex: Reducing the analysis to a single channel}

\begin{figure}[h]
 \centering
\includegraphics[width=1\linewidth]{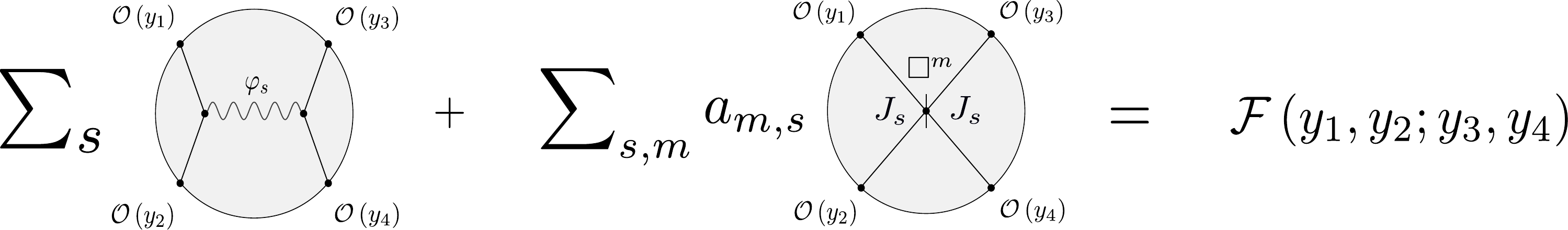}
 \caption{Simplified equation for the quartic vertex: All terms are expressed as conformal block expansions in the same channel. The LHS constitutes the bulk diagrams in the first column of figure \ref{fig::adscft2}, while the boundary term on the RHS is determined in the main text. The equation encodes the full amplitude, which can be obtained by exchanging $y_{2} \leftrightarrow y_{3},  y_{4}$ and adding the permuted equations to the original.}\label{fig::adscft3}
 \end{figure}

Instead of solving (\ref{qv1}) for the quartic vertex, which requires us to deal simultaneously with conformal blocks
 in different channels, since the operators inserted in the four-point function are identical we can instead solve the following equation 
 \begin{equation}
\label{qv2}
\sum_{m,s}a_{m,s}\;{\cal A}^{\text{cont.}}_{m,s}(y_1,y_2;y_3,y_4)+\sum\nolimits_s{\cal A}^{\text{exch.}}_{s}(y_1,y_2;y_3, y_4)={\cal F}(y_1,y_2;y_3,y_4),
\end{equation}
where ${\cal F}(y_1,y_2;y_3,y_4)$ is a sum of direct channel conformal blocks, and satisfies
 \begin{equation}
 \label{13oooo}
 {\cal F}(y_1,y_2;y_3,y_4)+{\cal F}(y_1,y_3;y_2,y_4)+ {\cal F}(y_1,y_4;y_3,y_2)=
 \langle {\cal O}(y_1){\cal O}(y_2){\cal O}(y_3){\cal O}(y_4)\rangle_{\text{conn.}}.
 \end{equation}
 This simplified problem is expressed diagrammatically in figure \ref{fig::adscft3}. The form of a suitable function ${\cal F}$ is simple to establish from the total connected CFT correlator, and we shall explain precisely how it can be obtained below. But let us first make the relation between equations \eqref{qv1} and \eqref{qv2} for the quartic vertex more concrete: If a vertex that solves \eqref{qv2} is found, then by the property \eqref{13oooo} of ${\cal F}$, equation \eqref{qv1} is clearly satisfied. Further, the non-trivial quartic vertex is completely determined by \eqref{qv1}. Thus any solution of \eqref{qv2} would give the only solution of \eqref{qv1}, and these two problems therefore are equivalent. Solving \eqref{qv2} for the vertex instead greatly simplifies the problem, since each term is expanded in one and the same channel. 

It should be noted that equation \eqref{13oooo} does not define ${\cal F}$ unambiguously. Moreover, through equation \eqref{qv2} the different ${\cal F}$ result in different s-channel amplitudes ${\cal A}^{\text{cont.}}\left(y_1,y_2;y_3,y_4\right)$. However, since the total contact amplitude
\begin{equation}
 {\cal A}^{\text{cont.}}(y_1,y_2;y_3,y_4)+{\cal A}^{\text{cont.}}(y_1,y_3;y_2,y_4)+ {\cal A}^{\text{cont.}}(y_1,y_4;y_3,y_2)
\end{equation}
is unique, we can conclude that any effect on the solution for quartic vertex ${\cal V}$ introduced by this ambiguity would vanish due to the Bose-symmetrisation over the external legs.

In fact, we shall use this freedom to our advantage: As observed in section \ref{sec::CFTWitten}, the contact amplitudes ${\cal A}^{\text{cont.}}_{m,s}(y_1,y_2;y_3,y_4)$ generated by individual vertices in the basis \eqref{quartbasis} do not receive contributions from conformal blocks of single-trace operators in the direct channel. It is therefore natural to try and choose ${\cal F}$ so that it cancels the single trace contributions to equation \eqref{qv2} generated by the exchange Witten diagrams. This can be achieved by noting that the first two terms of the connected four-point function
\begin{align}\label{4ptconnected}
&\langle \mathcal{O}\left(y_1\right) \mathcal{O}\left(y_2\right) \mathcal{O}\left(y_3\right) \mathcal{O}\left(y_4\right) \rangle_{\text{conn.}} =   \frac{4}{N} \frac{1}{\left(y^2_{12} y^{2}_{34}\right)^{d-2}} \left\{u^{\frac{d}{2}-1}+\left(\frac{u}{v}\right)^{\frac{d}{2}-1}+u^{\frac{d}{2}-1}\left(\frac{u}{v}\right)^{\frac{d}{2}-1}\right\},
\end{align}
 contain only single-trace conformal blocks in the direct-channel decomposition
\begin{align}\label{4ptconnectedAB}
  \text{A}+\text{B}\equiv \frac{4}{N} \frac{1}{\left(y^2_{12} y^{2}_{34}\right)^{d-2}} \left\{u^{\frac{d}{2}-1}+\left(\frac{u}{v}\right)^{\frac{d}{2}-1}\right\}= \frac{1}{\left(y^2_{12} y^{2}_{34}\right)^{d-2}}  \sum\nolimits_s c^2_s\;G_{s+d-2\,,\, s}\left(u,v\right),
\end{align}
while the remaining term receives contributions only from double-trace conformal blocks 
\begin{align}\label{4ptconnectedC}
\text{C}\equiv    \frac{4}{N} \frac{1}{\left(y^2_{12} y^{2}_{34}\right)^{d-2}} \left\{u^{\frac{d}{2}-1}\left(\frac{u}{v}\right)^{\frac{d}{2}-1}\right\}= \frac{1}{\left(y^2_{12} y^{2}_{34}\right)^{d-2}}   \sum\nolimits_{n,s} \bar{c}^2_{n,s}\;G_{\Delta_{n,s}\,,\, s}\left(u,v\right),
\end{align}
where $\bar{c}^2_{n,s}$ is the ${\cal O}\left(1/N\right)$ part of the double-trace conformal block coefficient \eqref{dtope}.

Moreover, as noted in section \ref{sec::CFTWitten}, the total single-trace contribution to the s-channel exchange amplitudes in equation \eqref{qv2}   precisely coincides with A+B. From the ansatz 
\begin{equation*}
{\cal F}=a\cdot (\text{A}+\text{B})+c\cdot \text{C},
\end{equation*}
it is therefore clear that the cancellation of single trace conformal blocks in (\ref{qv2}) requires $a=1$. The value of $c$ can then be fixed by noting that A, B and C transform into each other under $y_{2} \leftrightarrow y_{3}$ and $y_{2} \leftrightarrow y_{4}$:\footnote{This can straightforwardly be seen from figure \ref{fig::wick}, where A, B and C are respectively the first, second and third diagrams on the second line.}
\begin{align}
y_{2} &\leftrightarrow y_{3}:  \qquad \left(\text{A},\text{B},\text{C}\right)  \rightarrow \left(\text{C},\text{B},\text{A}\right),  \\ \nonumber
y_{2} &\leftrightarrow y_{4}:  \qquad \left(\text{A},\text{B},\text{C}\right)  \rightarrow \left(\text{A},\text{C},\text{B}\right),
\end{align} 
Then since the total correlator \eqref{4ptconnected} is equal to A+B+C, the definition (\ref{13oooo}) demands that $2a+c=1$. Our choice $a=1$ therefore requires $c=-1$.\footnote{The representation with $c=0$ and $a=\tfrac{1}{2}$ was already derived in \cite{Diaz:2006nm}.} The RHS of (\ref{qv2}) then reads
\begin{equation}
\label{13oooo1}
{\cal F}(y_1,y_2;y_3,y_4) = \frac{1}{\left(y^2_{12} y^{2}_{34}\right)^{d-2}}\left\{\,  \sum\nolimits_s c^2_s\;G_{s+d-2\,,\, s}\left(u,v\right)-\bar{c}^2_{n,s}\;G_{\Delta_{n,s}\,,\, s}\left(u,v\right)\right\}.
\end{equation}
In the sequel, taking $d=3$ we solve equation (\ref{qv2}) for the quartic vertex in AdS$_4$ using the form \eqref{13oooo1} above for ${\cal F}(y_1,y_2;y_3,y_4)$. 
\subsubsection*{Solving for the vertex}
\begin{figure}[h]
 \centering
\includegraphics[width=0.8\linewidth]{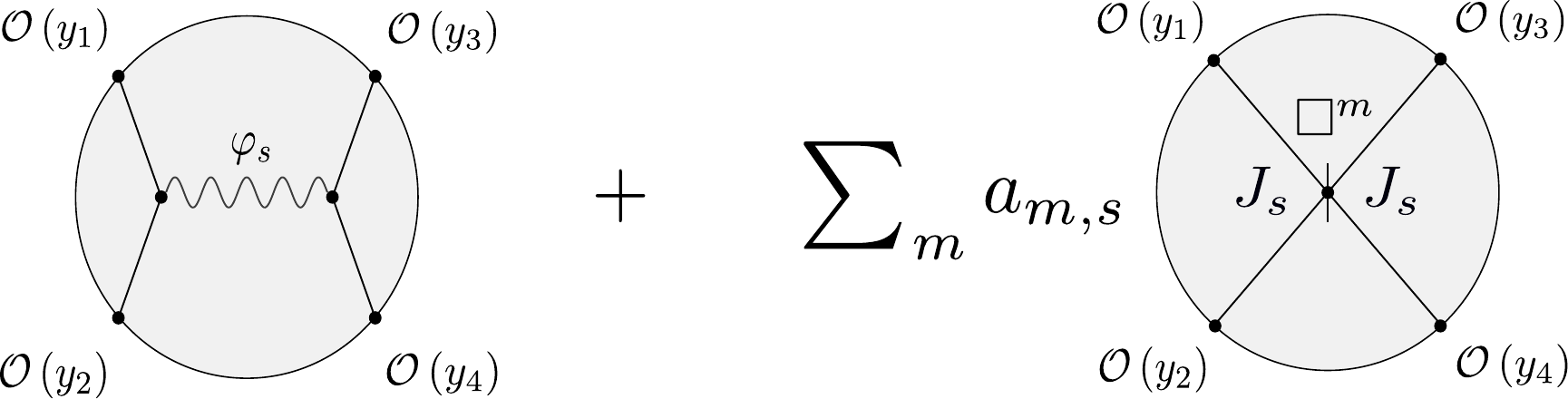}
 \caption{In AdS$_4$, the amplitudes of the above Witten diagrams generate only spin-$s$ conformal blocks. To determine the coefficients $a_{m,s}$ in the derivative expansion \eqref{quartsol} of the scalar quartic vertex, we identify them with the spin-$s$ contribution of the term \eqref{13oooo1cont} in the dual field theory on the RHS of equation \eqref{qv2}.}\label{fig::adscft4}
 \end{figure}
 
 As explained in section \ref{sec::combining}, we make the following ansatz for the quartic vertex
\begin{equation}
{\cal V}= \sum\nolimits_{m,s} a_{m,s}\; {\cal V}_{m,s}, \qquad  a_{m,s} \in \mathbb{R} \label{quartsol}
\end{equation}
composed of vertices ${\cal V}_{m,s}$ from the basis \eqref{quartbasis}, whose contribution to equation \eqref{qv2} is given by \eqref{ansatzamp}.

Extracting the vertex coefficients $a_{m,s}$ from equation (\ref{qv2}) is most straightforward using the contour integral representation of the conformal block expansion. In this formalism, our choice \eqref{13oooo1} for ${\cal F}$ in $d=3$ takes the form
\begin{equation} \label{13oooo1cont}
{\cal F}(y_1,y_2;y_3,y_4) =  \frac{1}{y^2_{12} y^{2}_{34}} \sum\nolimits_s \int^{\infty}_{-\infty} d\nu \; \left[\,p_{{\cal J}_s}\left(\nu\right) - \bar{p}_{{\cal O}^{(2)}_{s}}\left(\nu\right)\right]\kappa_s\left(\nu\right)G_{\tfrac{3}{2}+i\nu,s}\left(u,v\right),
\end{equation}
where $p_{{\cal J}_s}\left(\nu\right)$ is given by \eqref{pwhs} for the contribution of spin-$s$ operators ${\cal J}_s$ in the direct channel, while $\bar{p}_{{\cal O}^{(2)}_{s}}\left(\nu\right)$ gives the ${\cal O}\left(1/N\right)$ part of the double-trace operator contributions \eqref{directp}. The bulk amplitudes in (\ref{qv2}) are given by \eqref{cbexch} and \eqref{contactpwe} in this formalism. Although the functions $p_{s}\left(\nu\right)$ were originally introduced in a field theory context by equation \eqref{constr}, their definition extends naturally to the bulk amplitudes: For the total sum of the exchange amplitudes in (\ref{qv2}), we have
\begin{equation}
p^{\text{exch.}}_{s}\left(\nu\right) = \frac{2^{8-s}}{N \Gamma\left(s\right)^2} \frac{1}{\nu ^2+(s-\tfrac{1}{2})^2},
\end{equation}
to which only the spin-$s$ exchange contributes.\footnote{This is because in the specific case of AdS$_4$, spin-$s$ exchanges only generate spin-$s$ conformal blocks.} Likewise, for the contact amplitude of the quartic ansatz,
\begin{equation}
p^{\text{cont.}}_{s}\left(\nu\right) = \sum\nolimits_m a_{m,s} \left(-1\right)^m\left(\nu^2+s+\tfrac{9}{4}\right)^m,
\end{equation}
since the basis vertices ${\cal V}_{m,s}$ also generate only spin-$s$ contributions.

The coefficients $a_{m,s}$ for each fixed $s$ can now be determined for every $m = 0, 1, 2, ... $ by equating the total bulk and boundary $p_{s}\left(\nu\right)$ in equation (\ref{qv2}). In this way, the $a_{m,s}$ are given implicitly by 
\begin{align}\nonumber
&\sum\nolimits_m a_{m,s}\left(-1\right)^m\left(\nu^2+s+\tfrac{9}{4}\right)^m = p_{{\cal J}_s}\left(\nu\right)  - \frac{2^{8-s}}{N \Gamma\left(s\right)^2} \frac{1}{\nu ^2+(s-\tfrac{1}{2})^2} - \bar{p}_{{\cal O}^{(2)}_{s}}\left(\nu\right)
\\ \nonumber
&\hspace*{5.5cm} = \frac{2^{8-s}}{N} \frac{1}{\nu ^2+(s-\tfrac{1}{2})^2}\left[\frac{\pi}{\Gamma \left(\frac{2 s-2 i \nu +1}{4}\right)^2 \Gamma \left(\frac{2 s+2 i \nu +1}{4} \right)^2} - \frac{1}{ \Gamma\left(s\right)^2}\right] \\ \label{answervertex}
 &\hspace*{1.5cm} - \frac{1}{N} \frac{ \left(-1\right)^{\tfrac{s}{2}} \pi ^{\frac{3}{2}} 2^{s+5} \Gamma \left(s+\frac{3}{2}\right)\Gamma \left(\frac{s}{2}+\frac{1}{2}\right)}{\sqrt{2}\Gamma \left(\frac{s}{2}+1\right)\Gamma\left(s+1\right)\Gamma \left(\frac{3}{4}-\frac{i \nu }{2}\right) \Gamma \left(\frac{3}{4}+\frac{i \nu }{2}\right) \Gamma \left(s+\tfrac{1}{2}+ i \nu \right) \Gamma \left(s+\tfrac{1}{2}- i \nu \right)}.
\end{align}
 Then by setting $z = -\left(\nu^2+s+\tfrac{9}{4}\right)$, one can establish a generating function:
\begin{align} \nonumber
&a_s(z)\equiv\sum\nolimits_m a_{m,s}\;z^m. 
\end{align}
 This determines the quartic vertex \eqref{quartsol} of the parity even scalar in the minimal bosonic higher-spin theory on AdS$_4$ dual to the free $O(N)$ vector model in the form
\begin{equation}
{\cal V}= \sum\nolimits_{s\in 2\mathbb{N}} J_{s}\left(x, \partial_u\right) a_s(\Box)J_{s}\left(x, u\right),  \label{quartverts}
\end{equation}
and is one of the main results of this paper. In the next section we examine the properties of the vertex, particularly in the context of the locality of interactions in higher-spin theory. But first, let us make the following comments:

The solution (\ref{quartverts}) that we found contains vertices which are
not in the basis (\ref{quartbasis}) due to the terms with $m>k$. The fact that these extra vertices are
redundant just means that it should be possible to express them in terms of vertices
with $m\le k$ using the freedom of integration by parts and the free equations of motion. 
We used these extra vertices to avoid  conversions of conformal blocks between different channels and,
eventually, to present the result in a more concise form. 

Since we solve for the coefficients $a_{m,s}$ at the level of the integrand of the $\nu$ contour integral in \eqref{answervertex}, na\"ively it may seem that the vertex is not defined uniquely: One is free to add functions of $\nu$ that are entire within the contour. In doing so, this would change the form  of $a_{m,s}$ and thus the solution for the quartic vertex. However, the addition of such terms does not change the four-point amplitude and therefore the different vertices obtained in this manner would differ only by trivial terms which vanish on-free-shell, e.g. by the terms which can be accounted for by the redundancy of the representation
(\ref{quartsol}) with unbounded $m$.

Next, let us comment on the presence of anomalous dimensions in the bulk four-point amplitudes: As observed in section \ref{sec::CFTWitten}, the conformal block decompositions of the bulk four-point amplitudes contain contributions from double-trace operators with anomalous dimensions. These manifest themselves in the presence of double poles in the integrands of (\ref{cbexch}) and (\ref{contactpwe}) for $\nu = - i\left(2\left(d-2\right) +2n+s - \tfrac{d}{2}\right)$, $n = 0, 1, 2, ... $, which are the values of $\nu$ that yield double-trace operator contributions. Owing to the double-poles, evaluating the $\nu$-integral using the Cauchy theorem produces not only the ``non-anomalous'' double-trace conformal blocks $G_{\Delta_{n,s},s}$, but also their derivatives with respect to the dimension 
  $(\partial/\partial \Delta_{n,s})G_{\Delta_{n,s},s}$. These two terms together 
 indicate that it is not $G_{\Delta_{n,s},s}$ that is present in the conformal block decomposition,
 but rather the conformal block $G_{\Delta_{n,s}+\gamma_{n,s},s}$ with an anomalous dimension.  
 Indeed, this can be seen by expanding $G_{\Delta_{n,s}+\gamma_{n,s},s}$ as a Taylor series in $\gamma_{n,s}$,
 \begin{align}
 G_{\Delta_{n,s}+\gamma_{n,s},s} = G_{\Delta_{n,s},s} + \gamma_{n,s}\frac{\partial}{\partial \Delta_{n,s}}G_{\Delta_{n,s},s} + \mathcal{O}\left(1/N^2\right).
 \end{align}
The presence of anomalous dimensions is natural in the bulk theory of interacting higher spin fields. But at  the same time, the boundary theory is free
 and the double-trace operators ${\cal O}_{n,s}$ do not receive any anomalous dimensions. This is consistent
 with our (\ref{directp}) and (\ref{pwhs}) for the contour integral representation of the CFT four-point function, where the integrands contain only single poles in $\nu$. Thus, the holographic duality between higher spin theories and free $O(N)$
  vector models requires a rather delicate cancelation of anomalous conformal blocks in Witten diagrams.
  The quartic vertex that we have just found solves (\ref{qv1}), which implies, in particular, that
this cancellation indeed takes place.

In this respect, the contour integral representation for the conformal block decomposition
turned out to be a powerful tool: First of all, it allowed us to solve for a quartic vertex by
performing simple algebraic manipulations with the functions $p_s(\nu)$, instead of solving
an infinite system of linear equations balancing explicit conformal block coefficients of the
bulk and boundary results. Moreover, this representation allowed us to treat conformal
blocks with anomalous dimensions on the same footing as the non-anomalous ones, simply
by controlling the degree of the pole in $\nu$ at the associated point.

As a final comment, general consistency of the higher-spin holographic duality should imply that the vertex we have obtained is consistent with the Noether procedure. Further, to instead derive the vertex we found via the Noether procedure would involve going to quintic order in the fields. Indeed, the lowest order in the coupling constant where the quartic vertex for scalar fields contributes to the Noether consistency condition is
$g^3$, unlike other quartic vertices for fields of non-zero spin which first appear at the order $g^2$. 
The reason for this is that the scalar field does not posses its own undeformed gauge transformation.

\section{Locality}
\label{sec::locality}
Let us now turn to the issue of locality, which we outlined in the introduction. As mentioned there, while local theories are the standard arena of quantum field theory, locality is a debatable issue in higher-spin theories and has been the subject of recent investigations \cite{Vasiliev:2015wma,Kessel:2015kna}.
The standard approach to address the question of higher-spin interactions is to construct them
perturbatively using the Noether procedure \cite{Berends:1984rq} (which can be formalised as a BRST cohomological problem \cite{Barnich:1993vg,Stasheff:1997fe,Henneaux:1997bm}). In this approach,
one makes a general ansatz for {deformations of both the action and its gauge transformations}
and then selects consistent pairs {of such deformations} by demanding gauge invariance order by order in a coupling constant.
However, at quartic and higher orders in the number of fields this procedure is constraining only if the sought for vertices and deformations of gauge transformations are local \cite{Barnich:1993vg}. Otherwise any cubic interaction vertex consistent until this level, can be made
consistent with the gauge symmetry at all higher orders by an appropriate non-local deformation.
From this perspective, it seems reasonable to narrow the search for vertices by focusing
attention on local deformations only. 

In higher-spin theory, it is crucial to broaden the investigation to field theories which are ``quasilocal''\footnote{
This terminology is standard in functional renormalisation group literature (see e.g. \cite{Rosten:2010vm}), but in higher-spin literature, the term ``perturbatively local'' is also sometimes used (see e.g. \cite{Bekaert:2010hw}) and is essentially a synonym.} in the sense that
they possess a perturbative expansion (in powers of fields and their derivatives) where each individual term in the total Lagrangian is local, though the total number of derivatives may be unbounded in the full series.
Indeed, it is a well-established fact that higher-spin theory in dimension four and higher requires a spectrum of fields with unbounded spin and, consequently, interactions with an unbounded number of derivatives (since the maximal number of derivatives in a cubic vertex grows with the spin). In particular, Vasiliev's theory \cite{Vasiliev:1990en,Vasiliev:1992av,Vasiliev:2003ev} does contain infinitely many derivatives in 
its interaction terms. In order to resolve this apparent clash with locality, it appears necessary to define a weaker notion of locality that leaves room for vertices with infinitely many derivatives.

To get more intuition on how this generalised locality might be defined, let us for simplicity
concentrate on quartic interactions of scalars in flat space. It is natural to characterise interactions
by the corresponding $S$-matrix. It is well-known that local interactions give rise to amplitudes 
which are polynomial in Mandelstam variables. In turn, amplitudes for exchange diagrams, which are distinct from local quartic interactions, contain poles in Mandelstam variables. So, as a first requirement, one possibility is to demand that amplitudes of local contact interactions should be free
of poles.

A similar conclusion can be drawn by replacing the requirement of the amplitude to be polynomial
by the milder requirement that the coefficients in its Taylor series expansion should decrease fast enough.
To be more precise, one possibility is to demand an infinite radius of convergence
for this Taylor series. In this case the amplitude can be approximated
infinitely well by its Taylor series and for any values of Mandelstam 
parameters. This means, in turn, that the amplitude
is infinitely close to a local one.
The requirement for the amplitude to have an infinite
radius of convergence, or in other words, to be an entire function, not only excludes poles, but also branch cuts in the amplitude. 
According to these arguments, we define 
 \textit{a quartic interaction in flat spacetime to be weakly local iff the associated four-point amplitude is an entire function of Mandelstam
variables}.\footnote{A more refined classification could distinguish between ``strongly local'' (polynomial 4-point amplitude, e.g. $s^n$ with $n\in\mathbb N$)
``weakly local'' (entire, e.g. $exp(-s)$),
``quasi local'' (Taylor series around zero, e.g. $1/(s-s_0)$ with $s_0\neq 0$)
and ``strongly non-local'' (Laurent series around zero, e.g. $1/s$) quartic interactions.
}

To extend this definition to AdS space we turn to Mellin amplitudes. This extension is based on
numerous results indicating the similarity of Mellin amplitudes for Witten diagrams in AdS with flat amplitudes
in Minkowski space\cite{Mack:2009mi,Penedones:2010ue,Paulos:2011ie}. In particular, it was shown that
vertices with $2n$ covariant derivatives produce polynomial Mellin amplitudes of degree $n$ in Mandelstam
variables, exactly as in flat space amplitudes. It was also shown that Mellin amplitudes corresponding to 
AdS exchange Witten diagrams
 contain 
poles. Based on these results, and repeating the arguments detailed above for flat space vertices, we will consider
that \textit{in AdS spacetime a quartic interaction is weakly local iff the associated four-point Mellin 
amplitude is an entire function of Mandelstam
variables}. 

As shown in \cite{Penedones:2010ue}, a \textit{4-point conformal correlator does not contain single trace conformal blocks in the conformal block decomposition iff the associated
Mellin amplitude is an entire function}. According to this criterion, polynomial Mellin amplitudes and non-polynomial entire ones are in the same class, so it is rather natural to define locality in such a way
that the corresponding bulk vertices fall into the same class as well. This gives yet another argument in support
of the above definition of weak locality.

These definitions can be naturally extended to vertices of any degree in the fields and to fields with spin.
For the former the associated amplitudes then depend on a larger number of Mandelstam (or Mellin) variables, while for the latter one introduces polarisation vectors. Such flat (respectively, AdS) vertices can be considered weakly local if the associated amplitude is given by an entire function of Mandelstam (respectively, Mellin) variables for any value of polarisation vectors.

\subsubsection*{Flat limit}
Our definitions of weak locality in flat or AdS spacetime given above are independent. In particular, in the AdS case this definition does not rely on the existence of a flat limit. It can therefore be applied to higher-spin theories, which are notorious for their lack of a smooth flat limit. However, since the flat limit was important in the original works establishing the role of Mellin amplitudes as an AdS counterpart of scattering amplitudes in flat space, to avoid possible confusions we discuss this limit in more detail in the following.
  
 For a clear flat limit to exist, certain conditions have to be satisfied. In the flat limit, generally we expect to find an action that contains terms with different numbers of derivatives. Therefore, firstly, the theory in AdS should possess two dimensionful parameters (say with the dimensions of length): the curvature radius $R$ and furthermore an extra parameter $l$. The parameter $l$ remains finite in the limit $R/l \rightarrow \infty$ and is necessary to compensate the dimensions carried by the partial derivatives. Secondly, the action should not contain positive powers of $R$, or at least it should be possible to remove such terms by an overall rescaling of the action. To satisfy these two conditions, an $n$-point vertex ${\cal V}_n$ with a well-defined flat limit should therefore admit the following schematic form
 \begin{align}
 \label{flatads1}
 {\cal V}_n = f_0(l/R) \phi_1 \dots \phi_n + f_1(l/R) \phi_1 \dots (l\nabla) \dots (l\nabla) \dots \phi_n+
 \dots,
 \end{align}
 where $f_i(l/R)$ are finite at zero, i.e. $f_i(0)< \infty$. 
 Its flat limit is then given by
  \begin{align}
 \label{flatads2}
 {\cal V}_n = f_0(0) \phi_1 \dots \phi_n + f_1(0)  \phi_1 \dots (l\partial) \dots (l\partial) \dots \phi_n+
 \dots .
 \end{align}
In \cite{Penedones:2010ue} it was shown that the Mellin amplitude $M(\delta_{ij})$ for the
vertex (\ref{flatads1}) reproduces the flat space
amplitude $A(p_i\cdot p_j,l)$ for the vertex (\ref{flatads2}) in the limit $R/l \rightarrow \infty$, up to a fixed pre-factor,
\begin{equation*}
\lim_{R\to \infty} M(p_i\cdot p_j R^2, l)\propto A(p_i\cdot p_j, l),
\end{equation*} 
where the $p_i$ denote momenta of ingoing and outgoing particles.

Higher-spin gravity theories on the other hand do not possess a smooth flat limit. Firstly, the only dimensionful constant in the theory is the curvature radius. This means
that every covariant derivative enters in a dimensionless combination $R\nabla$. Moreover, interactions
of higher-spin fields contain terms with an unbounded number of derivatives, and consequently 
with arbitrarily large powers of $R$. In this case positive powers of $R$ cannot be removed from the action
by an overall rescaling, which eventually results in the absence of a smooth flat limit. Let us note, however, that separate
cubic vertices do possess a smooth flat space limit since they contain a finite number of derivatives.
For these vertices, what remains in the flat space limit is only the term with the highest number of 
derivatives \cite{Bekaert:2005jf,Boulanger:2008tg}.

\subsubsection*{Testing locality of the holographic quartic vertex}
We will now apply the definitions introduced in the above to test locality of the quartic vertex we obtained holographically in section \ref{sec::quartic}, given by \eqref{quartverts} with coefficients (\ref{answervertex}). 
One way to do this would be to evaluate the associated Mellin amplitude explicitly and
check whether or not it is an entire function. However, it is instead possible to give a simple argument for the analyticity of the Mellin amplitude by studying the operator contributions to the conformal block expansion of its four-point contact amplitude: In section \ref{sec::CFTWitten}, we observed that the exchange diagrams contributing to the bulk computation of the scalar single-trace operator four-point function \eqref{full} account for the entire contribution from the single-trace operators in the OPE \eqref{ope}. The quartic vertex \eqref{quartverts} thus only generates contributions from double-trace operators, which, as explained earlier in this section, implies that the associated Mellin amplitude must be an entire function.

An alternative way to conclude the weak locality of the quartic vertex, which is more direct and intuitively clear, can be attained by drawing on the similarity of Mellin amplitudes to flat space scattering amplitudes. We establish a flat space analogue of our vertex by simply replacing covariant derivatives with flat ones via
 \begin{align}
 \label{flatads3}
R\nabla \quad \rightarrow \quad l\partial,
 \end{align}
 and then test its locality based on the definitions we gave for flat space amplitudes in the previous section. This can be used to investigate the locality of our vertex in AdS, since Mellin amplitudes and flat scattering amplitudes for vertices related by (\ref{flatads3}) differ only by terms containing subleading powers of the Mandelstam variables. Given that the convergence of the Taylor series depends on the
rate of decay of the highest power coefficients, it is natural to expect that the Mellin amplitude for the original vertex and the flat scattering amplitude for the vertex obtained by (\ref{flatads3}) are entire (or not) 
simultaneously. We would like to emphasise that the replacement (\ref{flatads3}) is not related, strictly speaking, to the flat limit and
can be applied even in the case when the flat limit does not exist.\footnote{For instance, there is no reason to expect that the flat analogue of our vertex obtained by (\ref{flatads3}) is consistent with the Noether procedure in flat spacetime. Of course, if the flat limit would exist, then the vertex in flat space-time would be consistent with the
Noether procedure.}

For further analysis, it is convenient to split the complete vertex \eqref{quartverts} into sectors of fixed spin,\footnote{The spin is denoted $\ell$ here in order to distinguish it from the Mandestam variable $s$.}
\begin{equation}
\label{answervertexloc0}
{\cal V} = \sum_\ell {\cal V}_\ell,
\end{equation}
where 
\begin{equation}
\label{answervertexloc3}
{\cal V}_\ell=
 J_\ell(x,\partial_u) a_\ell(\Box) J_\ell(x,u)=\sum_m a_{m,\ell}J_\ell(x,\partial_u) \Box^m J_\ell(x,u)\,.
\end{equation}
After establishing the flat space analogue of the above vertex via (\ref{flatads3}), to simplify the evaluation of the corresponding flat space four-point amplitude, we replace the traceless conserved current $J_{\ell}$ with the current (\ref{bbvd}). This replacement is valid since the latter current is an improvement of the former, and so differs only by terms which do not contribute to the amplitude.

We first establish the weak locality of the vertex (\ref{answervertexloc0}) in the spin-$\ell$ sector, by showing that the amplitude 
\begin{equation}
\label{answervertexloc4}
 \quad {\cal A}_\ell(s,t,u)=a_\ell(s)(t-u)^\ell=\sum_m a_{m,\ell}s^m (t-u)^\ell\,,
\end{equation}
generated by the flat space analogue of the vertex ${\cal V}_\ell$ is an entire function in the Mandelstam variable $s$. This can be done by demonstrating that $a_{\ell}\left(\nu\right)$ is an entire function of $\nu$: From studying its explicit form \eqref{answervertex}, the only pole that could possibly appear is at $\nu^2=-(\ell-1/2)^2$, from denominators of the two terms on the first line. It is straightforward to check that these poles cancel. Indeed, this is due to our special
choice of the ${\cal F}$ amplitude (\ref{13oooo1}), which was made precisely to cancel the single trace contributions on the two sides of (\ref{qv2}). The associated flat amplitude (\ref{answervertexloc4}) is therefore an entire
function of Mandelstam variable $s$.

Thus we have shown that in each sector of fixed spin-$\ell$ the vertex (\ref{answervertexloc0}) 
is weakly local, since the corresponding flat space amplitudes \eqref{answervertexloc4} are entire in the Mandelstam variable $s$. To conclude that the whole vertex is weakly local, it remains to show that the sum
$\sum_\ell {\cal A}_\ell$ has an infinite radius of convergence in $(t-u)$. This is indeed the case, since
 the dominant spin dependence of $a_\ell$ at large spin $\ell$ is $1/\Gamma^2(\ell)$, so the 
 series converges for any $(t-u)$ by the ratio test. 

\subsubsection*{Mellin amplitudes}

Let us briefly comment on the use of Mellin amplitudes. So far we only touched upon Mellin amplitudes at the final stage of our work to quantify locality or non-locality of
the quartic vertex. For the main part of our computation,
we used the more pedestrian approach of the conformal block decomposition.
In solving the same problem entirely in terms of Mellin amplitudes, one faces
the following two problems. The first problem is that Mellin amplitudes for spinning correlators
involve Mack polynomials \cite{Mack:2009mi}, which are very complicated.
This eventually makes computations intractable {at the present stage.}
Another problem is that
the correlators that one encounters in free theories do not admit a well-defined Mellin transform.
This can be illustrated by considering the term $f(u)\equiv u^{d/2-1}$ in the
correlator (\ref{4ptconnected}) and applying the Mellin transform,
\begin{equation*}
{\cal M}[f](z) \equiv \int_0^{\infty} u^{z-1} f(u) du.
\end{equation*}
The integral on the RHS diverges for any value of $z$, and that is why the Mellin transform ${\cal M}[f]$ of a power function is ill-defined.  Regularisation of the Mellin transform for power functions remains an open issue. However, a possible way out is that a function $\tau(z)$ may exist
whose inverse Mellin transform
\begin{equation*}
{\cal M}^{-1}[\tau](u)=\frac{1}{2\pi i} \int_{-i\infty}^{i\infty} u^{-z}\tau(z) dz.
\end{equation*}
is equal to the power function $f(u)$.

Such an example precisely appeared during our computations. Let us consider
\begin{equation*}
\label{Mellinx}
\chi(z)\equiv\frac{2^{-2z}}{z^2+(s-1/2)^2}\frac{\Gamma(z+1/2)\Gamma^2\left(\frac{2s+2z+3}{4}\right)}{\Gamma(z)\Gamma^2\left(\frac{2s+2z+1}{4}\right)(2z+2s+1)},
\end{equation*}
which can be obtained from $f_{{\cal J}_s}(\nu)=p_{{\cal J}_s}(\nu)\kappa_{s}(\nu)$ with
$p_{{\cal J}_s}(\nu)$ and $\kappa_{s}(\nu)$ given by (\ref{pwhs}) and (\ref{kappa}) respectively
by the change of a variable $z\equiv i\nu$, and dropping unimportant $z$ independent terms.
It has only one pole in the right half-plane located at $z=s-1/2$.
Closing the integration
contour in  the right half-plane and picking a residue at $z=s-1/2$, for the
inverse Mellin transform of $\chi(z) $ we find
\begin{equation}
\frac{1}{2\pi i} \int_{-i\infty}^{i\infty} u^{-z}\chi(z) dz \sim u^{-s+1/2}.
\end{equation}
In this sense, $\chi(z)$ can be viewed as the Mellin transform of the power function $u^{-s+1/2}$.
We leave further study of this interesting issue for future research.

\subsubsection*{Locality from CFT}
As noted in section \ref{sec::CFTWitten}, the exchange diagrams contributing to the bulk computation of the scalar single-trace operator four-point function \eqref{full} account for the entire contribution from the single-trace operators in the OPE \eqref{ope}. It is therefore clear that any quartic self interaction of the bulk scalar, to be consistent with the OPE in the dual field theory, can only generate double-trace operator contributions. While the four-point amplitude generated by a contact interaction that is local in the strong sense (that is, with a finite number of derivatives) will always only yield double-trace contributions to a CFT four-point function, it is not necessarily the case when such a vertex contains an unbounded number of derivatives. However, this is guaranteed by the crossing symmetry of the dual CFT, as it places strong constraints on the coefficients in the derivative expansion of the vertex \cite{Heemskerk:2009pn}, and ensures that it is weakly local. It would be interesting to investigate further the power of the OPE and crossing in CFTs admitting a large $N$ expansion in taming non-local bulk interactions: Indeed, it seems that when exchange Witten diagrams generate the entire single-trace contribution of their dual CFT correlators (which may well be always the case), these constraints ensure that quartic contact interactions will always satisfy some notion of locality.

\section{Summary and Outlook}

In this work we have used holography to shed some light on the elusive nature of quartic interactions in higher-spin theories on AdS space. In particular, via the identification of Witten diagrams and CFT correlation functions, we extracted and analysed quartic self-interaction of the scalar in type A minimal bosonic higher-spin theory on AdS$_4$. We found that while the interaction is non-local in the sense that the vertex is unbounded in its number of derivatives, it still satisfies a weaker notion of locality in that its four-point bulk amplitude behaves similarly to those of local vertices, and does not exhibit any pathologically non-local behaviour. The weak locality of our quartic vertex appears to be ensured by the holographic duality, due to OPE and crossing constraints from the dual CFT. This would suggest that the interactions of higher-spin gravity theories dual to weakly-coupled CFTs are to some degree local. 

Although we focused our study on the simplest example of the scalar self-interaction, we expect that the techniques employed in this case can be extended to allow for the holographic study of quartic vertices involving fields with non-zero spin, as well as higher-order vertices. It would also be interesting to develop further the technology of Mellin amplitudes, in order for them to be applied more effectively in the higher-spin context. This would allow our definition of weak locality in terms of the analyticity of Mellin amplitudes to be more directly applicable.

We hope that this work will provide a basis for further discussion of locality in higher-spin theories.

\subsection*{Acknowledgements}

We thank S.~Chester, E.~Joung, K.~Mkrtchyan, H.~Osborn, A.~C.~Petkou, E.~Skvortsov and M.~Taronna for useful discussions. We also thank A. L.~Fitzpatrick, V.~Gon\c{c}alves and J.~Kaplan for helpful correspondence.

The research of X.B. was supported by the Russian Science Foundation grant 14-42-00047 in association with Lebedev Physical Institute. The work of D.P. was supported by the  DFG grant HO 4261/2-1 ``Generalized dualities relating
gravitational theories in 4-dimensional Anti-de Sitter space with 3-dimensional conformal field
theories''. The work of J.E. and C.S. was partially supported by the European Science Foundation Holograv network (Holographic methods for strongly coupled systems). 

\appendix
\section{Notation and conventions}
\label{appendix::notation}
\subsection{Symmetric tensors in AdS}
Symmetric tensors in AdS can be managed in a compact way by encoding them in generating functions. For example, a symmetric rank-$r$ tensor $t_{\mu_1 ... \mu_r}$ is represented in this formalism by
\begin{equation}
t_{\mu_1 ... \mu_r}\left(x\right) \rightarrow t\left(x, u\right) = \frac{1}{r!} t_{\mu_1 ... \mu_r}\left(x\right) u^{\mu_1} ... u^{\mu_r},
\end{equation}
where we have introduced the constant auxiliary vector $u^{\mu}$. The original tensor can be recovered by taking derivatives $\partial/\partial{u^{\mu}}$.

In this formalism, tensor operations are replaced by an operator calculus. For example, the contraction of two rank-$r$ symmetric tensors $t_{\mu_1 ... \mu_r}$ and $s_{\mu_1 ... \mu_r}$ is implemented via
\begin{equation}
t_{\mu_1 ... \mu_r}\left(x\right)s^{\mu_1 ... \mu_r}\left(x\right) = \frac{1}{r!} \, t\left(x, \partial_u\right) s\left(x, u\right)\Big|_{u=0} = \frac{1}{r!} \, s\left(x, \partial_u\right) t\left(x, u\right)\Big|_{u=0}.
\end{equation} 
Likewise, the following operations can be represented: 
\begin{equation}
\text{divergence:} \quad \nabla \cdot \partial_u, \hspace*{1cm} \text{symmetrised gradient:} \quad u \cdot \nabla, \hspace*{1cm} \text{trace:} \quad \partial_u \cdot \partial_u.
\end{equation}
The symmetric metric $g_{\mu \nu}$ is denoted simply by $u^2$, and thus terms proportional to $u^2$ are pure trace. 

\subsection{Symmetric and traceless boundary tensors}
On the boundary we work with primary fields represented by symmetric and traceless tensors on $d$-dimensional flat Euclidean conformal space. Like a symmetric tensor in AdS, a symmetric and traceless boundary tensor $t_{i_1 ... i_r}\left(y\right)$ can be encoded in a polynomial of a constant auxiliary boundary vector $z^i$, 
\begin{equation}
t_{i_1 ... i_r}\left(y\right) \rightarrow t\left(y, z\right) = \frac{1}{r!} \,t_{i_1 ... i_r}\left(y\right) z^{i_1} ... z^{i_r},
\end{equation}
but with the additional requirement that $z^2 = 0$ to enforce tracelessness.

Contractions of symmetric and traceless tensors can be implement by the Thomas-D operator $\hat{\partial}_{z^i}$
\begin{align}
t_{i_1 ... i_r}\left(y\right)s^{i_1 ... i_r}\left(y\right) = \frac{1}{r!\left(\tfrac{d}{2}-1\right)_r} \, t(y, \hat{\partial}_z) s\left(y, z\right) = \frac{1}{r!\left(\tfrac{d}{2}-1\right)_r} \, s(y, \hat{\partial}_z) t\left(y, z\right),
\end{align}
and \cite{1926, Dobrev:1975ru,Costa:2011mg}
\begin{equation}
\hat{\partial}_{z^i} = \left(\tfrac{d}{2} - 1+ z \cdot \frac{\partial}{\partial z}\right)\frac{\partial}{\partial z^i} - \frac{1}{2} z_i \frac{\partial^2}{\partial z \cdot \partial z}.
\end{equation}

\section{AdS harmonic functions}
\label{appendix::harmonic}
The AdS$_{d+1}$ bi-tensorial harmonic functions $\Omega_{\nu,\ell}\left(x_1, u_1; x_2, u_2\right)$ are spin-$\ell$ eigenfunctions of the Laplace operator
\begin{equation}
\left(\Box_1 + \left(\tfrac{d}{2}\right)^2 + \nu^2+\ell \right) \Omega_{\nu,\ell}\left(x_1, u_1; x_2, u_2\right) = 0, \qquad \nu \in \mathbb{R},
\end{equation}
which are traceless and transverse
\begin{equation}
\left(\partial_{u_1} \cdot \partial_{u_1} \right) \Omega_{\nu,\ell}\left(x_1, u_1; x_2, u_2\right) = 0, \qquad \left(\nabla \cdot \partial_{u_1} \right) \Omega_{\nu,\ell}\left(x_1, u_1; x_2, u_2\right) = 0. \label{harmeom}
\end{equation}
These functions provide a complete basis for symmetric and traceless rank-$r$ bi-tensors in AdS$_{d+1}$, given by
\begin{equation}
\left\{\left(w_1 \cdot \nabla_1\right)^{\ell}\left(w_2 \cdot \nabla_2\right)^{\ell}\Omega_{\nu,\ell-r}\left(x_1, w_1; x_2, w_2\right) \,\big| \quad \nu \in \mathbb{R}, \quad \ell = 0, 1, ..., r\,\right\}, \label{stbasis}
\end{equation}
with completeness relation
\begin{align}
\left(w_1 \cdot w_2\right)^{r} \delta^{d+1}\left(x_1-x_2\right) = \sum^{r}_{\ell=0} \int^{\infty}_{-\infty} d\nu \; c_{r,r-\ell}(\nu) \,\left(w_1 \cdot \nabla_1 \right)^{r-\ell} \left(w_2 \cdot \nabla_2 \right)^{r-\ell} \Omega_{\nu, \ell}\left(x_1,w_1;x_2,w_2\right), \label{complete1}
\end{align}
where
\begin{equation}
c_{r,\ell}(\nu) = \frac{2^{\ell} \left(r-\ell+1\right)_{\ell}\left(\tfrac{d}{2}+r-\ell - \frac{1}{2}\right)_{\ell}}{\ell! \left(d+2r-2\ell -1\right)_{\ell} \left(\tfrac{d}{2}+r-\ell -i\nu\right)_{\ell}  \left(\tfrac{d}{2}+r-\ell +i\nu\right)_{\ell}}. \label{complete2}
\end{equation} 
In the above we used constant light-like auxiliary vectors $w_{1,2}$ to enforce tracelessness.

AdS harmonic bi-tensors have many useful properties, and we only presented those which were useful for our purposes here. For more information, the reader can see for example section 4.C of \cite{Penedones:2007ns}, and \cite{Costa:2014kfa,Bekaert:2014cea} for their application in a similar context.

\section{Complete Set of Quartic Vertices}
\label{appendix::quarticbasis}
Here we  justify the basis  of quartic vertices  chosen in (\ref{quartbasis}). 
For our purposes one needs to find a complete set of quartic vertices which
are independent on-free-shell. Moreover, the same vertex can be brought to 
different forms using integration by parts. All possible relations of this type
should also be taken into account. Counting quartic vertices up to
total derivatives and equations of motion is equivalent to counting the associated
4-point amplitudes. For simplicity, the following analysis will be performed in flat 
space. 

Independent flat space 4-point amplitudes have been counted in \cite{Heemskerk:2009pn}.
They are generated by monomials of the form
\begin{equation}
\label{verticesfromHPPS}
 s^kt^ku^m\quad \text{with integers}\quad  k\ge m\ge 0\,,
 \end{equation}
where $s$, $t$ and $u$ are the Mandelstam variables
\begin{equation*}
s=(p_1+p_2)^2, \quad t=(p_1+p_3)^2, \quad u=(p_2+p_3)^2\,.
\end{equation*}
In massless case $(p_i)^2=0$ and one has
\begin{equation*}
s=2\,p_1\cdot p_2=2\,p_3\cdot p_4, \quad t=2\,p_1\cdot p_3=2\,p_2\cdot p_4, \quad u=2\,p_2\cdot p_3=2\,p_1\cdot p_4\,.
\end{equation*}

Our goal now is to replace the basis (\ref{verticesfromHPPS}) with an equivalent one,
but better suited for our purposes. Let us first consider the vertices in position space
\begin{equation}
\label{verticesequivalent1}
\Box_{12}^m \tilde{J}_{\mu_1\cdots\mu_\ell}\big(\phi(x_1),\phi(x_2)\big)\tilde{J}^{\mu_1\cdots\mu_\ell}\big(\phi(x_3),\phi(x_4)\big),
\end{equation}
with $\Box_{12}=(\partial_{x_1}+\partial_{x_2})^2$ and
\begin{equation}
\label{bbvd}
\tilde{J}_{\mu_1\cdots\mu_\ell}\big(\phi(x_1),\phi(x_2)\big)\equiv \phi(x_1)
\overleftrightarrow\partial_{\mu_1}\cdots\overleftrightarrow\partial_{\mu_\ell}
\phi(x_2)\,,
\end{equation}
are the conserved currents of \cite{Berends:1985xx}, with $\overleftrightarrow\partial=\overleftarrow\partial_{x_1}-\overrightarrow\partial_{x_2}$.
The amplitude in momentum space associated to the vertex \eqref{verticesequivalent1} reads
\begin{align}
(p_1+p_2)^m\left[(p_1-p_2)\cdot (p_3-p_4)\right]^l=s^m (t-u)^\ell.
\end{align}
For $\ell=2k$, this amplitude contains a term $s^m t^k u^k$. Up to a Bose symmetry transformation
this reproduces the form of the vertices 
 (\ref{verticesfromHPPS}).
Thus, we conclude that vertices (\ref{verticesequivalent1})
with $\ell=2k$ and $k\ge m\ge 0$ generate the basis of quartic vertices.

In order to write compact formulae involving symmetrised multi-indices, in this appendix and in the following ones we will make use of Vasiliev notation e.g. explained in footnote 2 of \cite{Vasilev:2011xf}.

For our purposes it is more convenient to use not the conserved current $\tilde{J}$ (\ref{bbvd}), but its improved traceless version \cite{Anselmi:1999bb}\footnote{Note that the normalisation here is such that it is consistent with the normalisation of \eqref{bbvd}.}
\begin{eqnarray} \label{anselmi}
J_{\mu(l)}\big(\phi(x_1),\phi(x_2)\big)=  \frac{\sqrt{\pi}\;\Gamma\left(d+s-2\right)}{2^{s+d-3}\Gamma\left(\tfrac{d-1}{2}\right)\Gamma\left(s+\tfrac{d}{2}-1\right)} \sum_{k=0}^l a_k (\partial_\mu)^k \phi(x_1) (\partial_\mu)^{l-k}\phi(x_2)- \text{traces},
\end{eqnarray}
where 
\begin{equation*}
a_k=(-1)^k \frac{l!}{k! (l-k)!}\frac{(\delta)_l}{(\delta)_k(\delta)_{l-k}}\,.
\end{equation*}
The improvement \eqref{anselmi} can always be done when the scalar field is conformal.
Analogously to (\ref{verticesequivalent1}) the flat space amplitude
corresponding to  a vertex
\begin{equation}
\label{verticesequivalent2}
{\cal V}_{n,\ell}\equiv\Box_{12}^m J_{\mu(\ell)}\big(\phi(x_1),\phi(x_2)\big)J^{\mu(\ell)}\big(\phi(x_3),\phi(x_4)\big)
\end{equation}
with $\ell=2k$ generates a term $s^mt^ku^k$. Thus, vertices (\ref{verticesequivalent2})
with $\ell=2k$ and  $k\ge m\ge 0$ can be used as a basis for quartic interactions.

\section{Double Trace Operators}
\label{appendix::doubletrace}
In this appendix we construct primary operators bilinear in two scalar primary operators
${\cal O}_1$ and ${\cal O}_2$
of dimensions $\Delta_1$ and $\Delta_2$. We use the convention where the conformal algebra is given by
\begin{align}
[M_{ij},P_k]=i(\eta_{ik}P_j-\eta_{jk}P_i), & \qquad [M_{ij},K_k]=i(\eta_{ik}K_j-\eta_{jk}K_i),\\
[M_{ij},D]=0, & \qquad [P_i,K_j]=-2(\eta_{ij}D+i M_{ij}),\\
[D,P_i]=P_i, & \qquad [D,K_i]=-K_i.
\end{align}
Primary operators ${\cal O}_1$ and ${\cal O}_2$ are annihilated by the special conformal 
generator  $K_i$
\begin{equation}
\label{primary}
K_i{\cal O}_1=0, \qquad K_i{\cal O}_2=0,
\end{equation}
and have well-defined scaling dimensions 
\begin{equation}
\label{dimensions}
D{\cal O}_1=\Delta_1{\cal O}_1,\qquad D{\cal O}_2=\Delta_2{\cal O}_2.
\end{equation}
Our goal is to construct a full set of symmetric higher spin primary operators.
It is well known that these operators have a schematic form\footnote{When $\mathcal{O}_1$ and $\mathcal{O}_2$ are identical operators, $\mathcal{O}_2 = \mathcal{O}_1 = \mathcal{O}$, for concision we denote the corresponding double-trace operators by $[{\cal O}{\cal O}]_{n,s} = {\cal O}^{(2)}_{n,s}$. This is employed throughout the main text, and in appendix \ref{appendix::cn0}.}
\begin{equation}
\label{schem1}
[{\cal O}_1{\cal O}_2]_{n,s} = {\cal O}_1\partial_{i(s)}\Box^n {\cal O}_2 +\dots,
\end{equation}
where ellipsis denote terms, which are required to make $[{\cal O}_1{\cal O}_2]_{n,s}$
primary. The double trace operator $[{\cal O}_1{\cal O}_2]_{n,s}$ has spin $s$ and
scaling dimension $\Delta=\Delta_1+\Delta_2+2n+s$.

Our goal here is to specify implicit terms in (\ref{schem1}). To this end we make the 
most general ansatz
\begin{equation}
\label{ansatz}
[{\cal O}_1{\cal O}_2]_{n,s} = \sum_{s_1,b_1,b_2} a_{n,s}(s_1,s_2;b_1,b_2,b_{12})
\partial_{i(s_1)}\Box^{b_1}\partial^{j(b_{12})}{\cal O}_1 \partial_{i(s_2)}\Box^{b_2}\partial_{j(b_{12})}{\cal O}_2-\text{traces},
\end{equation}
and specify $a_{n,s}$ from the requirement  that $[{\cal O}_1{\cal O}_2]_{n,s}$
is primary. 
In this sum $s=s_1+s_2$ and $n=b_1+b_2+b_{12}$, so there are only three independent
summations.

\subsection{The action of the conformal boost generator}

Let us replace derivatives with the momentum operator according to $P_i=i\partial_i$
and contract free indices with the auxiliary traceless symmetric tensor $V^{i(s)}$ to ensure that
$[{\cal O}_1{\cal O}_2]_{n,s} $ is traceless and symmetric. A typical term from the sum (\ref{ansatz})
will be denoted as
\begin{equation}
\label{oneterm}
T_{n,s}(s_1,s_2;b_1,b_2,b_{12})=V^{i(s_1)j(s_2)}(\eta^{ij})^{b_1}(\eta^{ij})^{b_{12}}(\eta^{jj})^{b_2}P_{i(s_1+2b_1+b_{12})}{\cal O}_1 P_{j(s_2+2b_2+b_{12})}{\cal O}_2.
\end{equation}

Using the conformal algebra, (\ref{primary}) and (\ref{dimensions}) one can show that
\begin{equation}
\label{specialconformal}
K_j P_{i(n)}{\cal O}_1=2n (\Delta_1+n-1)\eta_{ij}P_{i(n-1)}{\cal O}_1-n(n-1)\eta_{ii}P_j P_{i(n-3)}{\cal O}_1.
\end{equation}
Employing (\ref{specialconformal}) we can evaluate that
\begin{align}
\notag
K_j T_{n,s}(s_1,s_2;b_1,b_2,b_{12})
&=2[\Delta_1+s_1+2b_1+b_{12}-1]s_1 C_V(s_1-1,s_2;b_1,b_2,b_{12})\\
\notag
&+2b_1[2\Delta_1+2b_1-d] C_1(s_1+1,s_2;b_1-1,b_2,b_{12})\\
\notag
&-b_{12}(b_{12}-1)C_{1}(s_1+1,s_2;b_1,b_2+1,b_{12}-2)\\
\notag
&-2s_1 b_{12} C_1(s_1,s_2+1;b_1,b_2,b_{12}-1)\\
\notag
&+2[\Delta_1+s_1+2b_1+b_{12}-1]b_{12}C_2(s_1,s_2+1;b_1,b_2,b_{12}-1)\\
\notag
&+2[\Delta_2+s_2+2b_2+b_{12}-1]s_2 C_V(s_1,s_2-1;b_1,b_2,b_{12})\\
\notag
&+2b_2[2\Delta_2+2b_2-d] C_2(s_1,s_2+1;b_1,b_2-1,b_{12})\\
\notag
&-b_{12}(b_{12}-1)C_{2}(s_1,s_2+1;b_1+1,b_2,b_{12}-2)\\
\notag
&-2s_2 b_{12} C_2(s_1+1,s_2;b_1,b_2,b_{12}-1)\\
\label{Kaction}
&+2[\Delta_2+s_2+2b_2+b_{12}-1]b_{12}C_1(s_1+1,s_2;b_1,b_2,b_{12}-1),
\end{align}
where 
\begin{align}
\notag
&C_V(s_1-1,s_2;b_1,b_2,b_{12})\\
\notag
&\quad=V^{ji(s_1-1)k(s_2)}(\eta^{ii})^{b_1}(\eta^{ik})^{b_2}(\eta^{kk})^{b_2}P_{i(s_1+2b_1+b_{12}-1)}{\cal O}_1 P_{k(s_2+2b_2+b_{12})}{\cal O}_2,\\
\notag
&C_{1}(s_1+1,s_2;b_1-1,b_2,b_{12})\\
\notag
&\quad=V^{i(s_1)k(s_2)}(\eta^{ii})^{b_1-1}(\eta^{ik})^{b_{12}}(\eta^{kk})^{b_2}P_j P_{i(s_1+2b_1+b_{12}-2)}{\cal O}_1P_{k(s_2+2b_2+b_{12})}{\cal O}_2,\\
\label{structures}
&C_2(s_1,s_2+1;b_1,b_2,b_{12}-1)\notag\\
&\quad=V^{i(s_1)k(s_2)}(\eta^{ii})^{b_1}(\eta^{ik})^{b_{12}-1}(\eta^{kk})^{b_2}P_{i(s_1+2b_1+b_{12}-1)}{\cal O}_1P_{j}P_{k(s_2+2b_2+b_{12}-1)}{\cal O}_2.
\end{align}
The only difference of this formula with the one found by \cite{Fitzpatrick:2011dm} is that we
do not symmetrise the index carried by $K_j$ with the remaining ones and for this reason
we have more independent structures on the RHS of (\ref{Kaction}).

\subsection{Imposing that the double-trace operator is primary}

Having understood how the special conformal generator $K_j$ acts on each term of the 
ansatz (\ref{ansatz}) we act with $K_j$ on the complete expression. 
Using (\ref{Kaction}), $K_i [{\cal O}_1{\cal O}_2]_{n,s}$ can be expressed in terms
of  three independent structures $C_V$, $C_1$ and $C_2$ defined in (\ref{structures}).
Setting the prefactors for each of these structures to zero we obtain three equations on $a_{n,s}$.
 The equation corresponding to $C_V$ reads
\begin{align}
\notag
&a_{n,s}(s_1,s_2;b_1,b_2,b_{12}) 2[\Delta_1+s_1+2b_1+b_{12}-1]s_1\\
\label{cv}
&\qquad+
 a_{n,s}(s_1-1,s_2+1;b_1,b_2,b_{12})2[\Delta_2+s_2+2b_2+b_{12}](s_2+1)=0.
\end{align}
It can be used to fix the dependence of $a_{n,s}$ on $s_1$ for fixed $b_1$ and $b_2$
\begin{equation}
\label{sdep}
a_{n,s}(s_1,s_2;b_1,b_2,b_{12})=(-1)^{s_2} \frac{s!}{s_2! s_1!}\frac{(\Delta_1+2b_1+b_{12}+s_1)_{s_2}}{(\Delta_2+2b_2+b_{12})_{s_2}}a_{n,s}(s,0;b_1,b_2,b_{12}).
\end{equation}

Another equation, which sets the prefactor of $C_1$ to zero reads
\begin{align}
\notag
&a_{n,s}(s_1,s_2;b_1,b_2,b_{12})2b_1 (2\Delta_1+2b_1-d)\\
\notag
&\quad-a_{n,s}(s_1,s_2;b_1-1,b_2-1,b_{12}+2)(b_{12}+2)(b_{12}+1)\\
\notag
&\quad\quad-a_{n,s}(s_1+1,s_2-1;b_1-1,b_2,b_{12}+1)2(s_1+1)(b_{12}+1)\\
\label{2}
&\quad\quad\quad+a_{n,s}(s_1,s_2;b_1-1,b_2,b_{12}+1)2(\Delta_2+s_2+2b_2+b_{12})(b_{12}+1)=0.
\end{align}
The third term contains $a_{n,s}(s_1+1,s_2-1;b_1-1,b_2,b_{12}+1)$, which can be 
expressed in terms of $a_{n,s}(s_1,s_2;b_1-1,b_2,b_{12}+1)$ by (\ref{cv}). This results into
\begin{align}
\notag
&a_{n,s}(s_1,s_2;b_1,b_2,b_{12})2b_1 (2\Delta_1+2b_1-d)\\
\notag
&\quad-a_{n,s}(s_1,s_2;b_1-1,b_2-1,b_{12}+2)(b_{12}+2)(b_{12}+1)\\
\notag
&\quad\quad+a_{n,s}(s_1,s_2;b_1-1,b_2,b_{12}+1)2(b_{12}+1)(\Delta_2+s_2+2b_2+b_{12})
\\
\label{rec1}
&\qquad\qquad\qquad\qquad\qquad\qquad\qquad\qquad\qquad\times \frac{\Delta_1+s+2b_1+b_{12}-1}{\Delta_1+s_1+2b_1+b_{12}-1}=0,
\end{align}
which relates $a_{n,s}$'s for the same values of arguments $s_1$ and $s_2$ and different
values of $b_1$, $b_2$ and $b_{12}$. This equation
together with the one obtained in a similar way for $C_2$, namely, 
\begin{align}
\notag
&a_{n,s}(s_1,s_2;b_1,b_2,b_{12})2b_2 (2\Delta_2+2b_2-d)\\
\notag
&\quad-a_{n,s}(s_1,s_2;b_1-1,b_2-1,b_{12}+2)(b_{12}+2)(b_{12}+1)\\
\notag
&\quad\quad+a_{n,s}(s_1,s_2;b_1,b_2-1,b_{12}+1)2(b_{12}+1)(\Delta_1+s_1+2b_1+b_{12})
\\
\label{rec2}
&\qquad\qquad\qquad\qquad\qquad\qquad\qquad\qquad\qquad\times \frac{\Delta_2+s+2b_2+b_{12}-1}{\Delta_2+s_2+2b_2+b_{12}-1}=0,
\end{align}
defines  $b$-dependence of $a_{n,s}$ for fixed $s_1$ and $s_2$.

\subsection{Determining the dependence on the wave operator}
Now we will solve the recurrence equations (\ref{rec1}), (\ref{rec2}) with
a boundary condition
\begin{equation}
\label{boundcond}
a_{n,s}(s_1,s_2;b_1,b_2,b_{12})=0, \quad \text{for $b_1<0$ or $b_2<0$ or $b_1+b_2>n$}.
\end{equation}
More precisely, we are going to express the unknown coefficients in terms of 
$a_{n,s}(s_1,s_2;0,0,n)$. For brevity, in the following we will keep only the arguments
$b_1$ and $b_2$ of $a_{n,s}$ explicit.

First, we consider (\ref{rec1}) for $b_2=0$. Then the second term drops out due to 
the boundary condition and the recurrence equation simplifies. It allows to express
$a(b_1,0)$ with arbitrary $b_1$ in terms of $a(0,0)$
\begin{equation}
\label{b2eq0}
a(b_1,0)= (-1)^{b_1}\frac{n!}{b_1! b_2!}\frac{(\Delta_2+s_2+n-b_1)_{b_1}(\Delta_1+s+n)_{b_1}}{2^{b_1}(\Delta_1+1-h)_{b_1} (\Delta_1+s_1+n)_{b_1}}a(0,0).
\end{equation}

Then we use (\ref{rec2}) to specify $b_2$-dependence in the following way. First, we express $a(b_1,1)$
for any $b_1$
in terms of $a(b_1,0)$ and $a(b_1-1,0)$, which are known. Then we solve for
 $a(b_1,2)$ in terms of $a(b_1,1)$ and $a(b_1-1,1)$, which have been
determined at the previous step. This process can be continued to define $a(b_1,b_2)$
for any $b_2$ in terms of $a(0,0)$. Performing this procedure up to $b_2=3$ we conjecture 
\begin{align}
\notag
& a(b_1,b_2)=\left(-\frac{1}{2}\right)^{b_1+b_2}\frac{n!}{b_1!b_2!b_{12}!}\\
\notag
&\quad\times
\frac{(\Delta_1+s+n)_{b_1}(\Delta_2+s+n-b_1)_{b_2}}{(\Delta_1+1-h)_{b_1}(\Delta_2+1-h)_{b_2}(\Delta_1+s_1+n)_{b_1-b_2}(\Delta_2+s_2+n)_{b_2-b_1}}\\
\label{result}
&\quad\quad\quad \quad\times \sum_{k=0}^{b_2}\frac{b_2!}{k!(b_2-k)!}\frac{(b_1-k+1)_k(\Delta_1+b_1-h-k+1)_k}{(\Delta_2+s+n-b_1)_k(\Delta_1+s+n+b_1-k)_k}a(0,0).
\end{align}
It can then be checked that (\ref{result}) satisfies (\ref{rec1}) and (\ref{rec2}). 

The sum appearing in 
(\ref{result})
 can also be rewritten as
 \begin{align*}
&\sum_{k=0}^{b_2}\frac{b_2!}{k!(b_2-k)!}\frac{(b_1-k+1)_k(\Delta_1+b_1-h-k+1)_k}{(\Delta_2+s+n-b_1)_k(\Delta_1+s+n+b_1-k)_k}\\
& \quad\quad \quad= {}_3F_2(-b_1,-b_2,-(\Delta_1+b_1-h);-(\Delta_1+s+n+b_1-1),\Delta_2+s+n-b_1;1).
\end{align*}
Unfortunately, from particular examples one can see that the hypergeometric sum
entering here cannot be factorized into a product of Pochhammer symbols.

Let us note that one can solve (\ref{rec1}), (\ref{rec2}) differently: first solve for $a(0,b_2)$ 
from (\ref{rec2}) and then define $a(b_1,b_2)$ employing (\ref{rec1}). This leads to 
\begin{align}
\notag
& a(b_1,b_2)=\left(-\frac{1}{2}\right)^{b_1+b_2}\frac{n!}{b_1!b_2!b_{12}!}\\
\notag
&\quad\times
\frac{(\Delta_2+s+n)_{b_2}(\Delta_1+s+n-b_2)_{b_1}}{(\Delta_1+1-h)_{b_1}(\Delta_2+1-h)_{b_2}(\Delta_1+s_1+n)_{b_1-b_2}(\Delta_2+s_2+n)_{b_2-b_1}}\\
\label{resultalt}
&\quad\quad\quad \quad\times \sum_{k=0}^{b_1}\frac{b_1!}{k!(b_1-k)!}\frac{(b_2-k+1)_k(\Delta_2+b_2-h-k+1)_k}{(\Delta_1+s+n-b_2)_k(\Delta_2+s+n+b_2-k)_k}a(0,0).
\end{align}
The two expressions (\ref{result}) and (\ref{resultalt}) are not manifestly equal. However, 
from particular cases one
can see that they coincide. For example, from (\ref{result}) one finds 
\begin{equation}
\label{assym1}
a(1,1)=\frac{n(n-1)}{4}\frac{(\Delta_1+s+n)(\Delta_2+s+n-1)}{(\Delta_1+1-h)(\Delta_2+1-h)}
\left(1+\frac{(\Delta_1+1-h)}{(\Delta_2+s+n-1)(\Delta_1+s+n)}\right),
\end{equation} 
while  (\ref{resultalt}) entails 
\begin{equation}
\label{assym2}
a(1,1)=\frac{n(n-1)}{4}\frac{(\Delta_2+s+n)(\Delta_1+s+n-1)}{(\Delta_1+1-h)(\Delta_2+1-h)}
\left(1+\frac{(\Delta_2+1-h)}{(\Delta_1+s+n-1)(\Delta_2+s+n)}\right).
\end{equation} 
It is not hard to check that (\ref{assym1}) and (\ref{assym2}) are equal, despite it is not
manifest.

Eventually, combining $s$- and $b$-dependences we find as a final result
\begin{align}
\notag
&a_{n,s}(s_1,s_2;b_1,b_2,b_{12})=\frac{(-1)^{s_2+b_1+b_2}}{2^{b_1+b_2}}\frac{s!}{s_1!s_2!}\frac{(\Delta_1+s_1+2b_1+b_{12})_{s_2}}{(\Delta_2+2b_2+b_{12})_{s_2}}\\
\notag
&\quad\times \frac{n!}{b_1!b_2!b_{12}!}
\frac{(\Delta_1+s+n)_{b_1}(\Delta_2+s+n-b_1)_{b_2}}{(\Delta_1+1-h)_{b_1}(\Delta_2+1-h)_{b_2}(\Delta_1+s_1+n)_{b_1-b_2}(\Delta_2+s_2+n)_{b_2-b_1}}\\
\label{completedep}
&\qquad\quad \times  \sum_{k=0}^{b_2}\frac{b_2!}{k!(b_2-k)!}\frac{(b_1-k+1)_k(\Delta_1+b_1-h-k+1)_k}{(\Delta_2+s+n-b_1)_k(\Delta_1+s+n+b_1-k)_k}
a_{n,s}(s,0;0,0,n)\,,
\end{align}
where $a_{n,s}(s,0;0,0,n)$ is an arbitrary factor.

\section{Computation of OPE coefficients}
\label{appendix::cn0}
In this Appendix we use the explicit form of double trace operators ${\cal O}^{(2)}_{n,0}$
to compute the 3-point function 
\begin{equation}
\label{n03pt}
\langle {\cal O}(x){\cal O}(y) {\cal O}^{(2)}_{n,0}(z)\rangle,
\end{equation}
where the single-trace operator ${\cal O}=\,:\phi^a\phi_a:$ is of dimension $\Delta =d-2$ and the double-trace operator ${\cal O}^{(2)}_{n,0}$ is defined by (\ref{ansatz}) and
(\ref{completedep}). 
By the conformal symmetry the 3-point function of these operators is fixed to be
\begin{equation}
\label{n0symmetry}
\langle {\cal O}(x){\cal O}(y) {\cal O}^{(2)}_{n,0}(z)\rangle= C_{{\cal O}{\cal O}{\cal O}^{(2)}_{n,0}}\frac{\big(|x-y|^2\big)^{n}}{\big(|y-z|^2\big)^{\Delta+n}\big(|x-z|^2\big)^{\Delta+n}},
\end{equation}
where $C_{{\cal O}{\cal O}{\cal O}^{(2)}_{n,0}}$ remains to be found. We find $C_{{\cal O}{\cal O}{\cal O}^{(2)}_{n,0}}$
below by explicitly performing Wick contractions.

To fix the normalisation of ${\cal O}^{(2)}_{n,0}$  we compute the 2-point function
\begin{equation}
\label{n02pt}
\langle{\cal O}^{(2)}_{n,0}(x){\cal O}^{(2)}_{n,0}(y)\rangle.
\end{equation}
Again, it is fixed by the conformal symmetry to be
\begin{equation}
\label{n02ptsymm}
\langle{\cal O}^{(2)}_{n,0}(x){\cal O}^{(2)}_{n,0}(y)\rangle=\frac{C_{{\cal O}^{(2)}_{n,0}}}{\big(|x-y|^2\big)^{2\Delta+2n}}.
\end{equation}
After having computed both $C_{{\cal O}{\cal O}{\cal O}^{(2)}_{n,0}}$ and $C_{{\cal O}^{(2)}_{n,0}}$ we will obtain
the coefficient 
\begin{equation}
\label{confbldecno}
c^2_{n,0}=\frac{1}{ 4N^2}\frac{C^2_{{\cal O}{\cal O}{\cal O}^{(2)}_{n,0}}}{C_{{\cal O}^{(2)}_{n,0}}}
\end{equation}
of the conformal block decomposition of $\langle {\cal O}{\cal O}{\cal O}{\cal O}\rangle$ 
corresponding to a conformal block
with spin zero and dimension $2\Delta+2n$.

\subsection{Three-point function}

Performing Wick contractions we find
\begin{align}
\notag
&\frac{1}{4N^2}\langle {\cal O}(x){\cal O}(y){\cal O}^{(2)}_{n,0}(z)\rangle\\
\notag
& \quad =\frac{1}{4N^2}\sum_{b_1,b_2=0}^{b_1+b_2=n} a(b_1,b_2)\langle :\phi^a(x)\phi_a(x):\,:\phi^b(y)\phi_b(y): 
\\
\notag
&\qquad\qquad\qquad\qquad\qquad : \Box^{b_1}\partial^{i(n-b_1-b_2)}\phi^c(z)\phi_c(z)\Box^{b_2}\partial_{i(n-b_1-b_2)}\phi^d(z)\phi_d(z):\rangle\\
\notag
&\quad =\sum_{b_1,b_2=0}^{b_1+b_2=n} a(b_1,b_2)\Big( \Box^{b_1}\partial^{i(n-b_1-b_2)}[D^2(x-z)]\Box^{b_2} \partial_{i(n-b_1-b_2)}
[D^2(y-z)]\\
\notag
&\qquad\qquad +\frac{2}{N} \Box^{b_1} \partial^{i(n-b_1-b_2)}[D(x-z)D(y-z)]\Box^{b_2}\partial_{i(n-b_1-b_2)
}[D(x-z)D(y-z)]\\
\label{3ptn01}
&\qquad \qquad \qquad \qquad \qquad \qquad \qquad \qquad \qquad \qquad 
\qquad \qquad \qquad \qquad\qquad\quad  +(x\leftrightarrow y) \Big),
\end{align}
where
\begin{equation*}
D(x-y)=\frac{1}{|x-y|^{\Delta}}
\end{equation*}
and all differential operators act on variable $z$.
Here we added $1/4N^2$ to take into account the normalisation of ${\cal O}$, as in (\ref{confbldecno}).
The first line on the RHS of (\ref{3ptn01})
 and its
partner under $x\leftrightarrow y$  gives the $O(1)$ contribution to $c_{n,0}$. The
remaining terms produce contributions of order $O(1/N)$.

\subsubsection{Disconnected part of the 3-point function}
Let us first focus on the order $O(1)$ contribution
\begin{equation}
\label{n0o1}
\sum_{b_1,b_2=0}^{b_1+b_2=n} a(b_1,b_2) \Box^{b_1}\partial^{i(n-b_1-b_2)}[D^2(x-z)]\Box^{b_2} \partial_{i(n-b_1-b_2)}
[D^2(y-z)].
\end{equation}
The general formula (\ref{n0symmetry}) requires that the result should contain
$(|x-y|^2)^{n}$. The only way that $|x-y|^2$ can appear from (\ref{n0o1}) is
when derivatives acting on $D^2(x-z)$ and $D^2(y-z)$ are contracted with
each other. Indeed, this produces $(x-z)\cdot(y-z)$, which in turn gives
\begin{equation}
\label{howtoxminusy}
(z-x)\cdot(z-y)=\frac{1}{2}\left[(z-x)^2+(z-y)^2-(x-y)^2\right].
\end{equation}
With this observation  it is not hard to see that the only 
term that produces $(|x-y|^2)^{n}$ required by (\ref{n0symmetry}) is the one with $b_1=0$
and $b_2=0$. From now on we will focus only on terms that are capable to produce
 $(|x-y|^2)^{n}$. The remaining terms will not be written explicitly.
 One finds
\begin{align}
\notag
\partial^{i(n)}\left(\frac{1}{|x-z|^{2\Delta}}\right) &\partial_{i(n)}\left(\frac{1}{|y-z|^{2\Delta}}\right)\\
\notag
&=
2^n(\Delta)_n\left(\frac{(z-x)^{i(n)}}{|x-z|^{2\Delta+2n}}+\dots \right)
2^n(\Delta)_n\left(\frac{(z-y)_{i(n)}}{|y-z|^{2\Delta+2n}}+\dots \right)\\
\notag
&=\left[2^n (\Delta)_n\right]^2\left(\frac{\big((z-x)\cdot(z-y)\big)^n}{|x-z|^{2\Delta+2n}|y-z|^{2\Delta+2n}}+\dots\right)\\
\label{3ptn0o1}
&=\left[2^n (\Delta)_n\right]^2\left(-\frac{1}{2}\right)^n\frac{\left((x-y)^2\right)^n}{\left((x-z)^2\right)^{\Delta+2n}\left((y-z)^2\right)^{\Delta+2n}}+\dots\,.
\end{align}
Combining this with its $(x\leftrightarrow y)$ partner just doubles the result.
So, for
the $O(1)$ part of the 3-point function we find
\begin{equation}
\label{evaluatedn03pt}
\frac{1}{4N^2}C_{{\cal O}{\cal O}{\cal O}^{(2)}_{n,0}}=
(-2)^n2\left[ (\Delta)_n\right]^2+O(1/N),
\end{equation}
where we also set $a(0,0)=1$.

\subsubsection{Connected part of the 3-point function}

Here we compute the $O(1/N)$ part of the 3-point function (\ref{3ptn01}). As previously,
we will focus only on terms with the highest power of $(x-z)\cdot(y-z)$, because
this is the only way one can produce $(|x-y|^2)^{n}$ in the numerator. We find that
\begin{align}\notag
& \Box^{b_1} \partial^{i(n-b_1-b_2)}[D(x-z)D(y-z)]\Box^{b_2}\partial_{i(n-b_1-b_2)
}[D(x-z)D(y-z)]\\
\notag
&=2^{b_1+b_2}\partial^{i(n-b_1-b_2)}\left[\partial^{k(b_1)}\frac{1}{|x-z|^\Delta}
\partial_{k(b_1)}\frac{1}{|y-z|^\Delta}\right]
\notag
\partial_{i(n-b_1-b_2)}\left[\partial^{l(b_2)}\frac{1}{|x-z|^\Delta}
\partial_{l(b_2)}\frac{1}{|y-z|^\Delta}\right]\\
\notag
&=2^{b_1+b_2}\partial^{i(n-b_1-b_2)}\left[\left[2^{b_1}\left(\Delta/2\right)_{b_1}\right]^2
\frac{\left[(z-x)\cdot (z-y)\right]^{b_1}}{|x-z|^{\Delta+2b_1}|y-z|^{\Delta+2b_1}}\right]\\
\notag
& \qquad \qquad\qquad\qquad\times 
\partial_{i(n-b_1-b_2)}\left[\left[2^{b_2}\left({\Delta}/{2}\right)_{b_2}\right]^2
\frac{\left[(z-x)\cdot (z-y)\right]^{b_2}}{|x-z|^{\Delta+2b_2}|y-z|^{\Delta+2b_2}}\right]+\dots\\
\notag
&=2^{3(b_1+b_2)}\left[\left({\Delta}/{2}\right)_{b_1}\right]^2\left[\left({\Delta}/{2}\right)_{b_2}\right]^2\left[(z-x)\cdot (z-y)\right]^{b_1+b_2}\\
\label{interm1}
&\times\partial^{i(n-b_1-b_2)} \left(\frac{1}{|x-z|^{\Delta+2b_1}|y-z|^{\Delta+2b_1}}\right)
\partial_{i(n-b_1-b_2)} \left(\frac{1}{|x-z|^{\Delta+2b_2}|y-z|^{\Delta+2b_2}}\right)+\dots.
\end{align}

Let us further simplify the last line
\begin{align}
\notag
\tilde{\cal B}\equiv \partial^{m(n-b_1-b_2)} &\left(\frac{1}{|x-z|^{\Delta+2b_1}|y-z|^{\Delta+2b_1}}\right)
\partial_{m(n-b_1-b_2)} \left(\frac{1}{|x-z|^{\Delta+2b_2}|y-z|^{\Delta+2b_2}}\right)\\
\notag
&= \sum_{t=0}^{n-b_1-b_2}\frac{(n-b_1-b_2)!}{t!(n-b_1-b_2-t)!} \partial^{m(t)}\frac{1}{|x-z|^{\Delta+2b_1}}
\partial^{m(n-b_1-b_2-t)} \frac{1}{|y-z|^{\Delta+2b_1}} \\
\notag
&\times
\sum_{r=0}^{n-b_1-b_2}\frac{(n-b_1-b_2)!}{r!(n-b_1-b_2-r)!} \partial_{m(n-b_1-b_2-r)}\frac{1}{|x-z|^{\Delta+2b_1}}
\partial_{m(r)} \frac{1}{|y-z|^{\Delta+2b_1}}.
\end{align}
As we are interested only in terms with the highest power of $(x-z)\cdot(y-z)$, from the
last line we should only keep the terms where derivatives acting on $|x-z|$ 
are contracted with those acting on $|y-z|$. First of all it requires that $t=r$. So we find
\begin{align}
\notag
\tilde{\cal B}=\sum_{t=0}^{n-b_1-b_2}\left(\frac{(n-b_1-b_2)!}{t!(n-b_1-b_2-t)!} \right)^2&\partial^{m(t)}\frac{1}{|x-z|^{\Delta+2b_1}}
\partial^{m(n-b_1-b_2-t)} \frac{1}{|y-z|^{\Delta+2b_1}} \\
\notag
&\times
 \partial_{m(n-b_1-b_2-t)}\frac{1}{|x-z|^{\Delta+2b_1}}
\partial_{m(t)} \frac{1}{|y-z|^{\Delta+2b_1}}+\dots.
\end{align}
According to our notations, in the last expression all $n-b_1-b_2$ derivatives with upper indices 
and all $n-b_1-b_2$ derivatives with lower indices are symmetrised before contraction. Keeping
only the terms that produce the highest power of $(x-z)\cdot (y-z)$ we obtain
\begin{align}
\notag
\tilde{\cal B}=\sum_{t=0}^{n-b_1-b_2}\frac{(n-b_1-b_2)!}{t!(n-b_1-b_2-t)!}& \partial^{m(t)}\frac{1}{|x-z|^{\Delta+2b_1}}
\partial_{m(t)} \frac{1}{|y-z|^{\Delta+2b_1}}
 \\
 \notag
\times
\partial^{r(n-b_1-b_2-t)} &\frac{1}{|y-z|^{\Delta+2b_1}} \partial_{r(n-b_1-b_2-t)}\frac{1}{|x-z|^{\Delta+2b_1}}+\dots .
\end{align}

Next we proceed by evaluating the derivatives
\begin{align*}
\tilde{\cal B} &= \sum_{t=0}^{n-b_1-b_2}\frac{(n-b_1-b_2)!}{t!(n-b_1-b_2-t)!}
\left[2^t \left({\Delta}/{2}+b_1\right)_t\right]
 \left[2^t \left({\Delta}/{2}+b_2\right)_t\right] \\
 & \qquad \times\left[2^{n-b_1-b_2-t} \left({\Delta}/{2}+b_1\right)_{n-b_1-b_2-t}\right]
 \left[2^{n-b_1-b_2-t} \left({\Delta}/{2}+b_2\right)_{n-b_1-b_2-t}\right]\\
 & \qquad
 \times\frac{\left[(z-x)\cdot (z-y)\right]^t}{|x-z|^{\Delta+2b_1+2t}|y-z|^{\Delta+2b_2+2t}} \frac{\left[(z-x)\cdot (z-y)\right]^{n-b_1-b_2-t}}{|x-z|^{\Delta+2n-2b_1-2t}|y-z|^{\Delta+2n-2b_1-2t}}+\dots\\
 &= 4^{b_{12}}\sum_{t=0}^{b_{12}}\frac{b_{12}!}{t!(b_{12}-t)!}  \left({\Delta}/{2}+b_1\right)_t
 \left({\Delta}/{2}+b_2\right)_t\\
 & \qquad\qquad\qquad\qquad \times  \left({\Delta}/{2}+b_1\right)_{b_{12}-t}
 \left({\Delta}/{2}+b_2\right)_{b_{12}-t}
 \frac{\left[(z-x)\cdot (z-y)\right]^n}{|x-z|^{2\Delta+2n}|y-z|^{2\Delta+2n}}+\dots.
\end{align*}

Replacing $(z-x)\cdot(z-y)$ with $(x-y)^2$ according to (\ref{howtoxminusy}) we find
\begin{align}
\notag
&\Box^{b_1} \partial^{m(n-b_1-b_2)}[D(x-z)D(y-z)]\Box^{b_2}\partial_{m(n-b_1-b_2)
}[D(x-z)D(y-z)]\\
\notag
&\qquad=(-1)^n 2^{-b_{12}}4^{b_1+b_2}\left[\left({\Delta}/{2}\right)_{b_1}\right]^2\left[\left({\Delta}/{2}\right)_{b_2}\right]^2 \sum_{t=0}^{b_{12}}\frac{4^{b_{12}}b_{12}!}{t!(b_{12}-t)!}  \left({\Delta}/{2}+b_1\right)_t
 \left({\Delta}/{2}+b_2\right)_t\\
 \label{interm2}
 & \qquad\qquad\qquad\times \left({\Delta}/{2}+b_1\right)_{b_{12}-t}
 \left({\Delta}/{2}+b_2\right)_{b_{12}-t}
 \frac{\left[(z-x)\cdot (z-y)\right]^n}{|x-z|^{2\Delta+2n}|y-z|^{2\Delta+2n}}+\dots.
\end{align}
Eq. (\ref{interm2}) specifies the contribution of a single term with fixed $b_1$ and $b_2$ 
to the complete 3-point function (\ref{3ptn01}). 
Summing over $b_1$ and $b_2$ we encounter the following
expression
\begin{align}
\notag
{\cal B}&\equiv (-1)^n \sum_{b_1,b_2=0}^{b_1+b_2=n} a(b_1,b_2)2^{-b_{12}}4^{b_1+b_2}\left[\left({\Delta}/{2}\right)_{b_1}\right]^2\left[\left({\Delta}/{2}\right)_{b_2}\right]^2\\
\notag
&\times  \sum_{t=0}^{b_{12}}\frac{4^{b_{12}}b_{12}!}{t!(b_{12}-t)!}  \left({\Delta}/{2}+b_1\right)_t
 \left({\Delta}/{2}+b_2\right)_t\left({\Delta}/{2}+b_1\right)_{b_{12}-t}
 \left({\Delta}/{2}+b_2\right)_{b_{12}-t},
\end{align} 
where $a(b_1,b_2)$ is given by (\ref{completedep}) for $s=0$. Performing this sum
 with Mathematica for $\Delta=d-2$ and $n=0,1,2,3$ we conjecture that
\begin{equation*}
{\cal B} = 2^n \left(\Delta\right)_n (\Delta/2)_na(0,0).
\end{equation*}
Restoring the $2/N$ factor from (\ref{3ptn01}) and adding the $x\leftrightarrow y$ piece
 for the complete $O(1/N)$ part of the 3-point function we find
\begin{equation}
\label{evaluatedn-13pt}
\frac{1}{4N^2}C_{{\cal O}{\cal O}{\cal O}^{(2)}_{n,0}}=
\frac{4}{N} {\cal B}+O(1)=\frac{4}{N}2^n \left(\Delta\right)_n (\Delta/2)_n +O(1),
\end{equation}
where we set $a(0,0)=1$.

Finally, combining with the $O(1)$ contribution (\ref{evaluatedn03pt}) we obtain
\begin{equation}
\label{evaluatedfull3pt}
\frac{1}{4N^2}C_{{\cal O}{\cal O}{\cal O}^{(2)}_{n,0}}=
2(-2)^n\left[ (\Delta)_n\right]^2+\frac{4}{N}2^n \left(\Delta\right)_n (\Delta/2)_n.
\end{equation}

\subsection{Two-point function}
In this Section we evaluate the $O(1)$ part of the 2-point function (\ref{n02ptsymm}). This gives
\begin{align}
\notag
&\langle{\cal O}^{(2)}_{n,0}(x){\cal O}^{(2)}_{n,0}(y)\rangle\\
\notag
&\qquad\qquad=
\langle : \sum_{b_1,b_2=0}^{b_1+b_2=n}a(b_1,b_2)\Box^{b_1}\partial^{m(n-b_1-b_2)}\phi^a(x)\phi_a(x)\Box^{b_2}\partial_{m(n-b_1-b_2)}\phi^b(x)\phi_b(x): \\
\notag
&\qquad\qquad\qquad\sum_{b'_1,b'_2=0}^{b'_1+b'_2=n} a(b'_1,b'_2): \Box^{b_1}\partial^{m(n-b_1-b_2)}\phi^c(y)\phi_c(y)\Box^{b_2}\partial_{m(n-b_1-b_2)}\phi^d(y)\phi_d(y):\rangle\\
\notag
&\qquad\qquad=8N^2 \sum_{b_1,b_2=0}^{b_1+b_2=n}\sum_{b'_1,b'_2=0}^{b'_1+b'_2=n}a(b_1,b_2)a(b'_1,b'_2)
\Box_x^{b_1} \partial_x^{m(b_{12})}\Box_y^{b'_1}\partial_y^{n(b'_{12})}D^2(x-y)\\
\notag
&\qquad\qquad\qquad\qquad\qquad\qquad\qquad\qquad\qquad\times \Box_x^{b_2} \partial^x_{m(b_{12})}\Box_y^{b'_2}\partial^y_{n(b'_{12})}D^2(x-y)+O(1/N)\\
\notag
&\qquad\qquad=8N^2 \sum_{b_1,b_2=0}^{b_1+b_2=n}\sum_{b'_1,b'_2=0}^{b'_1+b'_2=n}a(b_1,b_2)a(b'_1,b'_2)
\partial_x^{m(b_{12}+b'_{12})}\Box_x^{b_1+b'_1} D^2(x-y)\\
\label{2ptn0simpl}
&\qquad \qquad\qquad\qquad\qquad\qquad\qquad\qquad\qquad\times \partial^x_{m(b_{12}+b'_{12})}\Box_x^{b_2+b'_2} D^2(x-y)+O(1/N).
\end{align}
Here we supplied differential operators with extra indices $x$ and $y$, which specify on
which variable these differential operators act upon. Then we eliminated differential operators
acting on $y$ using that $\Box_x=\Box_y$ and $\partial_x=-\partial_y$. For brevity, we omit
the index $x$ in computations below.

Let us focus on one term with fixed $b_1$, $b_2$, $b'_1$ and $b'_2$. The Laplacians can
be evaluated using that
\begin{equation*}
\Box^{n}r^{k}= 2^{2n}(k/2-n+1)_{n}(d/2-n+k/2)_{n}r^{k-2n}.
\end{equation*}
We find 
\begin{align}
\notag
\tilde{\cal C}&\equiv \partial_x^{m(b_{12}+b'_{12})}\Box_x^{b_1+b'_1} D^2(x-y)\partial^x_{m(b_{12}+b'_{12})}\Box_x^{b_2+b'_2} D^2(x-y)\\
\notag
&=2^{2(b_1+b'_1+b_2+b'_2)}\left(\Delta\right)_{b_1+b_2}(\Delta+1-d/2)_{b_1+b_2}\left(\Delta\right)_{b_2+b'_2}
(\Delta+1-d/2)_{b_2+b'_2}\\
\label{ctilde}
&\qquad\qquad \times 
\partial^{m(b_{12}+b'_{12})}\left(\frac{1}{|x-y|^{2\Delta+2b_1+2b'_1}}\right)
\partial_{m(b_{12}+b'_{12})}\left(\frac{1}{|x-y|^{2\Delta+2b_2+2b'_2}}\right).
\end{align}

The last contraction is of the form 
\begin{equation*}
\partial_{i(n)}\frac{1}{r^{\Delta_1}}\partial^{i(n)}\frac{1}{r^{\Delta_2}}.
\end{equation*}
To evaluate it we use
\begin{equation}
\label{d1r}
\partial_{j(n)}\frac{1}{r^{\Delta_1}}=\frac{1}{r^{\Delta_1+2n}}\sum_{m=0}^{[n/2]} \frac{n! (-1)^{n-m}}{(n-2m)! 2^m m!}
2^{n-m}(\Delta_1/2)_{n-m} x_{j(n-2m)} (\eta_{jj})^m.
\end{equation}
Next, let us define a function $\chi$ that computes the contractions of $x$'s and $\eta$'s
\begin{align}
\notag
&\chi(i,k,l)\cdot(r^2)^{l+k-i}\equiv x^{j(l+2k-2i)} (\eta^{jj})^i (\eta_{jj})^k x_{j(l)}=\frac{(r^2)^{l+k-i}}{(2k+l-2i+1)_{2i}}\\
\label{chi}
&\quad \times \sum_{m=0}^i 4^{i-m} \frac{i!}{m! (i-m)!}(k-i+m+1)_{i-m}(d/2+k+l-i)_{i-m} (l-2m+1)_{2m}.
\end{align}
The last equality can be obtained, for example, from (D.3) in \cite{Bekaert:2014cea}.

Let us introduce another function $\psi(\Delta_1,\Delta_2,n)$ 
\begin{align*}
\psi(\Delta_1,\Delta_2,n) &\equiv \sum_{m=0}^{[n/2]} \frac{n! (-1)^{n-m}}{(n-2m)! 2^m m!}
2^{n-m}(\Delta_1/2)_{n-m}\\
& \qquad\qquad\times\sum_{k=0}^{[n/2]} \frac{n! (-1)^{n-k}}{(n-2k)! 2^k k!}
2^{n-k}(\Delta_2/2)_{n-k}\chi(m,n-2k,k).
\end{align*}
It is not hard to see
from  (\ref{d1r}) and (\ref{chi}) that we introduced $\psi(\Delta_1,\Delta_2,n)$ so that 
\begin{equation}
\partial_{i(n)}\frac{1}{r^{\Delta_1}}\partial^{i(n)}\frac{1}{r^{\Delta_2}}=
\psi(\Delta_1,\Delta_2,n)\frac{1}{r^{\Delta_1+\Delta_2+2n}}.
\end{equation}

In these terms $\tilde {\cal C}$ (\ref{ctilde}) can be rewritten as
\begin{align}
\notag
\tilde {\cal C} = 2^{2(b_1+b'_1+b_2+b'_2)}\left(\Delta\right)_{b_1+b_2}(\Delta+1-d/2)_{b_1+b_2}\left(\Delta\right)_{b_2+b'_2}
(\Delta+1-d/2)_{b_2+b'_2}\\
\label{ctilde1}
\times \frac{\psi(2\Delta+2b_1+2b'_1,2\Delta+2b_2+2b'_2,b_{12}+b'_{12})}{|x-y|^{4\Delta+4n}}.
\end{align}
This formula specifies a contribution to $C_{{\cal O}^{(2)}_{n,0}}$ from a single term in (\ref{2ptn0simpl})
with fixed $b_1$, $b_2$, $b'_1$ and $b'_2$. The complete contribution requires to perform
the following summation
\begin{align}
\notag
{\cal C}\equiv \sum_{b_1,b_2=0}^{b_1+b_2=n}\sum_{b'_1,b'_2=0}^{b'_1+b'_2=n}a(b_1,b_2)a(b'_1,b'_2)
\psi(2\Delta+2b_1+2b'_1,2\Delta+2b_2+2b'_2,b_{12}+b'_{12})\\
\notag
\times 2^{2(b_1+b'_1+b_2+b'_2)}\left(\Delta\right)_{b_1+b_2}(\Delta+1-d/2)_{b_1+b_2}\left(\Delta\right)_{b_2+b'_2}
(\Delta+1-d/2)_{b_2+b'_2},
\end{align}
where $a(b_1,b_2)$ is given by (\ref{completedep}) with $s=0$. Evaluating this sum
with Mathematica for $n=1,2,3,4$ we conjecture that
\begin{equation}
\notag
{\cal C} = 2^{2n} n! \left[(\Delta)_n\right]^2(d/2)_n\frac{(2\Delta-d/2+n)_n (2\Delta-d+1+n)_n}{\left[(\Delta-d/2+1)_n\right]^2},
\end{equation}
where $a(0,0)$ was set to one.
Thus, the $O(N^2)$ part of the two point function is given by
\begin{equation}
C_{{\cal O}^{(2)}_{n,0}}= 8N^2{\cal C}+ O(N).
\end{equation}

\subsection{Conformal block coefficient}

Let us summarise the results of our computation
\begin{align*}
C_{{\cal O}{\cal O}{\cal O}^{(2)}_{n,0}}&=8N^2{\cal A}+8N^2 \cdot \frac{2}{N}{\cal B},\\
C_{{\cal O}^{(2)}_{n,0}}&=8N^2{\cal C}+O(N),
\end{align*}
where 
\begin{align*}
{\cal A}&=(-2)^n\left[ (\Delta)_n\right]^2,\\
{\cal B}&= 2^n \left(\Delta\right)_n (\Delta/2)_n,\\
{\cal C}&=2^{2n} n! \left[(\Delta)_n\right]^2(d/2)_n\frac{(2\Delta-d/2+n)_n (2\Delta-d+1+n)_n}{\left[(\Delta-d/2+1)_n\right]^2}.
\end{align*}

Having found the 3- and the 2-point functions (\ref{n03pt}), (\ref{n02pt}) we can extract the
conformal block coefficient $c_{n,0}$ according to (\ref{confbldecno}). The $O(N)$ part of $C_{{\cal O}^{(2)}_{n,0}}$, which we did not evaluate explicitly
can be extracted from the requirement that $c_{n,0}$ 
contains only terms of orders $O(1)$ and $O(1/N)$. This gives
\begin{align}
\notag
c^2_{n,0} &= 2\frac{{\cal A}^2}{{\cal C}}\left[1+\frac{1}{N} \frac{2{\cal B}}{{\cal A}}\right]\\
\label{completen0}
&=2\frac{\left[(\Delta)_n\right]^2 \left[(1+\Delta-d/2)_n\right]^2}{n!(d/2)_n(2\Delta-d/2+n)_n(2\Delta-d+1+n)_n}\left[1+(-1)^n\frac{2}{N} \frac{(\Delta/2)_n}{(\Delta)_n}\right]\\ \nonumber
&=2\frac{\left[(d-2)_n\right]^2 \left[(\tfrac{d}{2}-1)_n\right]^2}{n!(d/2)_n(\tfrac{3d}{2}-4+n)_n(d-3+n)_n}\left[1+(-1)^n\frac{2}{N} \frac{(\tfrac{d}{2}-1)_n}{(d-2)_n}\right],
\end{align}
where in the final equality we inserted $\Delta = d-2$. This agrees with the result of \cite{Dolan:2000ut} for $d=4$ and $\Delta=2$, as does the analogous calculation of the OPE coefficients for $s$ fixed and $n = 0, 1$. For the latter we obtain
\begin{align}\label{complete0s}
&c^2_{0, s} = \frac{\left[\left(-1\right)^s+1\right] 2^{s} \left(d-2\right)^2_{s}}{s!  \left(2d +s-5\right)_{s}} \left(1+\frac{4}{N} \frac{\Gamma\left(s\right)}{2^{s}\Gamma\left(\frac{s}{2}\right)} \frac{\left(\frac{d}{2}-1\right)_{\tfrac{s}{2}}}{\left(\frac{d-1}{2}\right)_{\tfrac{s}{2}}\left(d-2\right)_{\tfrac{s}{2}}}\right),
\end{align}
and 
\begin{align} \label{complete1s}
&c^2_{1, s} = \frac{\left[\left(-1\right)^s+1\right] 2^{s-2}\left(d-2\right) \left(d-2\right)^2_{s+1}}{s! \left(s+\tfrac{d}{2}\right) \left(2d +s-3\right)_{s} \left(\tfrac{3d}{2}-3 +s\right)} \left(1-\frac{4}{N} \frac{\Gamma\left(s\right)}{2^{s}\Gamma\left(\frac{s}{2}\right)} \frac{\left(\frac{d}{2}-1\right)_{1+\tfrac{s}{2}}}{\left(\frac{d-1}{2}\right)_{\tfrac{s}{2}}\left(d-2\right)_{1+\tfrac{s}{2}}}\right).
\end{align}
We have not displayed the lengthy derivation of the above for concision, since they were obtained in the same way as for the $s=0$, fixed $n$ calculation above. Based on the explicit results \eqref{completen0}, \eqref{complete0s} and \eqref{complete1s}, a conjecture for the form for general $n$ and $s$ follows quite naturally:
\begin{align}
&c^2_{n, s} = \frac{\left[\left(-1\right)^s+1\right] 2^{s}\left(\tfrac{d}{2}-1\right)^{2}_n \left(d-2\right)^2_{s+n}}{s! n! \left(s+\tfrac{d}{2}\right)_n \left(d-3 + n \right)_n \left(2d + 2n +s-5\right)_{s} \left(\tfrac{3d}{2}-4 + n +s\right)_n} \\ \nonumber 
&\hspace*{5cm} \times \left(1+\left(-1\right)^n\frac{4}{N} \frac{\Gamma\left(s\right)}{2^{s}\Gamma\left(\frac{s}{2}\right)} \frac{\left(\frac{d}{2}-1\right)_{n+\tfrac{s}{2}}}{\left(\frac{d-1}{2}\right)_{\tfrac{s}{2}}\left(d-2\right)_{n+\tfrac{s}{2}}}\right),
\end{align}
which agrees with the results available in $d = 4$  \cite{Dolan:2000ut}, and for $N=\infty$ in general dimensions \cite{Fitzpatrick:2011dm}.

\providecommand{\href}[2]{#2}\begingroup\raggedright\endgroup

\end{document}